\documentclass[11pt, a4paper, logo, nonumbering]{deepmind}

\usepackage[bibstyle=numeric,citestyle=authoryear,sorting=nyt,natbib=true,uniquename=false]{biblatex}
\usepackage{blindtext}
\usepackage{verbatim}
\usepackage{xcolor}
\usepackage{graphicx}
\usepackage{subfig}
\usepackage{amsmath}
\usepackage{amssymb}
\usepackage{placeins}

\reportnumber{} % Leave blank if n/a

\renewcommand{\today}{}

\usepackage{multirow, array}
\newcommand{\R}{\mathbb{R}}

\newcommand{\oldVersion}{\textbf{[64]}}

\title{Modeling human reputation-seeking behavior in a spatio-temporally complex public good provision game}
\author[a]{Edward Hughes}
\author[a]{Tina O. Zhu}
\author[a]{Martin J. Chadwick}
\author[a]{Raphael Koster}
\author[a]{Antonio Garc{\'\i}a Casta{\~n}eda}
\author[a]{Charles Beattie}
\author[a,b]{Thore Graepel}
\author[a,c]{Matthew M. Botvinick}
\author[a]{Joel Z. Leibo}

\affil[a]{Google DeepMind, London, UK}
\affil[b]{Department of Computer Science, UCL, London, UK}
\affil[c]{Gatsby Computational Neuroscience Unit, UCL, London, UK}

\correspondingauthor{Some of this work was previously published on arxiv as \oldVersion. It reflects the interpretation of the underlying data favored by the present list of authors; it differs from the interpretation given to the same data in \oldVersion. It should be noted that all the experimental methods, statistical procedures, and quantitative results are identical to those in \oldVersion, and their descriptions and visualizations in figures have been reproduced here with permission from the authors of \oldVersion. \nocite{mckee2021multi}}

\begin{abstract}
Multi-agent reinforcement learning algorithms are useful for simulating social behavior in settings that are too complex for other theoretical approaches like game theory. However, they have not yet been empirically supported by laboratory experiments with real human participants. In this work we demonstrate how multi-agent reinforcement learning can model group behavior in a spatially and temporally complex public good provision game called Clean Up. We show that human groups succeed in Clean Up when they can see who is who and track reputations over time but fail under conditions of anonymity. A new multi-agent reinforcement learning model of reputation-based cooperation demonstrates the same difference between identifiable and anonymous conditions. Furthermore, both human groups and artificial agent groups solve the problem via turn-taking despite other options being available. Our results highlight the benefits of using multi-agent reinforcement learning to model human social behavior in complex environments.
\end{abstract}

\begin{document}

\maketitle
% \thispagestyle{firststyle}
% \ifthenelse{\boolean{shortarticle}}{\ifthenelse{\boolean{singlecolumn}}{\abscontentformatted}{\abscontent}}{}

\section{Significance}
% Please add a significance statement to explain the relevance of your work
Multi-agent reinforcement learning algorithms produce autonomous agents that can operate in complex virtual environments. This may be useful for simulating real-world social ecological systems. However, it is not known whether such artificial agents behave the same way humans would in the same settings. In this paper, we show that once endowed with an intrinsic motivation to build a good reputation, agents cooperate in a complex social dilemma similarly to humans. We also show that anonymity destroys cooperation for both human and artificial agents in this setting.

%-------------------------------------------------------------------------
\section{Introduction} 

Computational modeling methods based on multi-agent reinforcement learning (MARL) algorithms show great promise for their ability to advance our computational and theoretical understanding of cooperation \citep{wong2022deep}. There are two main reasons. First, MARL methods work in situations where there is substantial spatio-temporal complexity (e.g.~\citet{leibo2017multi} and~\citet{jaderberg2019human}). Second, MARL methods may capture phenomena related to learning and practice \citep{koster2022spurious}. Competing methodologies lack one or the other or both of these properties. Both classical and evolutionary game theory methods usually lack a formal concept of the environment in which the interactions take place and thus lack a way to scale up the spatio-temporal complexity of the situations they model \citep{luce1957games}. Evolutionary game theory features population-level changes in strategy over time, usually interpreted as arising from individuals doing payoff-biased imitation \citep{weibull1997evolutionary}, but does not incorporate processes that resemble individual-level trial and error, and thus cannot capture effects of practice as MARL does. Another approach, agent-based modeling with theory-based agent programs, can be used in situations with substantial spatio-temporal complexity \citep{lansing2005cooperation, janssen2006empirically, tesfatsion2021}. Some agent-based models feature adaptive agents and thus can capture some learning-related phenomena \citep{schill2019more}. However, MARL tends to scale better than theory-based agents in both the behavioral complexity it can achieve and the environmental complexity where it can operate.

\begin{figure*}[t]
    \centering
    \subfloat[(a)]{\includegraphics[height=5.65cm]{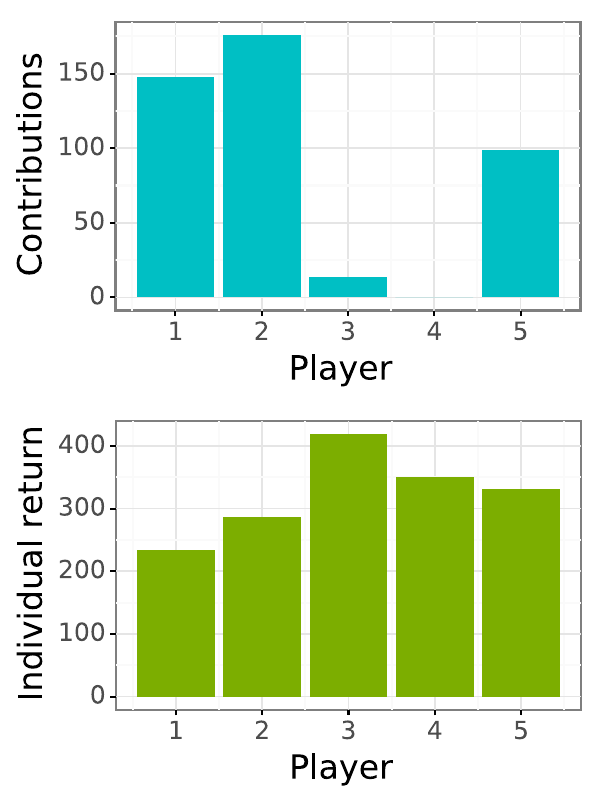}}
    ~ 
    \subfloat[(b)]{\includegraphics[height=5cm]{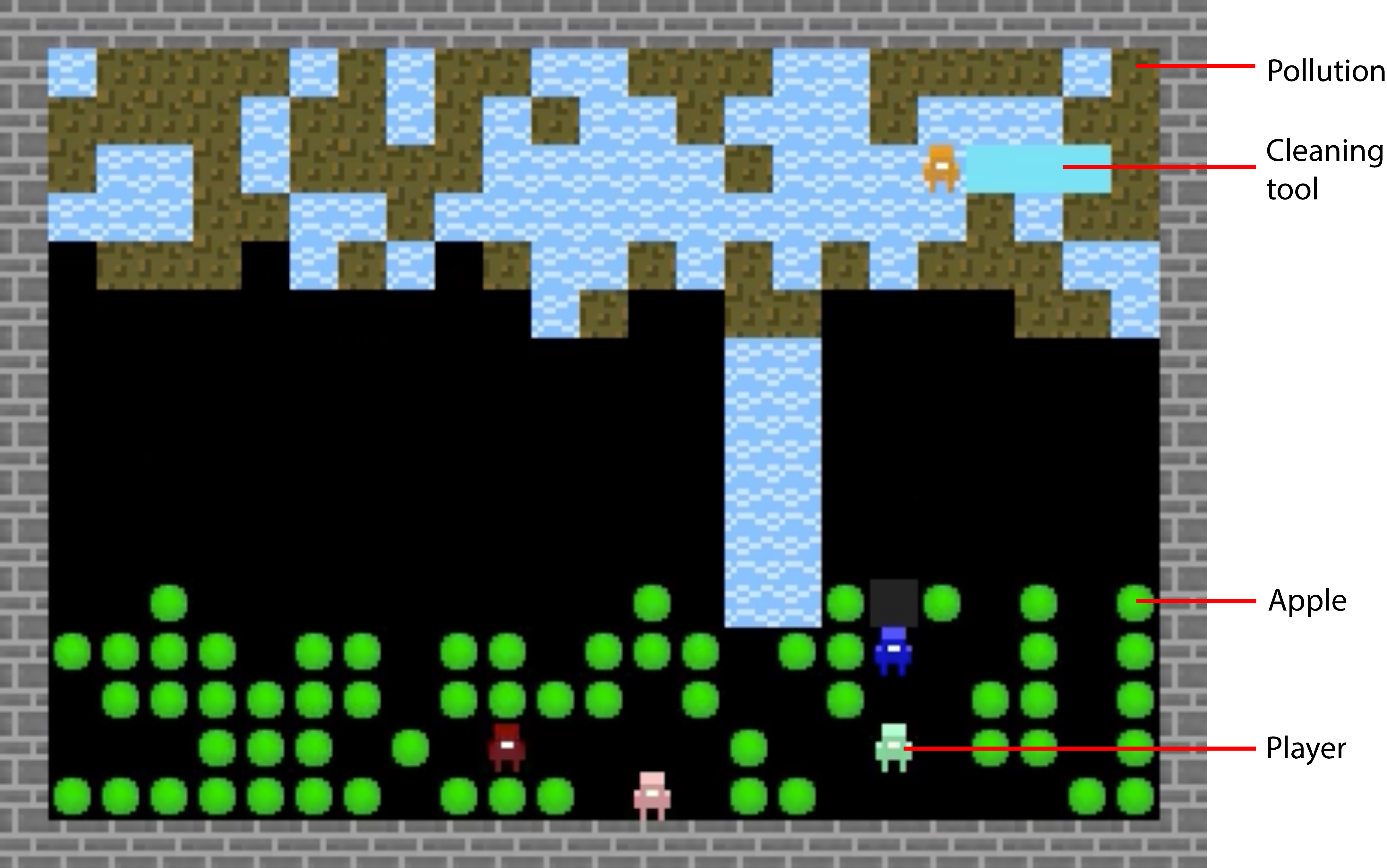}} \\
    ~
    %\subfloat[(c)]{\includegraphics[height=4.25cm]{main_figures/timecourse_human.pdf}}
    \caption{\small Figure \ref{fig:environment}: Clean Up is a spatio-temporally dynamic public good provision task which allows for the exploration of distributional, spatial, and temporal dynamics of collective action (i.e., \textit{how much}, \textit{where}, and \textit{when} individuals contribute to group efforts). (a) Variations in group member behavior and rewards emerge in each episode of Clean Up (sample episode from human data). Because of the spatial separation between the river and the orchard, there is a general tradeoff between contributions and individual return. (b) On each time step, group members ($n = 5)$ take actions in a shared environment. Individuals are rewarded for collecting apples in an orchard. Apples grow less quickly in the orchard the more pollution there is in the river. Pollution accumulates stochastically in the river at a constant rate. Individuals have the ability to clean pollution but must take time away from collecting apples in order to do so. Figure reproduced with permission from \oldVersion.} %(c) Group behavior unfolds on fast timescales (sample episode from human data). Individuals are able to respond both to environmental events and to the actions of their groupmates.}
    \label{fig:environment}
\end{figure*}

MARL discovers strategies its human users may not even know exist \citep{silver2017mastering, jaderberg2019human, baker2019emergent, vinyals2019grandmaster}. This property creates new research affordances in the computational study of social phenomena like cooperation. Instead of starting the modeling process with a set of \textit{a priori} interesting strategies in mind, e.g. tit-for-tat versus pure cooperate/defect \citep{axelrod1984evolution}, in MARL modeling the focus shifts to defining an environment where a property of interest may arise, and identifying interpretable manipulations that affect whether or not it does. Manipulations may be to the environment or to the agent's learning algorithm. For instance, many papers have shown that ``off-the-shelf'' agents, i.e. agents that are self-interested and \textit{tabula rasa}, do not learn to cooperate in social dilemmas unless modified in some way, e.g.~\citep{lerer2017maintaining, leibo2017multi, perolat2017multi}. The needed modifications are typically interpretable and easy to control in computational experiments: e.g. an antecedent preference for fairness \citep{hughes2018inequity} or reciprocity \citep{kleiman2016coordinate, lerer2017maintaining, foerster2018learning, eccles2019learning}. 

A particular environment called Clean Up has been the focus of a large number of different MARL studies \citep{hughes2018inequity, mckee2020social, baker2020emergent, eccles2019learning, jaques2019social, yang2020learning, radke2022importance, gemp2022d3c, leibo2021scalable, christofferson2022get, tilbury2022identity, kolle2023learning}. It was inspired by social dilemmas of public goods provision \citep{trawick2001successfully, ostrom1993coping, mollinga2003waterfront, janssen2007robustness}, though it has more spatio-temporal complexity than the abstracted public goods game typically used in behavioral game theory experiments \citep{camerer2003behavioral, janssen2010introducing}. This is because in Clean Up the agents are all embodied in a virtual world together. They don't face a binary choice like Cooperate or Defect or a single scalar number of how much to contribute. Instead they take in high-dimensional observations of part of their environment (typically RGB images) and issue low-level motor commands to make small micro-actions that have no meaning in themselves but, taken as a whole, may ``add up'' to something corresponding abstractly to cooperation, defection, or some particular level of contribution to the public good. In Clean Up, the aim is to collect apples from an orchard. Each group member receives one reward for every apple they collect. Apples can regrow after they are harvested; apple growth is driven by the cleanliness of a geographically separate river. The river fills up with pollution with a constant probability over time. As the proportion of the river filled with pollution increases, the growth rate of apples monotonically decreases. No apples grow at all once pollution levels exceed a threshold. Individuals can spend time cleaning the river to remove its pollution, an extended course of action that is analogous to making a contribution to the public good of size proportional to the amount of time they clean and their skill in doing so.

Clean Up was used in several studies of intrinsic motivations that promote cooperation. In the reinforcement learning community, the term intrinsic motivation is normally used to refer to any interpretable term added to a reward function \citep{singh2005intrinsically}. They are often chosen by hand based on some prior knowledge of how to get the agent to solve a specific task and they are mathematically equivalent to shaping \citep{sutton1998introduction}, but interpreted as corresponding to mechanisms inside the agent rather than out in the world. Intrinsic motivations are useful in MARL because they provide a way to make otherwise selfish and solipsistic agents care about one another. For example, one of the first ways discovered to get agents to cooperate in Clean Up was ``advantageous inequity aversion'' \citep{fehr1999theory}: it penalizes agents when their average rewards pass too far above the average of all other players. This has the effect of discouraging them from free riding by eating apples while other players clean, and encouraging them to clean so that others will eat and the average reward will rise to alleviate the ongoing penalty they face for having eaten in the past \citep{hughes2018inequity}. Another augmentation of the basic agent that leads to cooperation is called Social Value Orientation \citep{balliet2009social}. In this model agents seek a specific ratio between their own average reward and the average rewards of others, their social preference. Groups with heterogeneous social preferences cooperated more effectively in Clean Up, obtaining higher collective return than homogeneous groups \citep{mckee2020social}. Relatedly, random uncertain social preferences were also shown to be a cooperation-eliciting mechanism in Clean Up \citep{baker2020emergent}. Another study wrapped MARL with an outer loop of evolution controlling parameters of the MARL intrinsic motivation and found in accord with evolutionary theory that such altruistic intrinsic motivations did not evolve when evolutionary fitness was computed based solely on individual rewards but could evolve in a group-selection-like setting where group member fitnesses were all coupled with one another \citep{wang2019evolving}. In other work, an intrinsic motivation to encourage agents to become conditional cooperators by imitating the ``niceness'' level of their co-players was a cooperation-eliciting mechanism in Clean Up \citep{eccles2019learning}. Another paper showed that endowing agents with a taste for social influence: an extra reward when their actions had a large effect on the subsequent actions of others, also provided a path to cooperation in Clean Up \citep{jaques2019social}.

\begin{figure*}[t]
    \centering
    \includegraphics[height=9.65cm]{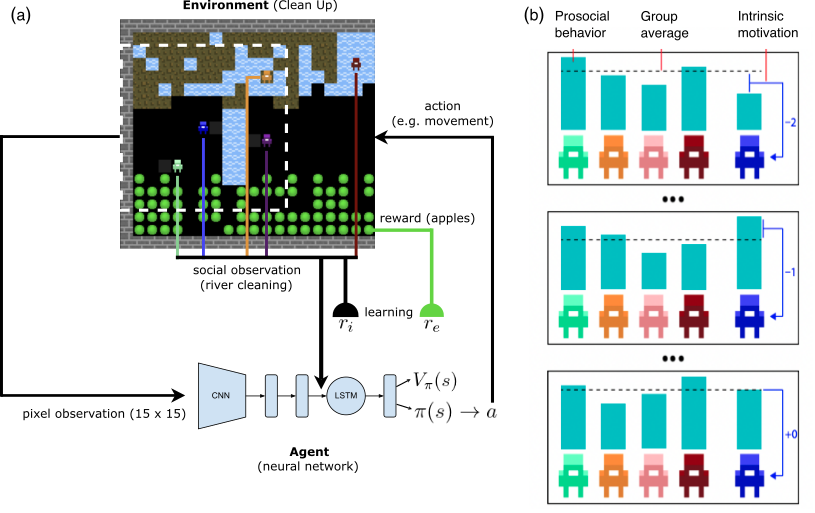}
    \caption{\small Figure \ref{fig:agent_architecture}: Agent architecture and training procedure (a) Each individual receives \emph{visual} and \emph{social} observations from the environment for the current timestep. The agent then uses the visual and social observations to select an action. Each group member also receives \emph{extrinsic reward} $r_e$ from the environment and computes an \emph{intrinsic reward} $r_i$ on the basis of the social observation. Visual observations are processed first using a convolutional neural network (CNN) followed by two linear layers. The resulting representation is processed with an LSTM, which maintains a state containing memory for past states. The social observation is concatenated with the linear layers' output before both are processed together by the LSTM. Finally, after the LSTM there is another linear layer followed by two output ``heads'' which predict the value $V_\pi(s)$ and policy $\pi(s)$. On each timestep the agent samples its policy to obtain its choice of action for the current step. There were nonlinear functions between each step of the computation, all of which were ReLUs. (b) The intrinsic reward for reputation primarily reflects an aversion to having a lower reputation than one's peers and secondarily an aversion to the situation where their peers take advantage of their efforts by themselves cooperating much less (free riding). The figure shows three example social observations, in the first, the focal agent contributed less than the group average. In the second it contributed more. And in the third example its contributions were equal to the group average.}
    \label{fig:agent_architecture}
\end{figure*}

Despite its growing popularity as a model of human behavior, MARL in Clean Up was never validated in an experiment with real human participants. This is a vitally important step in validating that the environment captures the dynamics of human cooperation. Here we describe such an experiment. Since this experiment will involve human participants, we have to carefully choose a cooperation-eliciting mechanism to study. Many of those studied so far are not manipulable in humans. They correspond to traits (e.g.~\citet{mckee2020social}) and internal drives (e.g.~\cite{hughes2018inequity}), which we can't easily manipulate in a one-hour experiment. However, at least one cooperation-eliciting mechanism appears easy to control experimentally: identifiability \citep{hardy2006nice, ariely2009doing, van2010cooperation, wedekind2000cooperation, berridge2009dissecting, izuma2008processing, nowak1998evolution, engelmann2009indirect, phan2010reputation, milinski2002reputation}. Reputation-based accounts of cooperation all agree that anonymity can destroy group cooperation while making individuals identifiable so they can track each another's reputations can facilitate it \citep{wedekind2000cooperation, semmann2004strategic, roberts2021benefits, bradley2018does}. In this paper we show that this effect holds when humans interact in Clean Up. We then construct a MARL model that captures the difference in the amount of contribution (cleaning) elicited in anonymous versus identifiable conditions. Furthermore, beyond matching the overall level of public good contribution, we show that the model also chooses to implement its cooperative strategy in the same way that humans do: by turn taking, as opposed to other possibilities admitted by the environment like territoriality.

A \textit{stylized fact} is an empirical regularity deemed important for candidate models to explain (see~\cite{hirschman2016stylized}). The most important contribution of this paper is to identify two stylized facts about human behavior in the spatio-temporally complex public good provision setting captured by Clean Up. They are (1) humans cooperate when they can see who is who and track public good contributions over time, and (2) anonymity destroys human cooperation in this setting, leading to low levels of public good provision. The MARL model we describe is the first to reproduce both stylized facts. It employs a mechanism which we interpret as a model of the human motivation to build a good reputation on the basis of being seen to perform altruistic actions. Notice that we do not claim this family of models is the only one that could reproduce the two critical stylized facts, nor do we claim reputation mechanisms are necessarily underpinning human cooperation in this setting. Future research may discover other models that fit the same stylized facts, and they may employ mechanisms other than reputation motivations. However, so far, no other such models have emerged. Most MARL algorithms do not cooperate in Clean Up at all (see~\citet{leibo2021scalable}). The algorithms that do cooperate are very unlikely to be affected by anonymity since the mechanisms by which they cooperate do not depend on identifying individuals (e.g.~\citet{hughes2018inequity, mckee2020social, eccles2019learning}). We suggest that building up a body of stylized facts to ground computational theorizing in real human behavior is a critical step in the maturing of the MARL research methodology. In the future, we will be able to compare alternative models to one another by the extent to which they either do or do not predict a body of stylized facts accumulated over time from multiple experiments.

%-------------------------------------------------------------------------
\section*{Constructing MARL models of social behavior}

A MARL model, like other agent-based models \citep{bonabeau2002agent}, consists of two interlocking parts: (1) a model of the environment and (2) a model of individual decision-making. For MARL, the first takes the form of a researcher-designed simulator: an interactive program that takes in a current environment state and agent actions, and outputs the next environment state as well as the observations of all agents and their instantaneous rewards. The model of individual decision-making is likewise conditioned on environment state. It is an agent that learns from its past experience using a form of trial and error. Importantly, in designing the simulator, the modeler attaches rewards to physical events, not to joint strategies. This allows much greater complexity to be achieved since the designer need not understand in advance how all mechanisms in the simulated world cache out in terms of payoffs accruing to joint strategies. MARL allows agents to explore their environment and potentially discover strategies the designer did not even know were possible \citep{hassabis2017artificial, leibo2019autocurricula, baker2019emergent}.

Agents interact with their environment by taking in observations and outputting actions. Each agent selects actions according to its policy, a mapping from observations to actions. Agents learn by changing their policy to improve it along any desired dimension, typically to improve its reward. The policy is stored in a neural network. Agents incrementally learn `from scratch’ how the world works and what they can do to earn more rewards. They accomplish this by tuning their network weights in such a way that the pixels they receive as observations are transformed into coherent actions through a sequence of steps called network layers \citep{sutton1998introduction}. Several learning agents can inhabit the same environment as one another. In this case the agents become interdependent because their actions affect one another \citep{leibo2017multi, wong2022deep}. 

Formally, MARL algorithms are applicable in a  model of strategic social situations called a Markov Game. Markov games generalize matrix games by introducing the concept of environment state. The environment's state influences payoffs jointly with the actions of the game's players and is itself influenced by them. In a partially-observable Markov game, players do not have direct access to the state though they may be able to infer it from the data they do have. 

%--------------------------------------------
~\newline\newline{\noindent\textbf{Definition: Partially-Observable Markov game} \citep{shapley1953stochastic, littman1994markov}. At each state $s \in \mathcal{S}$ of a Markov game, each player $i \in I = \{1, \dots, N\}$ takes an action $a_i \in \mathcal{A}_i$. On each step of simulation players receive their own $d$-dimensional partial observation of the state $o_i \in \R^d$, which is determined by the observation function $\mathcal{O} : \mathcal{S} \times I \rightarrow \R^d$. After the players' joint action $\vec{a} = (a_1, \dots , a_N)$, the state changes according to the stochastic transition function  $\mathcal{T} : \mathcal{S} \times \mathcal{A}_1 \times \! \cdots \! \times \mathcal{A}_N \rightarrow \Delta(\mathcal{S})$, where $\Delta(\mathcal{S})$ denotes the set of discrete probability distributions over $\mathcal{S}$. After each transition, each player $i$ receives a reward $r_i \in \R$ according to the reward function $\mathcal{R} : \mathcal{S} \times \mathcal{A}_1 \times \! \cdots \! \times \mathcal{A}_N \times \mathcal{S} \times I \rightarrow \R$.}

We train the agents in the model using independent multi-agent reinforcement learning \citep{tan1993multi}. Concretely, this means that each agent has their own individual deep neural network and their own independent stream of experience. There is no parameter sharing between agents.

In independent multi-agent reinforcement learning, each agent $i$ learns a policy $\pi_i$ intended to maximize its value from some initial state $s_0$ under the joint policy of all agents. The value for agent $i$ is defined to be $V_{\mathbf \pi}(s_0) = \mathbb{E}_{\mathbf{\pi}} \left( \sum_{t=0}^\infty \gamma_t r^t_i \right)$, where $\gamma$ is a discount factor and $r^t_i$ is a random variable representing the reward at time $t$ given the actions sampled from the stochastic policies and the stochastic transition function of the environment. We use $\gamma = 0.99$ for our experiments. Policy gradient algorithms perform gradient ascent on the parameters of a neural network defining a policy in order to maximize the value it achieves. The specific policy gradient algorithm we used here is called advantage actor-critic \citep{mnih2016asynchronous}.

\section*{The Clean Up environment}

The Clean Up environment draws inspiration from social dilemmas arising from public good provision \citep{trawick2001successfully, ostrom1993coping, mollinga2003waterfront, janssen2007robustness}. Spatio-temporally complex simulation environments are important in the study of social-ecological systems \citep{janssen2010introducing, janssen2010lab}. For example, studies of fisheries indicate that institutional rules for managing the commons are more often based on where, when, and how to harvest, not how much to harvest \citep{ostrom2005understanding}. 

Clean Up is a 2-dimensional, video game-like environment (Figure \ref{fig:environment}a; \citep{eccles2019learning, hughes2018inequity, jaques2019social, mckee2020social, wang2019evolving, kramar2020should}). Each episode of Clean Up places a group of individuals ($n = 5$) into a $23 \times 16$ 2D environment consisting of an orchard and a river. Group members are able to expend effort to contribute to a public good (Figure \ref{fig:environment}a), in this case by cleaning pollution which accumulates in the river with a constant probability over time. A clean river causes a high apple growth rate, which benefits everyone. The public good in Clean Up is the growth rate of the apples in the orchard.

The formal concept of a social dilemma requires a specific relationship between individual and collective ``rationality''. In order to verify that the specific implementation of Clean Up we used provides a social dilemma, we analyzed the reward implications of adopting policies that clean and eat versus policies that only eat (free riding). See Supplementary Information Section: Social Dilemma Analysis for details.

Individuals contribute to provision of the public good by cleaning the river. However, unlike single-decision public goods games, in Clean Up a contributing group member must move over to the river and then perform a sequence of actions which may be accomplished with varying degrees of skill (Figure \ref{fig:environment}b). It would take a long time for a single individual to clean the river on their own since pollution continues to accumulate, and it could only be accomplished by one who is skilled at moving around and aiming their cleaning tool. Additionally, group members must actively collect the benefits of the public good by physically moving over to the orchard and harvesting apples there. So it is not possible to clean and eat at the same time. Individuals must balance the costs and benefits of time spent harvesting apples versus contributing to the public good (Figure \ref{fig:environment}c). If others clean the river then it is possible to free ride on their efforts by spending all one's time in the orchard eating apples. 

We manipulated anonymity in our experiments with Clean Up. Supplementary Video 1 shows human behavior in the identifiable condition\footnote{The video of human gameplay in the identifiable condition is also available at\\ \url{https://youtu.be/ohQrN46n9sQ}.} and Supplementary Video 2 shows human behavior in the anonymous condition\footnote{The video of human gameplay in the anonymous condition is also available at\\ \url{https://youtu.be/AqCKDibiE9Q}.}. Note that these videos show the third-person view, it covers the whole map at once. Players only perceive an egocentric window centered on their avatar during gameplay. In the identifiable condition each players has a unique color so it is easy to distinguish them from the others. In the anonymous condition all players are assigned the same color. Everyone views their own avatar in a unique color which differs from the other players, even in the anonymous condition.

We assume common knowledge of the public good in the following sense. Agents receive a social observation on each timestep in addition to their visual observation. The social observation incorporates knowledge of what must be done to contribute to the public good, in this case, clean the river. It may be interpreted as reflecting a normative evaluation of each player's recent behavior \citep{manrique2021psychological}. In the identifiable condition, the social observation contains a measurement of the recent contribution levels of oneself and all other players (i.e. a vector of length 5). In the anonymous condition this information is absent (for humans) and corrupted for the computational model so that the same number of LSTM parameters can be used in both conditions, see Supplementary Information for details. There are screenshots in the supplementary information showing the view seen by the human participants in both the anonymous and identifiable conditions. Note that the same measurement of recent contribution levels used by the agents' social observation is visible in the human view of the identifiable condition while the anonymous view only highlights one's own recent contribution level.

%-------------------------------------------------------------------------
\section*{A MARL model of reputation-based cooperation in Clean Up}

The reinforcement learning agents in this model seek to maximize both their extrinsic reward (for Clean Up this is mostly apple consumption) and an intrinsic reward (internally generated incentives \citep{singh2005intrinsically} such as satisfaction of other-regarding preferences generated by social-cognitive mechanisms in the brain \citep{berridge2009dissecting, izuma2008processing}). In our model, agents are intrinsically motivated to manage their reputation. In particular, we posit that agents prefer to obtain for themselves a relatively higher reputation than others in their group.

This idea is called competitive altruism \citep{roberts1998competitive, barclay2004trustworthiness, hardy2006nice}. It can be justified by considering that conspicuously prosocial behavior may be seen as an investment in social capital that could pay off down the line, perhaps by inducing others to provide benefits to keep especially altruistic players in the group. Or alternatively, it may be justified from honest signalling theory \citep{zahavi1999handicap, gintis2001costly, roberts2021benefits}. In this case, competitive altruism is the idea that, just as a male peacock grows a large tail to demonstrate its underlying quality as a mate or ally since the tail is a handicap that all know would be unaffordable to a lower quality bird, humans compete to show themselves to be more altruistic than their peers. Altruism is by definition costly so the same logic as the peacock tail applies. Prior empirical work showed that competitive altruism explains human behavior in abstracted public goods dilemmas better than other models \citep{giardini2021competitive}. Our implementation motivates agents to contribute at least as much as the mean of the their group. 

In accord with competitive altruism-based accounts of human reputation-based cooperation, the internal motivation we posit serves to encourage agents to contribute relatively more to the public good than other group members. Specifically, we implement this in an analogous manner to other intrinsic motivation-based accounts of cooperation which posit aversions to relative differences between oneself and others~\citep{fehr1999theory, hughes2018inequity}. 

The overall reward signal $r$ is the sum of the extrinsic reward $r_e$ and the intrinsic reward $r_i$:

\begin{gather}
   r = r_e + r_i \, , \\
   r_i = - \alpha \cdot \textrm{max}(\bar{c} - c_{\textrm{self}}, 0) - \beta \cdot \textrm{max}(c_{\textrm{self}} - \bar{c}, 0) \, .
   \label{intrinsic_reward_function}
\end{gather}
\noindent Here $c_{\textrm{self}}$ is one's own contribution level (i.e., the amount of pollution cleaned from the river), $\bar{c}$ is an estimated or observed average of the group's contribution levels, and $\alpha$ and $\beta$ are scalar parameters. The $\alpha$ and $\beta$ parameters control how much $r_i$ is affected by one's contribution level falling behind and rising above the group average contribution, respectively. {For the experiments here, we parameterize the model with $\alpha \sim \mathcal{U}(2.4, 3.0)$ and $\beta \sim \mathcal{U}(0.16, 0.20)$.}

%This intrinsic motivation reflects an antecedent preference not to contribute less to the public good nor to contribute more than the other players. 

Since $\alpha > \beta$ this intrinsic motivation involves primarily an aversion to contributing less than other group members \citep{barclay2006partner}, and secondarily an aversion to others taking advantage of one's efforts \citep{rockenbach2006efficient}. It reflects a hypothesis that the positive effect on reputation of helping more is psychologically distinct from the negative effect of helping less \citep{barclay2013strategies}. The aversion associated with helping less than the norm is an aversion to being in the situation where others are more generous than oneself, which is bad under competitive altruism. The aversion associated with helping more than the norm is operative in the situation where you do more than your share of the work while others free ride.

It's important to note that the intrinsic reward is always negative (since $\alpha > 0$ and $\beta > 0$). The only way to get positive reward in Clean Up is to eat apples. Notice that at convergence the magnitude of the intrinsic reward is always small compared to the extrinsic reward (|$r_i| \ll |r_e|$), see Fig.~\ref{fig:agent_pseudoreward} and Fig.~\ref{fig:agent_reward_comparison}. Therefore, the intrinsic reward cannot be interpreted as a pure taste for altruism because agents at convergence do not derive much of their overall reward from cleaning. Rather, the intrinsic reward is better understood as having an effect of nudging the learning dynamics such that the population as a whole can explore a different part of the policy space and thereby discover cleaning. 

The present work concerns how MARL modeling may be validated by experiments with real human participants. The specific MARL model we evaluate uses agents endowed with an intrinsic motivation for reputation. First we show that the MARL model recapitulates the difference between treatment and control conditions manipulating anonymity in the human experiment. Second, we show that the model also captures a more detailed pattern we observed in the human data: that human groups solve Clean Up by partitioning time as opposed to space. That is, they take turns instead of picking territories.

\section*{Results}

%-----------------------------------
\subsection*{Effects of identifiability versus anonymity on contribution to the public good}

In both the behavioral experiment and in our computational model, we test the effects of reputation by comparing two conditions: (1) an \textit{identifiable} condition, in which group members are individually distinguishable and thus contribution behavior is public, and (2) an \textit{anonymous} condition, in which group members are largely indistinguishable and unable to perfectly monitor each other's contribution behavior.

We are interested in whether the MARL model recapitulates the difference between identifiable and anonymous conditions. After training, we simulated rounds of Clean Up with our computational model and compared its behavior against the behavior of groups of human participants (overall $N = 120$) in Clean Up. We parameterized the computational model with $N = 120$ reinforcement learning agents for each condition, allowing a similar group assortment as with the participant groups. For further details on the correspondence between the computational model and behavioral experiment design, see Materials and Methods.

Conditions of identifiability lead human groups to substantially increase their contribution levels. That is, the group contribution level is significantly higher in the identifiable condition than the anonymous condition, $p < 0.0001$ (repeated-measures ANOVA, Figure \ref{fig:human_results}a). Similarly echoing findings from prior experiments, collective return for human groups increased significantly in the identifiable condition, $p < 0.0001$ (repeated-measures ANOVA, Figure \ref{fig:human_results}a).

\begin{figure*}[!h]
    \centering
    \subfloat[(a)]{\includegraphics[height=4.85cm]{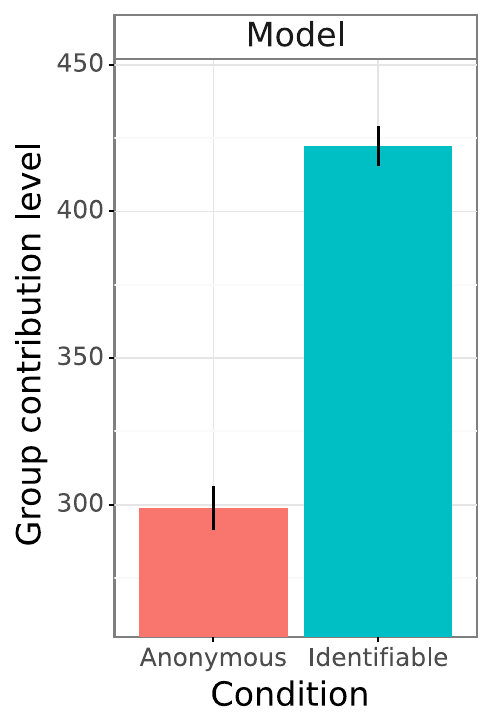}}
    \subfloat{\includegraphics[height=4.85cm]{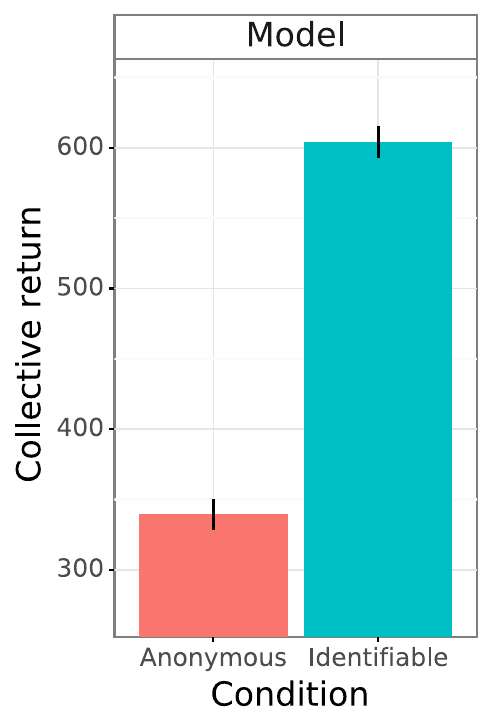}} \\
    \subfloat[(b)]{\includegraphics[height=4.5cm]{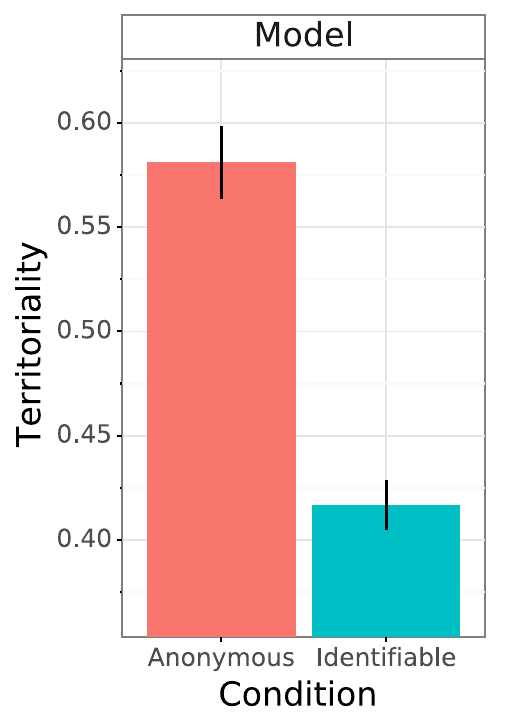}} \hspace{1em}
    \subfloat{\includegraphics[height=4.5cm]{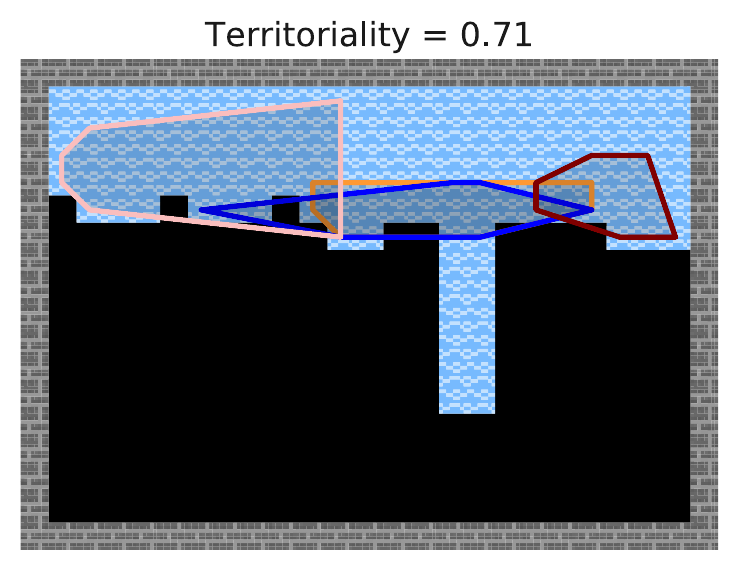}} \:
    \subfloat{\includegraphics[height=4.5cm]{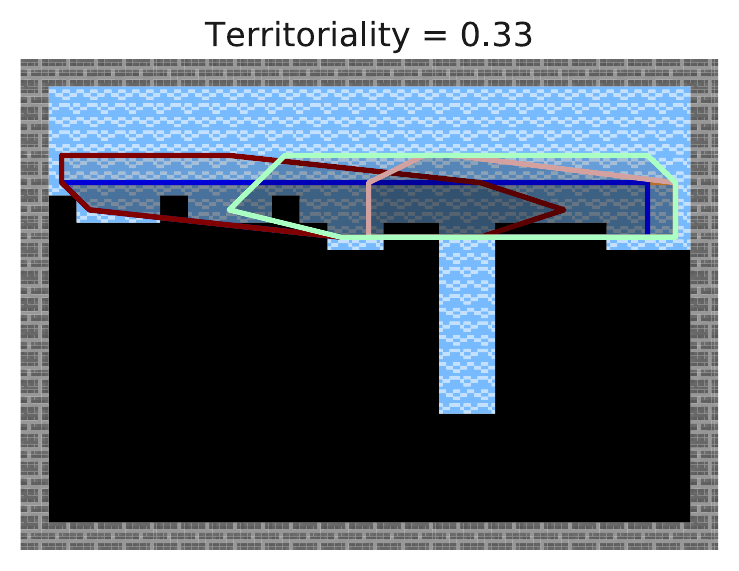}} \\
    \subfloat[(c)]{\includegraphics[height=4.5cm]{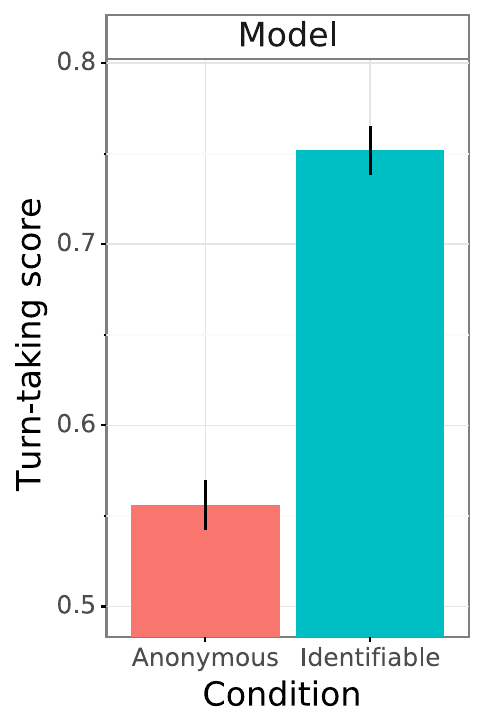}} \hspace{1em}
    \subfloat{\includegraphics[height=3.5cm]{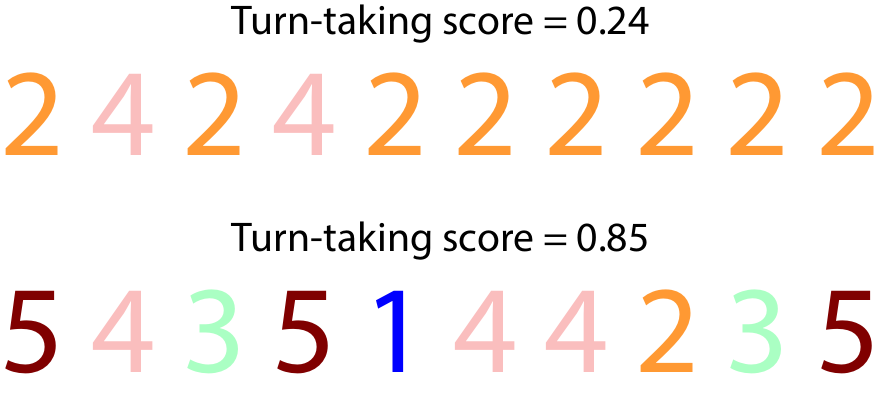}} \hspace{1em}
    \subfloat{\includegraphics[height=4.5cm]{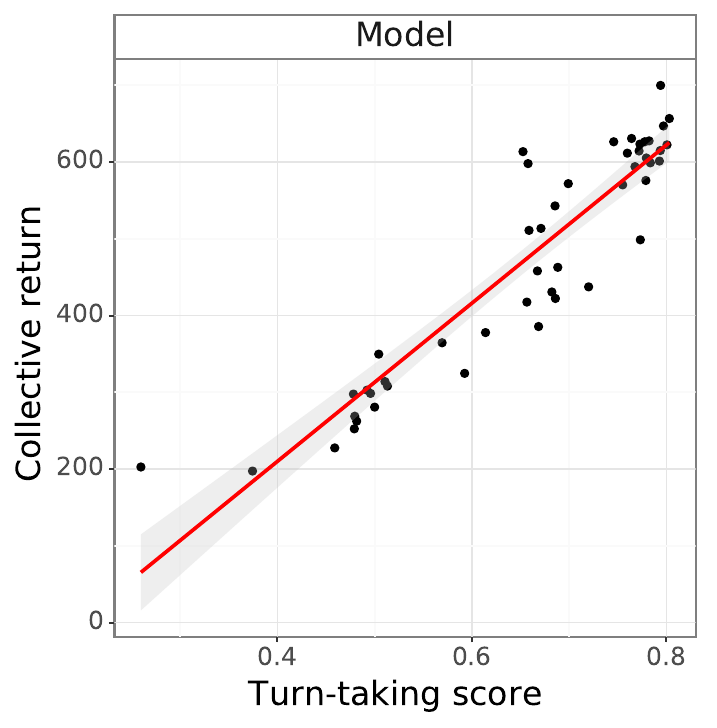}}
    \caption{\small Figure \ref{fig:model_results}: The intrinsic motivation for reputation substantially alters the behavior of deep reinforcement learning agents in Clean Up. Here we report the $F$ ratio, degrees of freedom, and $p$-value for each ANOVA. Errors bars reflect 95\% confidence intervals. (a) In the identifiable condition, when the motivation for reputation exerted a strong influence on behavior, there were significant increases in both average contribution level, $F(1,\:311) = 1090.7$ ($p < 0.0001$), and collective return, $F(1,\:311) = 2030.3$ ($p < 0.0001$). (b) Under conditions of identifiability, groups were significantly less territorial in the river, $F(1,\:311) = 432.0$ ($p < 0.0001$). To the right are example patterns of river ``territories'' from the model, showing a group that exhibited high territoriality ($\textrm{territoriality} = 0.71$) and a group that exhibited lower territoriality ($\textrm{territoriality} = 0.33$). (c) In contrast, groups were significantly \textit{more} reliant on turn taking in the identifiable condition, $F(1,\:311) = 758.3$ ($p < 0.0001$). Example patterns of turn taking from the model are included here, listing the identities of the group members that took the first ten turns of the episode (i.e., entering into the river to clean). The examples include a group that exhibited low turn taking ($\textrm{turn taking} = 0.24$) and a group that exhibited higher turn taking ($\textrm{turn taking} = 0.85$). These turn-taking scores were significantly correlated with {group performance, $\beta = 1030.3$, $p < 0.0001$.} Figure reproduced with permission from \oldVersion.}
    \label{fig:model_results}
\end{figure*}

Next we looked at the effect of the reputation motivation on group outcomes in our computational model. As expected, identifiability produced a significant increase in group contribution levels, $p < 0.0001$ (repeated-measures ANOVA, Figure \ref{fig:model_results}a). This increase in contribution levels led to significantly higher collective returns, $p < 0.0001$ (repeated-measures ANOVA, Figure \ref{fig:model_results}a). When motivated by reputation, group members in the model increase their cooperativeness, resulting in higher payoffs for the entire group.

%-----------------------------------
\subsection*{Temporal coordination not spatial coordination}

In a spatiotemporally complex setting like Clean Up our analysis need not stop once we've verified that the model matches the human results for \emph{how much} participants contribute to the public good. Part of the attraction of an environment like Clean Up---and for MARL methods in general---is that they make it possible to ask more nuanced questions about the where and when details of \emph{how} groups of players go about coordinating their contributions.

Groups of agents in the computational model appear to coordinate their efforts with a temporal rotation scheme. After measuring the extent to which group members' take ``turns'' entering and cleaning the river, we observe significantly greater turn taking under identifiable conditions than under anonymous conditions, $p < 0.0001$ (repeated-measures ANOVA, Figure \ref{fig:model_results}c). Example patterns of low and high turn taking from the computational model are presented in Figure \ref{fig:model_results}c. Further, results from the model confirm that this turn-taking strategy is associated with higher group performance. Across episodes, the more a group relied on a turn-taking rotation scheme, the higher {the collective return it received}, $p < 0.0001$ (linear regression, Figure \ref{fig:model_results}c). Furthermore, a mediation analysis showed that the extent of turn taking mediated the relationship between identifiability and group performance, $p < 0.0001$ (see supplementary information). We carried out a similar set of analyses to ask if they also used a territorial coordination strategy to achieve higher cooperation levels but found they did not.
Though there was a significant difference in territoriality between the two conditions, it did not explain the higher cooperation levels in the identifiable condition: there was significantly less territoriality (i.e. more spatial overlap) in the identifiable condition than in the anonymous condition, $p < 0.0001$ (repeated-measures ANOVA, Figure \ref{fig:model_results}b). Example patterns of high and low territoriality from the computational model are presented in \ref{fig:model_results}b.

\begin{figure*}[!h]
    \centering
    \subfloat[(a)]{\includegraphics[height=4.85cm]{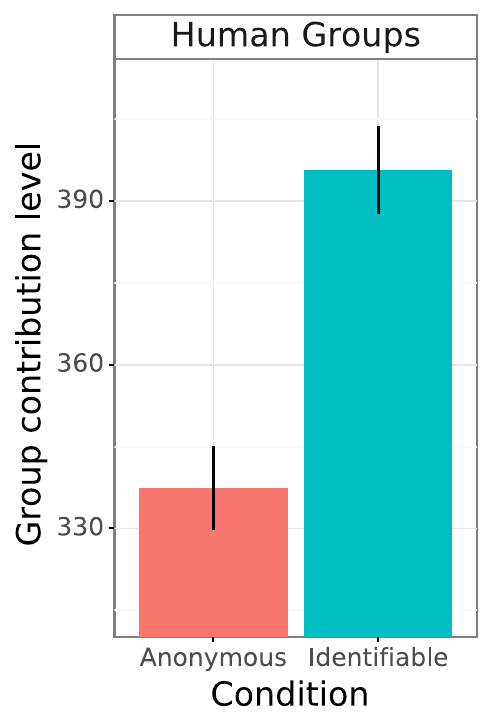}}
    \subfloat{\includegraphics[height=4.85cm]{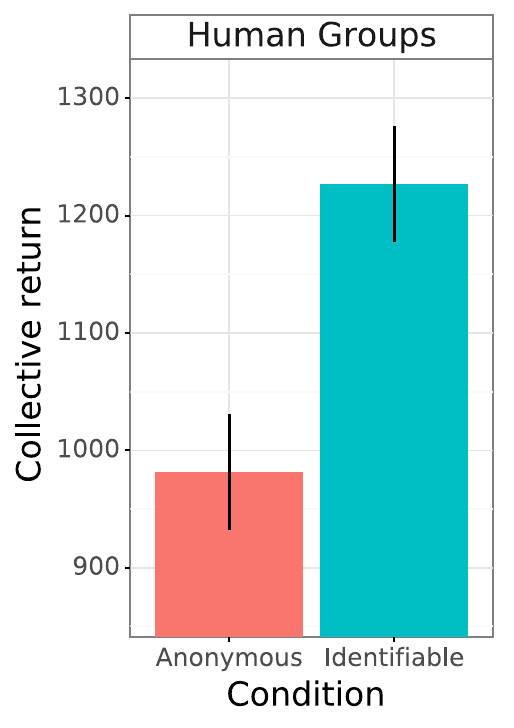}} \\
    \subfloat[(b)]{\includegraphics[height=4.5cm]{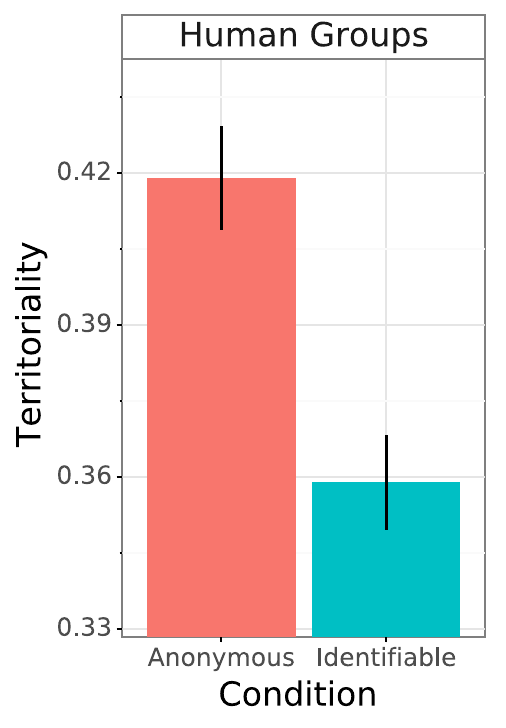}} \hspace{1em}
    \subfloat{\includegraphics[height=4.5cm]{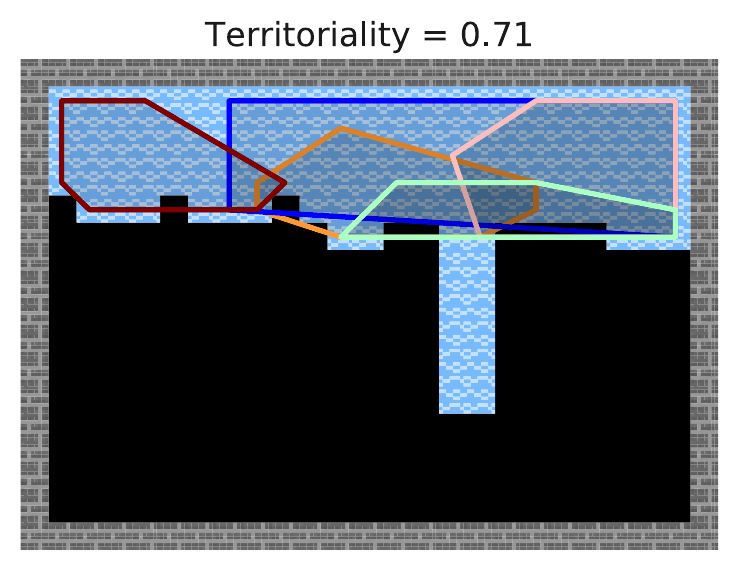}} \:
    \subfloat{\includegraphics[height=4.5cm]{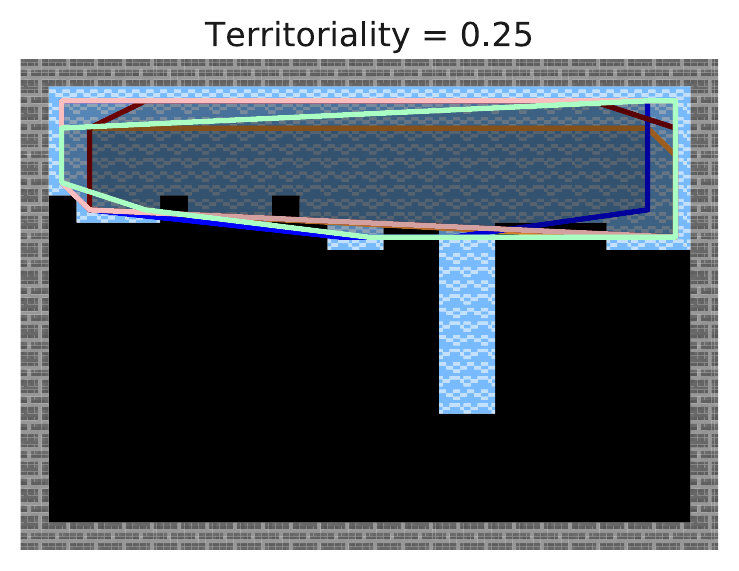}} \\
    \subfloat[(c)]{\includegraphics[height=4.5cm]{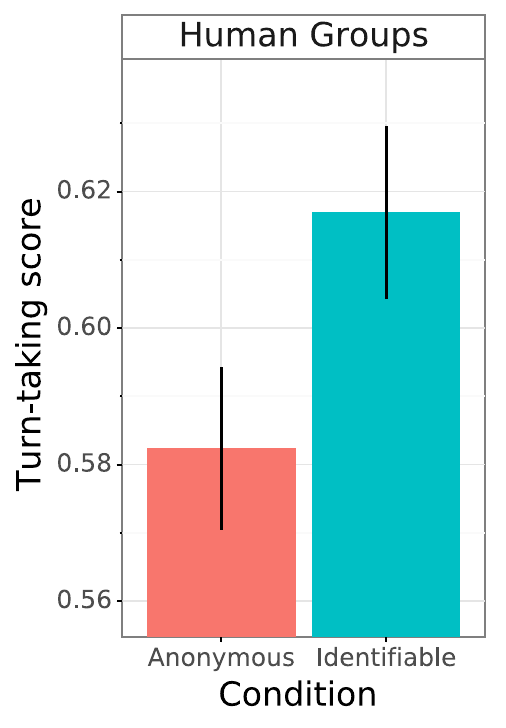}} \hspace{1em}
    \subfloat{\includegraphics[height=3.5cm]{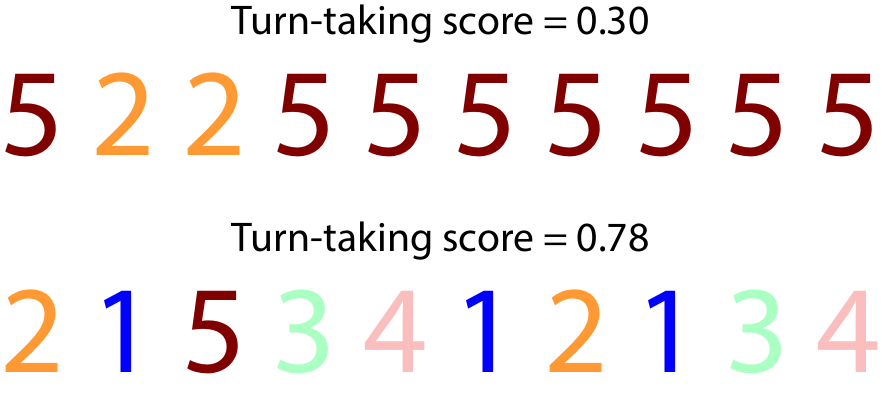}} \hspace{1em}
    \subfloat{\includegraphics[height=4.5cm]{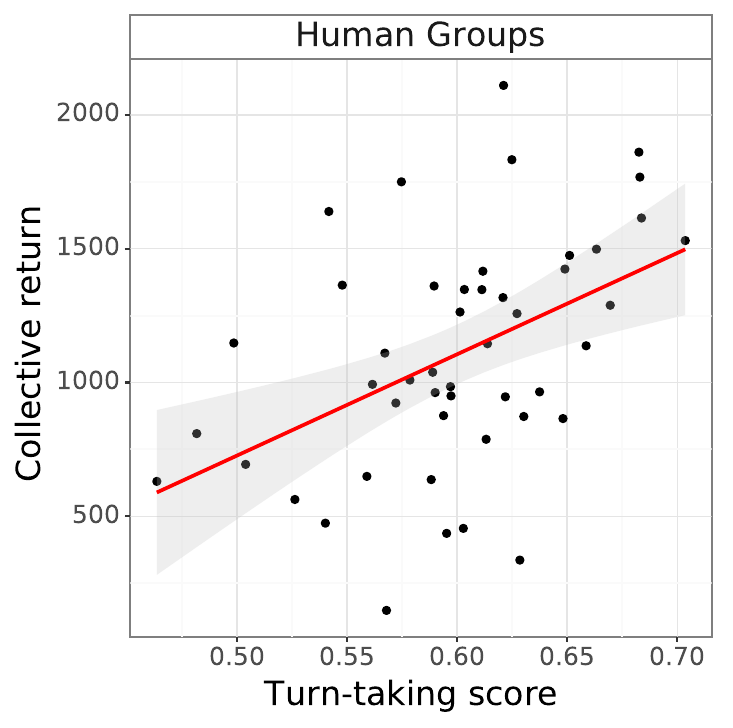}}
    \caption{\small Figure \ref{fig:human_results}: The intrinsic motivation for reputation substantially alters the behavior of human groups in Clean Up, matching the predictions made by our computational model. Errors bars reflect 95\% confidence intervals. (a) In the identifiable condition, when the motivation for reputation exerted a strong influence on behavior, there were significant increases in both average contribution level, $F(1,\:310) = 199.4$ ($p < 0.0001$), and collective return, $F(1,\:310) = 89.4$ ($p < 0.0001$). (b) Under conditions of identifiability, groups were significantly less territorial in the river, $F(1,\:310) = 138.4$ ($p < 0.0001$). To the right are example patterns of river ``territories'' among human groups, showing a group that exhibited high territoriality ($\textrm{territoriality} = 0.71$) and a group that exhibited lower territoriality ($\textrm{territoriality} = 0.25$). (c) In contrast, groups were significantly \textit{more} reliant on turn taking in the identifiable condition, $F(1,\:311) = 29.4$ ($p < 0.0001$). Example patterns of turn taking from the model are included here, listing the identities of the group members that took the first ten turns of the episode (i.e., entering into the river to clean). The examples include a group that exhibited low turn taking ($\textrm{turn taking} = 0.30$) and a group that exhibited higher turn taking ($\textrm{turn taking} = 0.78$). These turn-taking scores were significantly correlated with {group performance, $\beta = 3784.6$, $p = 0.0010$.} Figure reproduced with permission from \oldVersion.}
    \label{fig:human_results}
\end{figure*}

Results from the behavioral experiment with human participants accord with those of the computational model. In particular, participant groups also used turn-taking-like rotation schemes to organize their collective action to provide the public good. There was significantly greater turn taking in the identifiable condition than in the anonymous condition of the human behavioral experiment, $p < 0.0001$ (repeated-measures ANOVA, Figure \ref{fig:human_results}c). Example patterns of low and high turn taking from the human groups are provided in Figure \ref{fig:human_results}c. As in the model, rotating responsibility for contributions was associated with {improved group performance. The more a group relied on a turn-taking strategy, the higher the collective return it tended to achieve}, $p = 0.0010$ (linear regression, Figure \ref{fig:human_results}c) and the extent of turn taking mediated the relationship between identifiability and collective return, $p < 0.0001$ (mediation analysis; see supplementary information). Also in accord with the computational model, identifiability did not induce a territorial strategy for providing the public good: human groups exhibited significantly less territoriality in the identifiable condition than in the anonymous condition, $p < 0.0001$ (repeated-measures ANOVA, Figure \ref{fig:human_results}b). Example patterns of low and high territoriality from the human groups are provided in Figure \ref{fig:human_results}b.

%-------------------------------------------------------------------------
\section*{Discussion}

%--------------------------------------------

\subsection{Summary}

Multi-agent deep reinforcement learning algorithms promise to scale up theoretical social science so that it can address complex situations involving spatial and temporal dynamics. MARL methods enable  researchers to formulate models that go beyond merely selecting a discrete option from a menu of choices, the state of the art in most of behavioral game theory and social-ecological systems modeling \citep{elsawah2020eight}. In fact, MARL algorithms can even operate in situations where the researchers themselves do not know in advance how to program a behavior to solve the problem at hand. The learning algorithm can figure out a successful behavior on its own using trial and error. To realize this promise, it is necessary to develop methodology for grounding MARL models in laboratory experiments with real human participants \citep{poteete2010working}. We showed here how to conduct a laboratory experiment in a way that makes it a useful target for MARL modeling. Specifically, we showed that cooperation in the popular MARL public good provision environment, Clean Up, was drastically reduced when participants played in an anonymous mode where they could not recognize one another as individuals or track their contributions, versus a mode where individuals could all recognize one another and track reputations over time. We then proposed a particular MARL model featuring agents with an intrinsic motivation to obtain a good reputation and showed that it recapitulates the difference between anonymous and identifiable conditions. Furthermore, the model also captured the coordination mode used by the human participants. Both humans and artificial agents used a turn-taking strategy, not a territorial strategy.

\subsection{Implications of the anonymity effect for other MARL models of cooperation}

There are a number of different MARL-based accounts of how cooperation may emerge in Clean Up \citep{hughes2018inequity, mckee2020social, baker2020emergent, eccles2019learning, jaques2019social, christofferson2022get}. However, we believe none of them predict that anonymity diminishes cooperation as we observed in human behavior, except the reputation-based model we proposed here. For instance \citep{hughes2018inequity, mckee2020social, baker2020emergent} achieve cooperation by endowing agents with intrinsic motivations that depend on the overall distribution of rewards, not on the identities of the agents receiving them. This mechanism should be unaffected by anonymity. In \citep{eccles2019learning}, some agents are motivated to imitate the ``niceness'' level of the behavior of others, thereby implementing a kind of conditional cooperation. All that matters for this strategy is the niceness level of the others, i.e.~whether they clean or not; their identity has no effect. Likewise, the social influence motivation \citep{jaques2019social} and contracting model \citep{christofferson2022get} also do not explicitly depend on agents being able to track one another's identity.

\subsection{Limitations}

One limitation of the present work is the requirement that all agents have common knowledge of a norm expecting that they spend time cleaning the river. In the future it may be possible to remove this common knowledge assumption. \cite{vinitsky2023learning} studied the emergence of social norms with MARL using a mechanism where norms arise from a bandwagon effect and violators are sanctioned by decentralized third-party punishment, creating incentives for agents to follow the norm. In principle, this approach could generate spurious and arbitrary norms (e.g. silly rules or harmful rules as in ~\cite{koster2022spurious}). However, if combined with group selection to reject norms that do not lead to a high level of social welfare, then such a system might discover that any norm providing incentives to clean the river is a good route to success in Clean Up.

\subsection{Theoretical implications}

Are there phenomena recognized within game theory as being difficult to treat with classical methods where MARL could be helpful? We argue that \emph{focal points} are one such topic. Sometimes coordination may be achieved by considering, out of all the different equilibria that might arise, are any of them especially conspicuous or prominent? If most people align with one another in how they view the prominence of the various ways of coordinating, then they may coordinate by selecting the option they all view as most prominent. When this joint behavior is an equilibrium it is called a focal point. Prominence resolves questions of equilibrium selection by selecting a specific equilibrium called the focal point of the game in question (see \cite{gintis2014bounds}). There are many reasons a particular way of coordinating could be seen as prominent. In particular, the idea that the ``default'' strategy has high prominence (if one exists) has been with the field since Schelling first proposed the concept~\citep{schelling1960strategy}. Importantly, prominence is understood as a property that cannot be incorporated directly into a game representation's payoffs (e.g.~\cite{sugden2004economics, gintis2014bounds}). Therefore, the concept has remained difficult to treat with the standard tools of game theory. Using MARL however, we can operationally define one type of prominence in a way that makes it more amenable to study. Intuitively, the idea is to associate learnability with prominence. Specifically, we can say an equilibrium is a focal point relative to a population of agents if it is the equilibrium that is easiest for them to learn. For instance, the easiest equilibrium to discover may be the equilibrium that agents who learn by random trial and error stumble upon first. Equilibria that are prominent by virtue of their being easily learnable may be regarded as MARL analogs to the default option type of prominence in classical game theory.

This way of thinking about the focal point concept can be used to understand the behavior of both human and artificial agents in Clean Up. It is a striking result that both human and artificial agents coordinated by taking turns rather than by dividing the river into territories. We suggest that for Clean Up, turn taking may be regarded as more prominent in the sense we described above, namely that it is an easier coordinated strategy to learn. Viewing a cleaning player leave the river and return to the apple orchard to eat is a reliable signal that the river will shortly fill up with pollution and prevent apple growth unless another player takes action to prevent that from happening (by cleaning themself). Thus, when one is in the orchard, there is a fairly reliable correspondence between the event of observing another player leave the river and the diminishing of apple density and reward shortly after. The reliability of this signal makes it easy for both agents and humans to learn when to clean. By contrast, random exploration by agents in the early stages of training is very unlikely to produce the pattern of experience that would be needed to discover territoriality, simply because a random walk on a 2D space is unlikely to remain restricted to a small territory, and five random walks remaining restricted to five small territories smaller still \citep{lawler2018introduction}. The data that would guide agents toward territorial equilibria are unlikely to be generated by chance relative to the data that would guide agents toward turn-taking. Therefore territorial equilibria are less likely to emerge. Similar logic may apply for human players, though they have a different initial policy. Humans don’t start out by playing the game randomly, but they do have priors from previous experience playing video games. In many games it is useful to explore one's entire environment \citep{gazzard2011unlocking}. So humans may be also unlikely to stumble upon territorial solutions by chance, and thus they too may get no opportunity to learn that such a strategy could be effective.

%Keep in mind that one may also adopt a rigorously Popperian philosophy of science in which nothing can be said about non physical entities like focal points. In that case this work has no implications for how we may understand focal points since focal points do not exist.

\subsection{Conclusion}

The new approach introduced here has implications beyond the study of reputation. This framework draws our understanding of multi-agent reinforcement learning closer to our understanding of human cognition and behavior. {It expands the toolkit available to investigate and examine mechanisms of group cooperation \citep{dafoe2020open}.}
How does the temporal and spatial structure of human interaction affect our ability to solve collective action problems \citep{miller1992collective}? What intrinsic motivations can support the formation and maintenance of institutions \citep{north1991institutions, ostrom2005understanding}? Answers to these questions can help us scaffold collective action and strengthen cooperation in communities of humans and artificially intelligent agents.

\FloatBarrier

%-------------------------------------------------------------------------
%\section*{Materials and Methods}
\section{Methods}

%{In addition to this overview, precise details of the computational model and human behavioral experiment are provided in the supplementary information.}

\subsection*{Clean Up environment}

Clean Up is a public good provision game implemented in a 2-dimensional environment. At each timestep each player only sees a part of the environment centered on their avatar. The objective of the game is to collect apples from an orchard. A player gains one point for every two apples he or she collects. Apples can regrow after they are harvested; apple regrowth is driven by a geographically separate river that supplies water and nutrients. This river fills up with pollution with a constant probability over time. As the proportion of the river filled with pollution increases, the respawn rate of apples monotonically decreases. For sufficiently high pollution levels, no apples will respawn. Players have a tool which allows them to clean pollution from the river. The public good in Clean Up is the regrowth rate of the orchard; players contribute to the public good by using their cleaning tool to clean the river. Players also possess a ``ticketing'' tool, used in prior studies to instantiate a costly punishment mechanism. The ticketing tool allows players to punish the other group members by lowering their scores. More details are available in the supplementary information.

More information on the Clean Up environment, including exact parameters and additional analyses verifying its social dilemma structure, can be found in supplementary information.

%--------------------------------------------
\subsection*{The experiment with human participants}

The University College London Research Ethics Committee conducted ethical review for the project and approved the study protocol (CPB/2013/015). All participants provided informed consent for the study.

The experiment was completed by 120 participants (age: mean $m = 21.5$, standard deviation $sd = 2.3$; gender: 50 male, 70 female), drawn from the University College London psychology department participant pool. Participants were first individually instructed on the action controls and the environmental dynamics in Clean Up through a series of tutorial levels (for exact details, see supplementary information). Subsequently, participants were sorted into groups of $n = 5$ and progressed through {14} episodes of Clean Up: {seven episodes in each condition}. We used a counterbalanced, within-participant design, with half of the groups completing the identifiable task first and the anonymous task second, and the other half completing the anonymous task first and the identifiable task second. {Like the agents in the computational model, participants observe and act based on a local view of the Clean Up environment.} After finishing both conditions, participants completed post-task questionnaires. At the end of the experiment, participants were paid according to their performance in the task. Each point accrued was worth $\frac{1}{2}$ pence. {A detailed description of the experimental protocol (including instructions and comprehension checks) can be found in supplementary information.}

%--------------------------------------------
\subsection*{Multi-agent reinforcement learning details}

The inputs to each agent's neural network are the pixels representing their local view of the environment and temporally smoothed data on its own contributions and the contributions of its peers. The outputs from the network are a policy (a probability distribution over the next action to take in the environment) and a value function (an estimate of the agent's discounted future return under the policy). The network architecture consists of a convolutional neural network with $3 \times 3$ kernel, stride $1$ and $32$ output channels, a two-layer multi-layer perceptron with $64$ hidden units in each layer, a long short-term memory (LSTM) \citep{hochreiter1997long} of hidden size $128$, and linear layers for the policy logits and value function.

All agents had their own independent neural network. There was no parameter sharing between agents.

In the evaluation stage of the reinforcement learning experiment, we partitioned each population into $24$ groups of five agents at random. We assessed the performance and behavior of each group in seven episodes of Clean Up. Groups were assigned to the identifiable or anonymous condition based on the condition they experienced during training.

%--------------------------------------------
\subsection*{Distributed computation}

In the training stage of the reinforcement learning experiment, we used a distributed framework \citep{espeholt2018impala, mckee2022quantifying} to train a population of $120$ reinforcement learning agents for each condition. The parameters of the agents were stored on $120$ learner processes, each responsible for carrying out the policy gradient update for one agent. To generate experience for the agents, $2000$ parallel arenas were created. For each episode in each arena, $5$ agents were randomly sampled from the population and their parameters synchronized from their respective learners. At the end of each episode, trajectories for agents were forwarded to the respective learners. Each learner aggregated trajectories in batches of $10$ and processed these to update the parameters for the associated agent, unrolling the LSTM for $100$ steps to train the recurrent network. We augmented the advantage actor-critic algorithm with VTrace \citep{espeholt2018impala} to correct for off-policy trajectories. Each agent was trained using 100 million $(s_t^i, a_t^i, r_t^i)$ tuples.

We used the RMSProp optimizer \citep{tieleman2012lecture} to compute the gradient, using learning rate $0.000321$, epsilon $10^{-5}$, momentum $0$, and decay $0.99$. To encourage exploration, we used entropy regularizer (as in \cite{mnih2016asynchronous}) with entropy cost $0.00154$.

%\newpage

\part*{Supplementary Information\footnote{Note: the content of this supplementary information section is almost identical to \oldVersion, with permission from the original authors. It has however been updated here to fit our more recent understanding of the best way to interpret the computational model and harmonized with the main text of this paper.}}

%\part*{Supplementary Information\footnote{Version note: the content of this supplementary information section is almost identical to that of its previous version, which was published as \oldVersion. It has however been updated here to fit our more recent understanding of the best way to interpret the computational model and harmonized with the present (and final) version's main text.}}

%\footnote{Version note: the contents of this section are almost identical to their earlier version, which was previously published in \cite{mckee2021multi}. Since the present version reflects the model interpretation used in the main text here, this version should be considered the final definitive one.}

%\footnote{The  content in this section is almost identical between this version of the paper and the previous version, which was published on arxiv as \cite{mckee2021multi}. However, the version in this paper contains additional clarifications which were not in the older draft and thus this version should be considered the final one.}

%\renewcommand{\figurename}{Fig.}
% \renewcommand{\thefigure}{1}
\renewcommand\thefigure{S\arabic{figure}}
\setcounter{figure}{0} 

\section{Design of Computational Model}

We built our computational model using advantage actor-critic~\citep{mnih2016asynchronous}, a deep reinforcement learning algorithm. Within the algorithm, we formalize the overall reward signal $r$ as a combination of the intrinsic, social reward $r_i$ and extrinsic, environmental reward $r_e$:

\begin{gather}
   r = r_e + r_i \, ,\\
   \label{eqn:reward_function}
   r_i = - \alpha \cdot \textrm{max}(\bar{c} - c_{\textrm{self}}, 0) - \beta \cdot \textrm{max}(c_{\textrm{self}} - \bar{c}, 0) \, ,
\end{gather}

\noindent where $c_{\textrm{self}}$ is one's own contribution level (i.e., the amount of pollution cleaned from the river), $\bar{c}$ is an estimated or observed average of the group's contribution levels, and $\alpha$ and $\beta$ are scalar parameters.

\begin{figure*}[h]
    \centering
    \includegraphics[width=7.5cm]{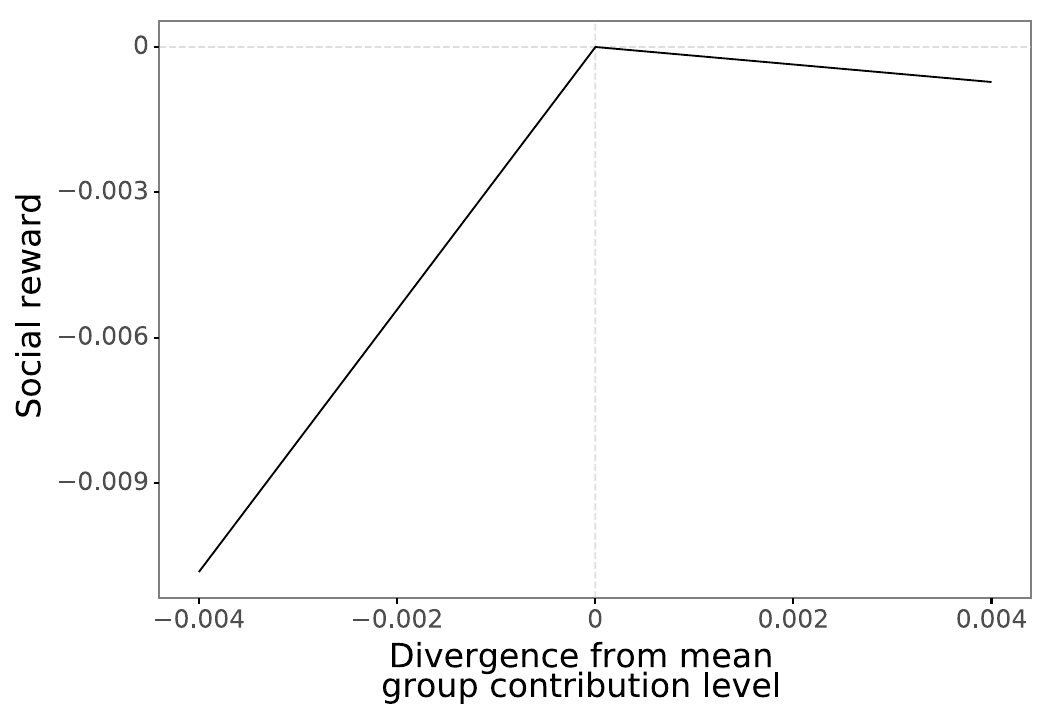}
    \caption{\small Figure \ref{fig:agent_pseudoreward}: Intrinsic motivation for each reinforcement learning agent varies as a function of $c_{\textrm{self}} - \bar{c}$. Here we show the empirical mean effect for the population of agents in the model, with $\alpha \approx 2.71$ and $\beta \approx 0.18$.}
    \label{fig:agent_pseudoreward}
\end{figure*}

The intrinsic motivation function was parameterized with $\alpha \sim \mathcal{U}(2.4, 3.0)$ and $\beta \sim \mathcal{U}(0.16, 0.20)$ (see Figure \ref{fig:agent_pseudoreward}). It therefore primarily represents an aversion to falling behind the group mean \citep{hardy2006nice}, as well as secondarily reflecting an aversion to being taken advantage of by free-riding others \citep{rockenbach2006efficient}.

\begin{figure}[h]
    \centering
    \includegraphics[width=7.5cm]{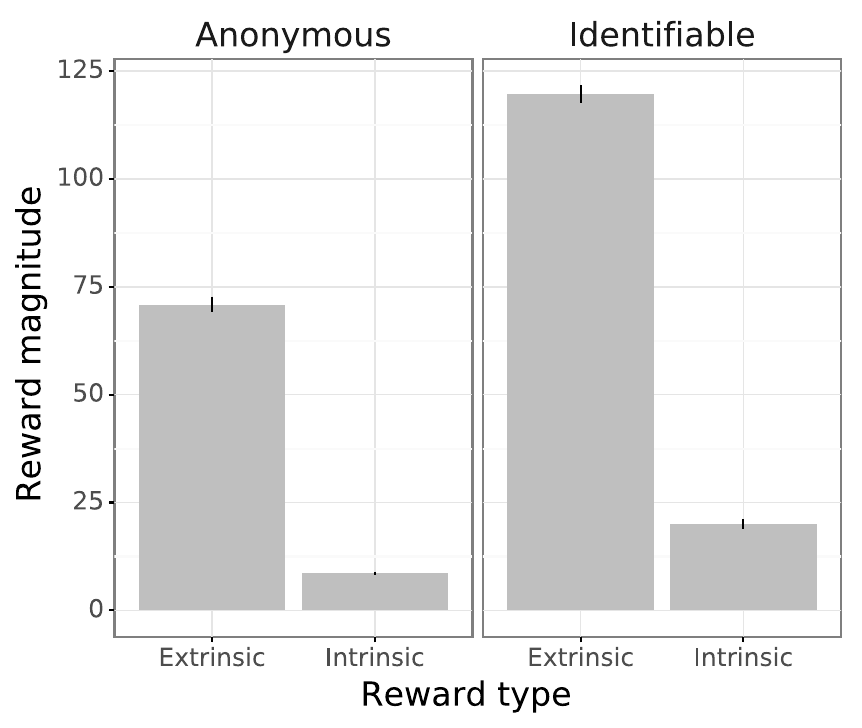}
    \caption{\small Figure \ref{fig:agent_reward_comparison}: Agents in the model do not live off intrinsic reward. A paired \textit{t}-test indicates that agents received significantly less intrinsic reward than extrinsic reward per episode across all episodes of the anonymous condition, $t(839) = 67.8$, $p < 0.0001$ and all episodes of the identifiable condition, $t(839) = 103.6$, $p < 0.0001$. Errors bar reflect 95\% confidence intervals.}
    \label{fig:agent_reward_comparison}
\end{figure}

A paired \textit{t}-test indicates that agents received significantly less intrinsic reward than extrinsic reward across all episodes of the anonymous condition, $t(839) = 67.8$, $p < 0.0001$ and all episodes of the identifiable condition, $t(839) = 103.6$, $p < 0.0001$ (Figure \ref{fig:agent_reward_comparison}).

We vary the intrinsic motivation parameters to better capture individual variability (cf., individual differences research; \citep{cronbach1957two}) and because population heterogeneity has important effects in multi-agent reinforcement learning \citep{mckee2020social, mckee2022quantifying}.

Let $q^t_j$ be the instantaneous index of agent contribution:
\begin{equation}
   q^t_j =     \left\{
\begin{array}{ll}
      1 & \text{agent $j$ contributed on timestep $t$} \, , \\
      0 & \text{otherwise} \, , \\
\end{array} 
\right. 
\end{equation}

\noindent and let $T$ be the number of timesteps in an episode ($T=1000$ for the reinforcement learning experiment). For agents $j = 1, \dots 5$, and the sequence of instantaneous agent contributions $\{q^t_j:t = 1, \dots T\}$, we update the temporally smoothed inputs ($c_{\textrm{self}}$ and $\bar{c}$) in Equation \ref{eqn:reward_function} as follows:

\begin{equation}
    c_j^t (c_j^{t-1}, q_j^t) = \lambda c^{t-1}_j + q^t_j\, ,
\end{equation}

\noindent where we choose smoothing factor $\lambda = 0.97$ and set $c_j^0 = 0$.

%To simulate the identifiable condition for the reinforcement learning agents, the environment provides contribution information for each group member as an input to the agent. To simulate the anonymous condition, the environment provides contribution information with a reduced visibility range. Contribution information about other agents is circumscribed by a visibility range $R = 9$, calculated using the $\ell^\infty$ norm. Behavior of agents falling outside of this range is not included in updates. This mimics the limited visual attention that participants marshal to track behavior in the anonymous condition (see also \cite{cohen2016bandwidth}). The provision of this altered contribution information reflects the ability of humans to track behavior of individual group members nearby on short timescales, even in the anonymous condition.

The environment provides contribution information for each group member currently in view as an input to the agent. In the anonymous condition the contribution information is corrupted by substantial noise. This mimics the limited visual attention that participants marshal to track behavior in the anonymous condition (see also \cite{cohen2016bandwidth}). It reflects the fact that humans can still track behavior of individual nearby group members on short timescales, even in the anonymous condition.

Overall, our computational model incorporates four assumptions:

\begin{enumerate}
    \item Behavior is motivated by the combination of extrinsic reward (externally determined incentives; e.g., the payoff dictated by the rules of a game or task) and intrinsic reward (internally generated incentives; e.g., satisfaction generated by social cognitive mechanisms in the brain \citep{berridge2009dissecting, izuma2008processing, singh2005intrinsically}).
    \item The intrinsic reward reflects primarily an aversion to having a lower reputation than one's peers \citep{barclay2006partner, hardy2006nice} (competitive altruism), and secondarily an aversion to peers taking advantage of one's efforts \citep{rockenbach2006efficient}.
    \item It is common knowledge among group members which choices or actions affect reputations, and in which direction \citep{milinski2002reputation, sherif1936psychology}.
    \item Human cooperation plays out over multiple timescales, and the intrinsic motivation for reputation can operate on the timescale of minutes \citep{izuma2008processing, phan2010reputation, saxe2008love}.
\end{enumerate}

\section{Experimental Design}

Cleanup is a partially observable Markov game \citep{littman1994markov}. A small set of parameters control the Cleanup environment. Environmental dynamics are defined by two functions:

\begin{equation}
   \textrm{Pr}_{\textrm{apple}}^t = \textrm{Pr}_{\textrm{apple}} \cdot \frac{H_{\textrm{depletion}} - F_{\textrm{polluted}}^t}{H_{\textrm{depletion}} - H_{\textrm{abundance}}} \, ,\label{eqn:apple_production_function}
\end{equation}

\begin{equation}
   \textrm{Pr}_{\textrm{pollution}}^t = \textrm{Pr}_{\textrm{pollution}} \cdot (F_{\textrm{polluted}}^t < H_{\textrm{depletion}}) \, . \label{pollution_production_function}
\end{equation}

Equations \ref{eqn:apple_production_function} and \ref{pollution_production_function} describe the probabilistic production functions for apple regrowth and pollution accumulation, respectively. In Equation \ref{eqn:apple_production_function}, $\textrm{Pr}_{\textrm{apple}}^t$ represents the probability of apple regrowth at time $t$, $\textrm{Pr}_{\textrm{apple}}$ reflects the underlying probability of apple regrowth when the river is sufficiently clean, and $F_{\textrm{polluted}}^t$ represents the fraction of the river that is filled with pollution at time $t$. $H_{\textrm{depletion}}$ reflects the proportion of the river filled with pollution above which apples can no longer regrow, and $H_{\textrm{abundance}}$ reflects the proportion of the river filled with pollution below which apples regrow with maximum probability. Equation \ref{pollution_production_function} describes the Bernoulli process that generates additional pollution in the river. $\textrm{Pr}_{\textrm{pollution}}^t$ reflects the probability that a new unit of pollution accrues in the river at time $t$, while $\textrm{Pr}_{\textrm{pollution}}$ represents the underlying probability that pollution accumulates if the river is not saturated with pollution.

We ran Cleanup with the parameter values listed in Section: Computational Model and Section: Human Behavioral Experiment. For the reinforcement learning experiment, we drew from canonical research using this task \citep{hughes2018inequity, mckee2020social} and largely carried over the established parameters. {Because of general differences between human and agent reaction times (cf. \cite{vinyals2019grandmaster}), this approach had to be adapted to develop a human behavioral research protocol instantiating a parallel social dilemma} (see Section: Social Dilemma Analysis).

For consistency with prior work on Cleanup (e.g.~\cite{hughes2018inequity}), group members also possess a ``ticketing'' tool which they can use to implement costly punishment \citep{fehr2002altruistic, henrich2006costly}. It costs four points to ticket another group member (sanction them); the group member receiving the ticket loses 40 points. We did not manipulate any properties of this sanctioning mechanism since our aim was to study a reputation mechanism that does not require sanctioning to work (equations S1--S4).

\subsection{Computational Model}
\label{sec:agent_env_params}

See \cite{hughes2018inequity, mckee2020social} for exact setup. The experiment was parameterized with episode length $T = 1000$ steps, cost of giving a ticket $-1$ and penalty for receiving a ticket $-50$, and the following environmental parameters:

\begin{itemize}
    \item $\textrm{Pr}_{\textrm{apple}} = 0.03\, $.
    \item $\textrm{Pr}_{\textrm{pollution}} = 0.5\, $.
    \item $H_{\textrm{abundance}} = 0.0\, $.
    \item $H_{\textrm{depletion}} = 0.32\, $.
\end{itemize}

\begin{figure}[ht]
    \centering
    \includegraphics[width=7.5cm]{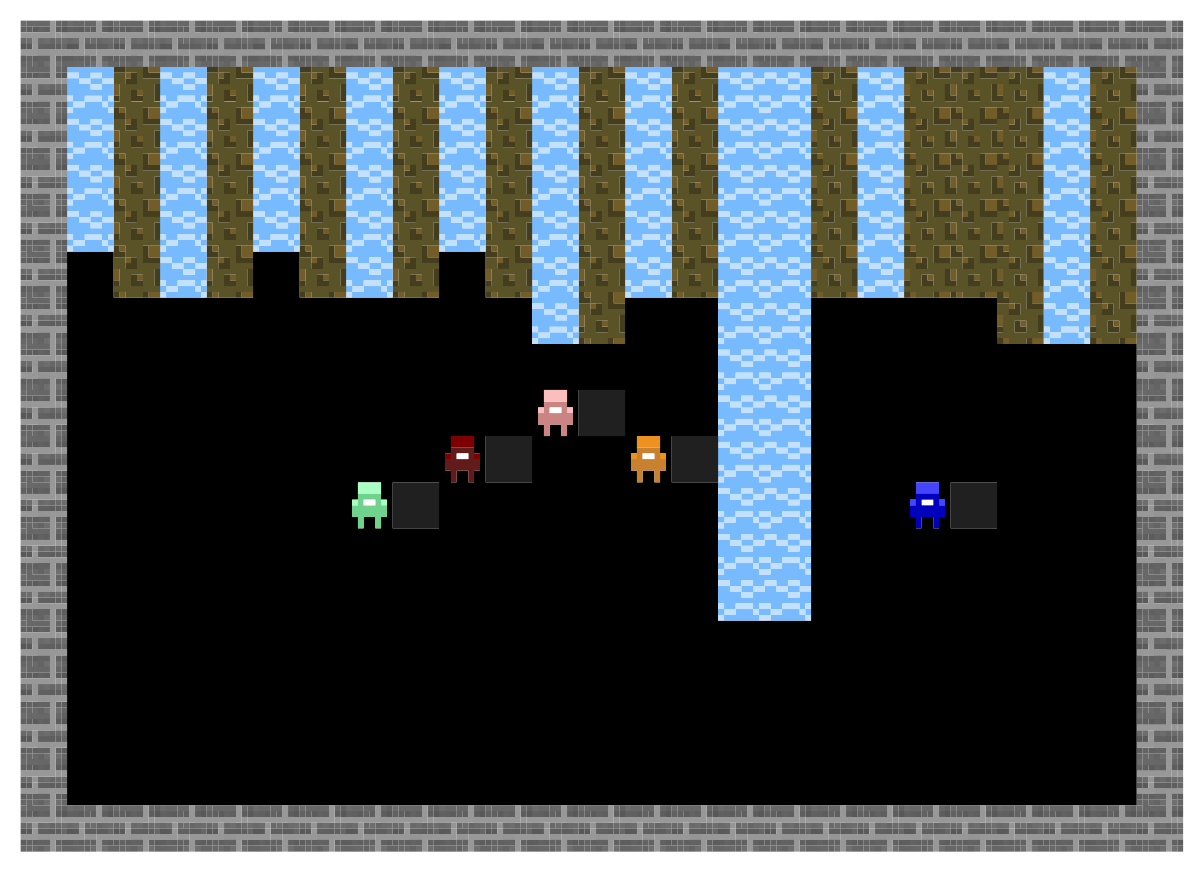}
    \caption{\small Figure \ref{fig:agent_training_map}: Example initial environmental conditions for each episode of agent training. Group members were randomly initialized across various positions in the middle of the environment.}
    \label{fig:agent_training_map}
\end{figure}

Following previous studies \citep{hughes2018inequity, mckee2020social}, each episode of reinforcement learning agent training began with river pollution at saturation and an empty orchard (Figure \ref{fig:agent_training_map}). During agent evaluation, each episode began with zero river pollution and a full orchard, matching the task conditions as experienced by human participants.

\subsection{Human Behavioral Experiment}
\label{sec:human_env_params}

Consistent with the design of video games with similar step rates, we implemented an input buffer to accept actions at most once every 100 ms. {In order to instantiate an analogous social dilemma for participants}, we selected the following set of environmental parameters for the human behavioral experiment:

\begin{itemize}
    \item $\textrm{Pr}_{\textrm{apple}} = 0.067\, $.
    \item $\textrm{Pr}_{\textrm{pollution}} = 0.6\, $.
    \item $H_{\textrm{abundance}} = 0.3\, $.
    \item $H_{\textrm{depletion}} = 0.6\, $.
\end{itemize}

\begin{figure}[!t]
    \centering
    \subfloat[]{\includegraphics[width=9.5cm]{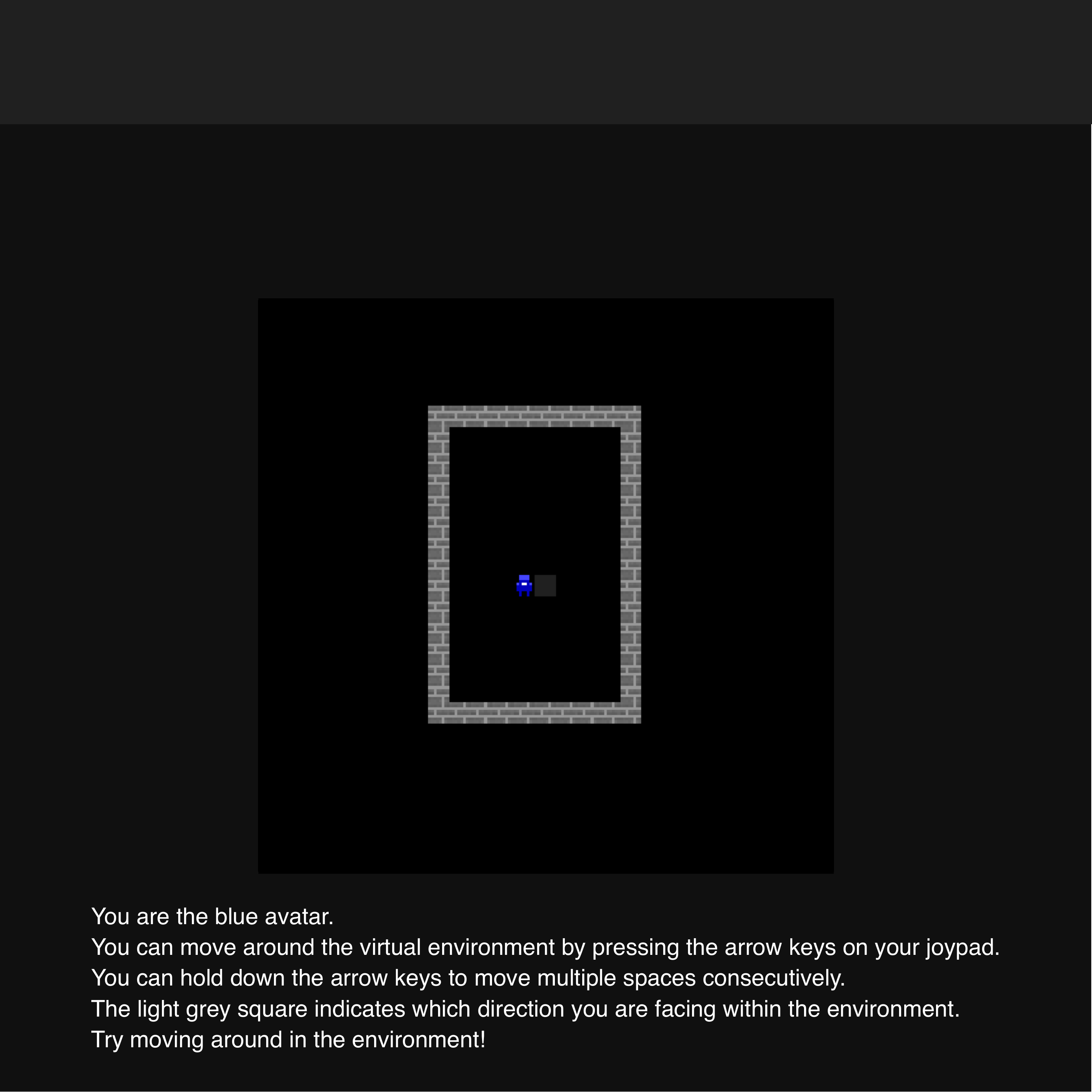}} \\
    \subfloat[]{\includegraphics[width=9.5cm]{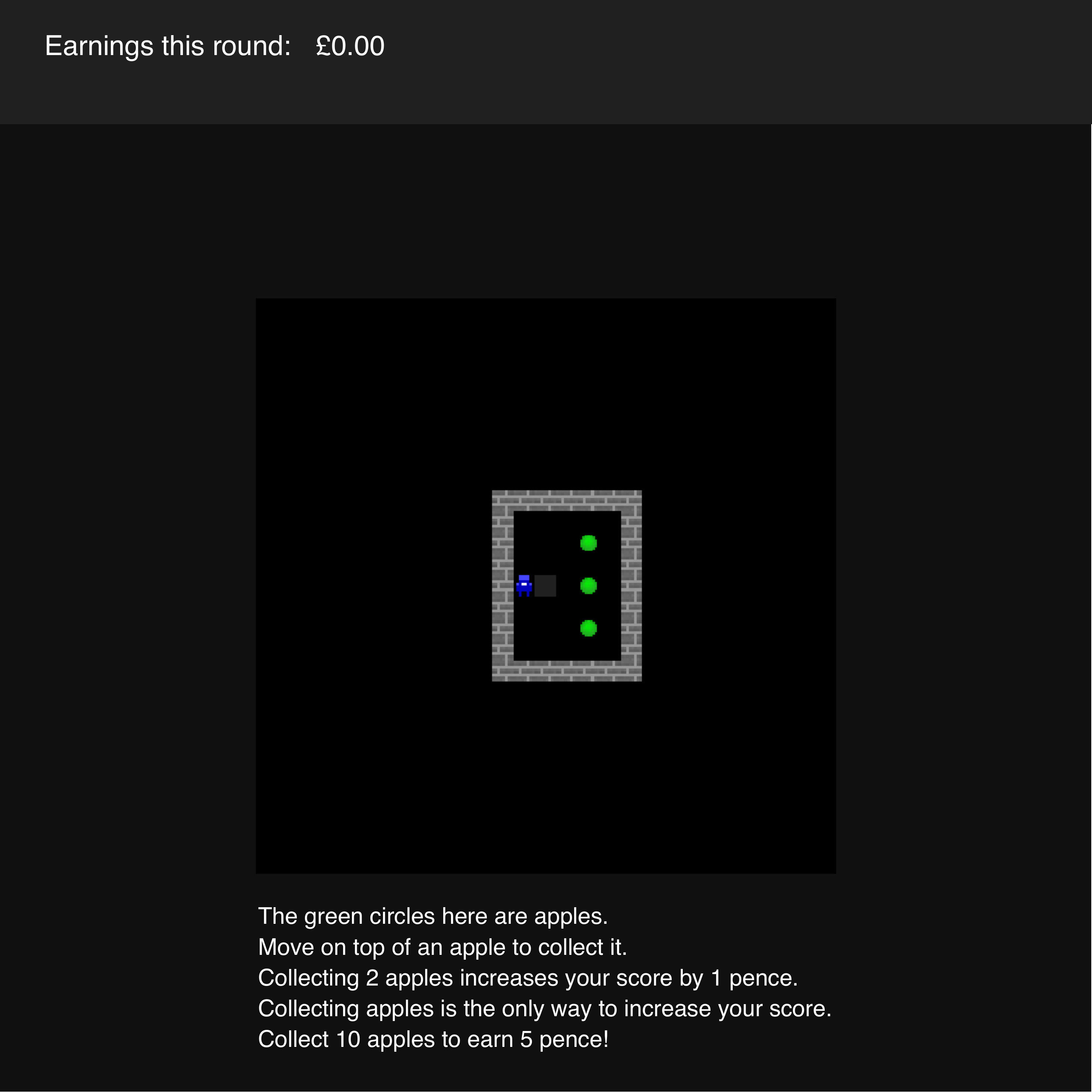}}
    \caption{\small Figure \ref{fig:tutorials_a_b}: a-b: Participants completed a number of tutorials to help them learn the controls for the task and the environmental dynamics of Cleanup.}
    \label{fig:tutorials_a_b}
\end{figure}

\begin{figure}[!t]
    \centering
    \subfloat[]{\includegraphics[width=9.5cm]{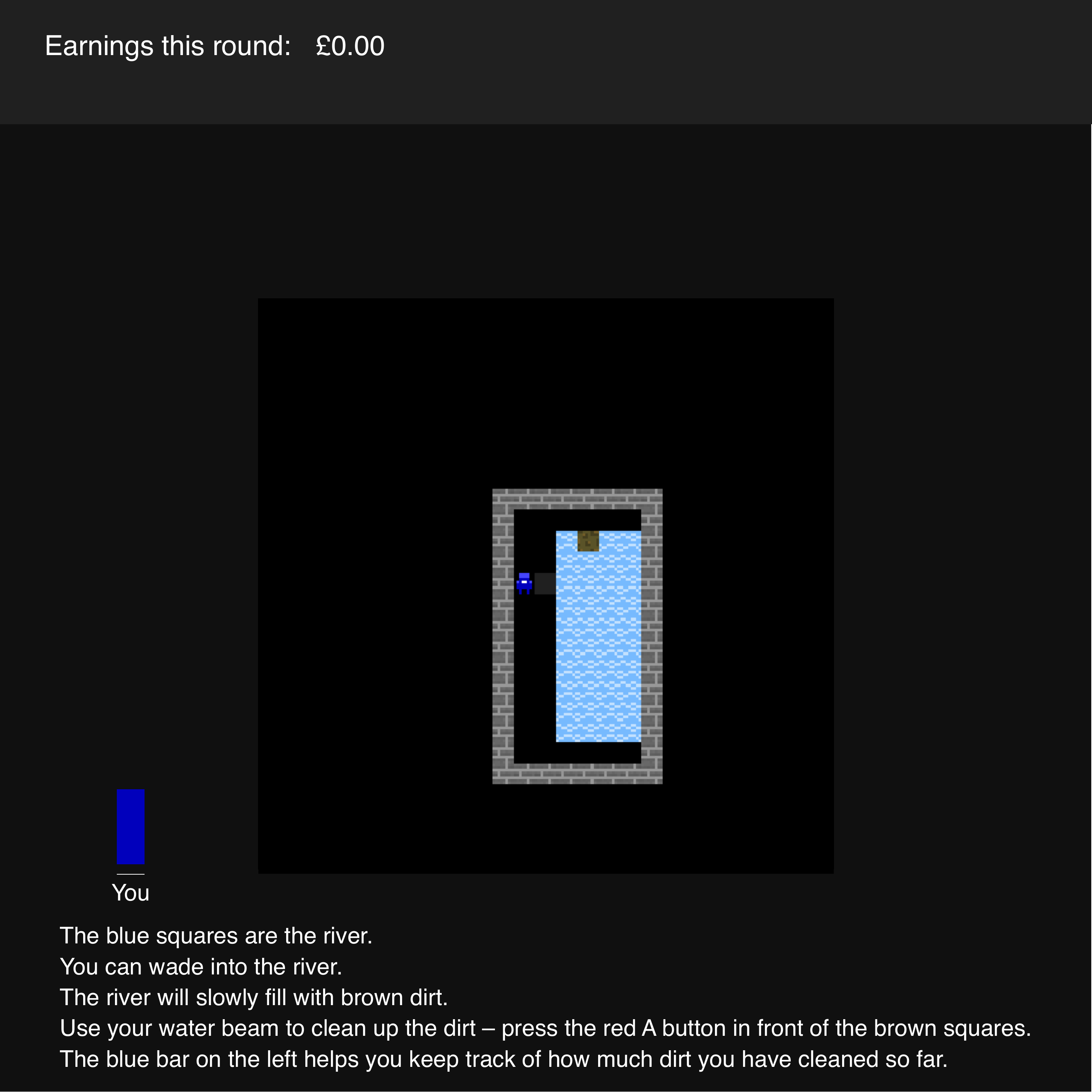}} \\
    \subfloat[]{\includegraphics[width=9.5cm]{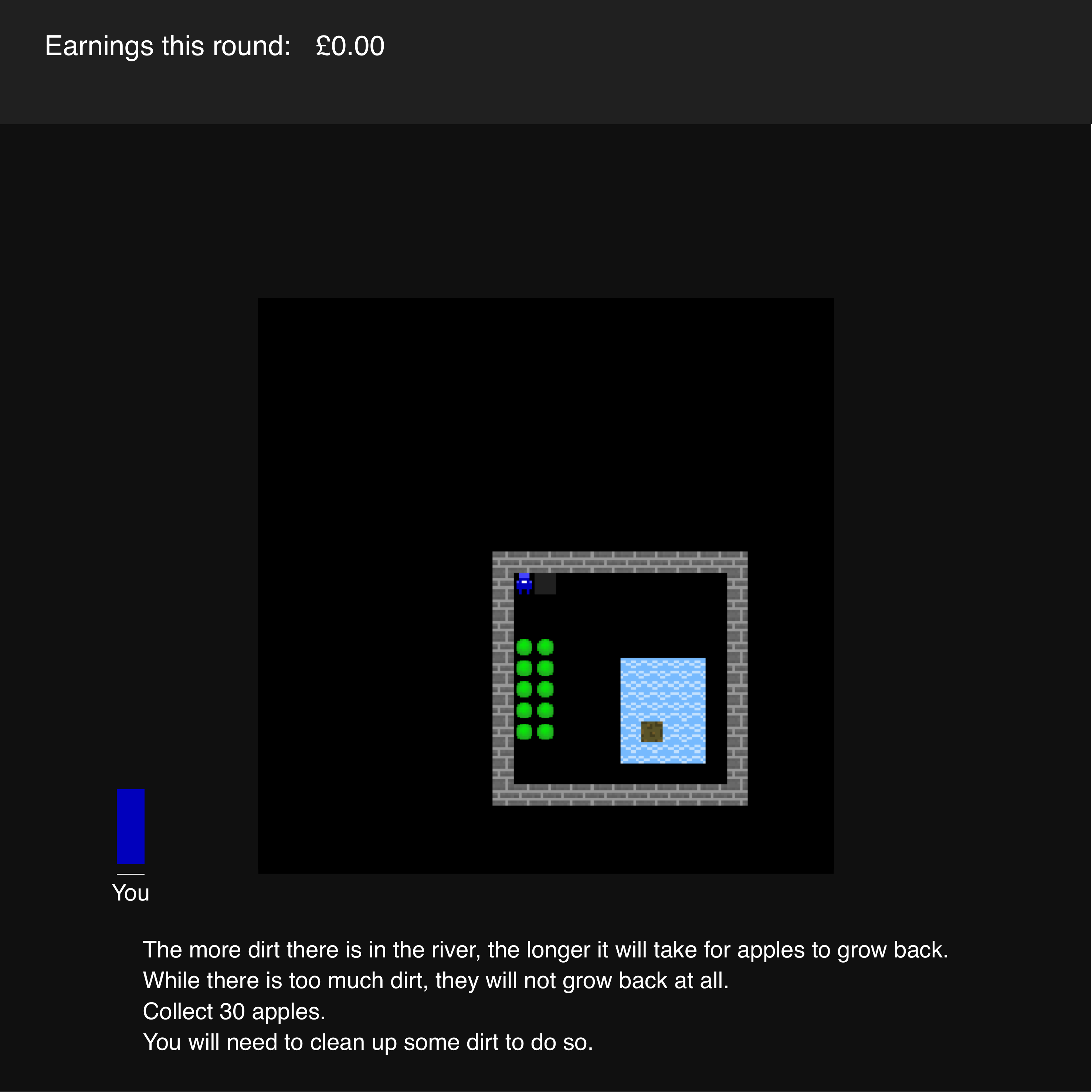}}
    \caption{\small Figure \ref{fig:tutorials_c_d}: a-b: Participants completed a number of tutorials to help them learn the controls for the task and the environmental dynamics of Cleanup.}
    \label{fig:tutorials_c_d}
\end{figure}

\begin{figure}[!t]
    \centering
    \subfloat[]{\includegraphics[width=9.5cm]{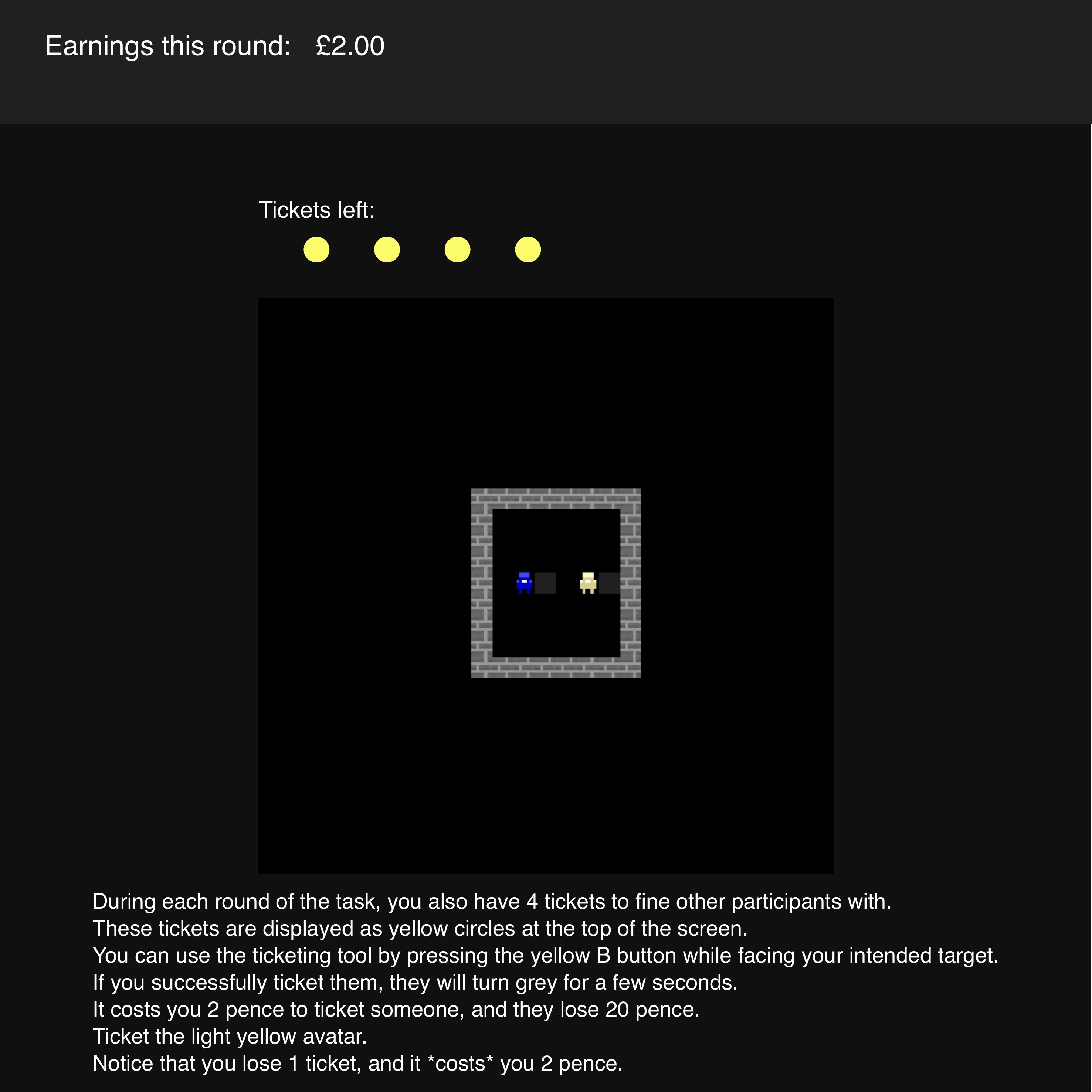}} \\
    \subfloat[]{\includegraphics[width=9.5cm]{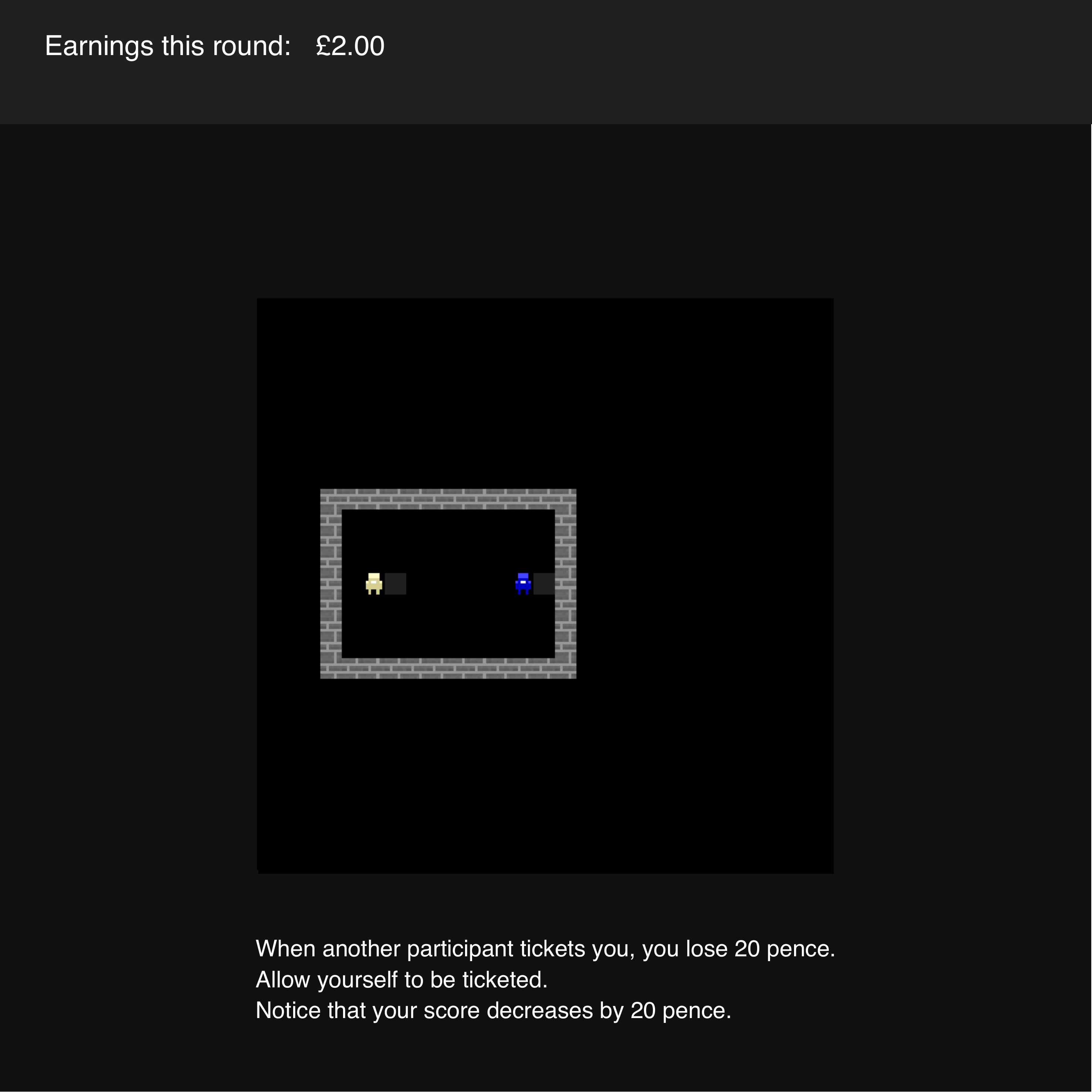}}
    \caption{\small Figure \ref{fig:tutorials_e_f}: a-b: Participants completed a number of tutorials to help them learn the controls for the task and the environmental dynamics of Cleanup.}
    \label{fig:tutorials_e_f}
\end{figure}

During the first stage of the experiment, participants receive a series of six tutorials on the action controls and the environmental dynamics for the Cleanup task (Figures \ref{fig:tutorials_a_b}-\ref{fig:tutorials_e_f}). The tutorials aimed to familiarize participants with (1) avatar movement, (2) apple collection, (3) pollution accumulation and the cleaning tool, (4) the effects of pollution on apple growth, (5) the ticketing tool and the cost of giving a ticket, and (6) the cost of receiving a ticket. Participants were subsequently instructed on the group nature of the task, including an explanation of the symmetry of information available about their own behavior to themselves and to their peers. Participants were also informed of the performance incentivization (i.e., the rules for receiving a bonus) at this stage. The tutorials described the river pollution as ``dirt'' to avoid explicitly priming participants with environmental concerns or pro-sustainability motives.

\begin{figure}[ht]
    \centering
    \subfloat[]{\includegraphics[width=7.5cm]{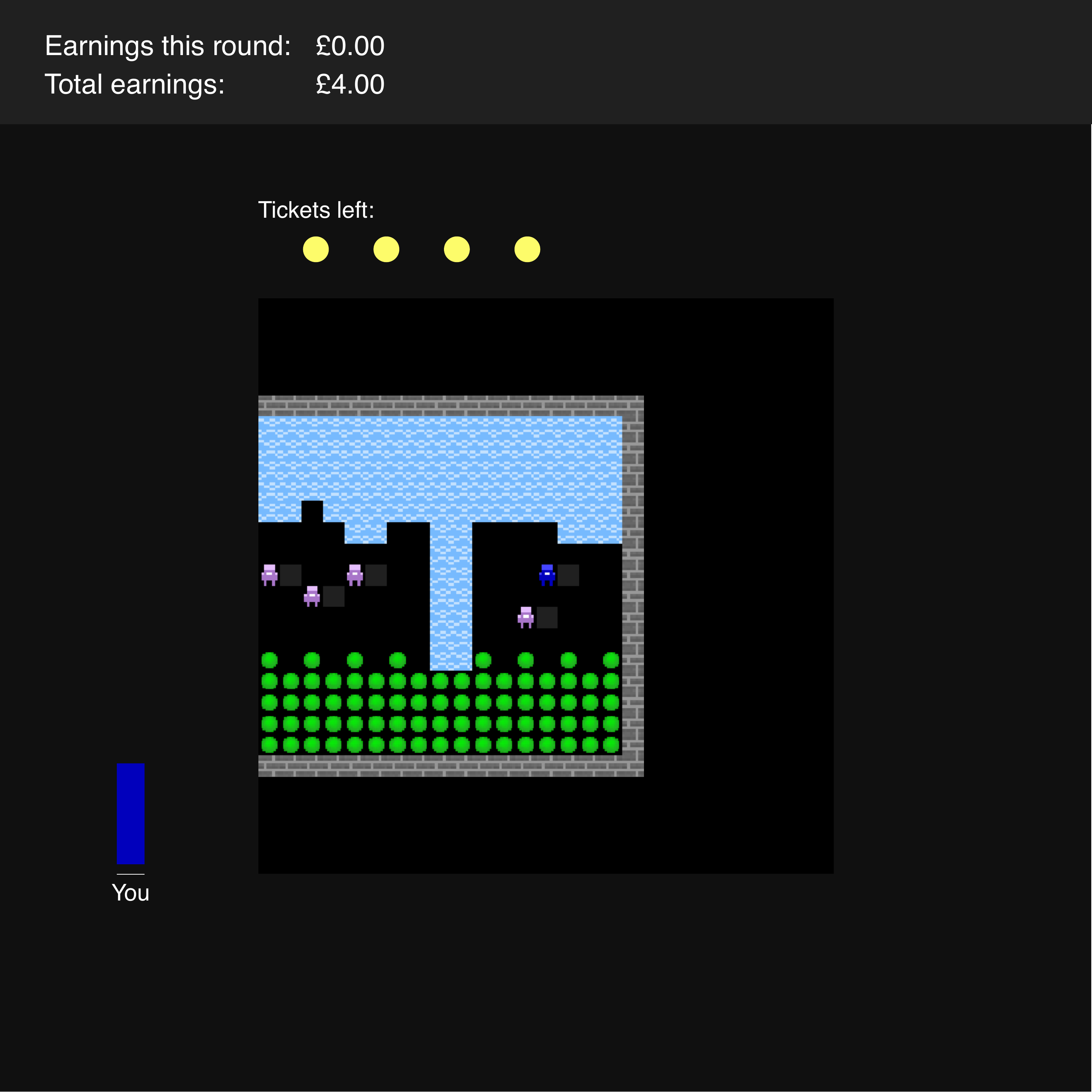}} \:
    \subfloat[]{\includegraphics[width=7.5cm]{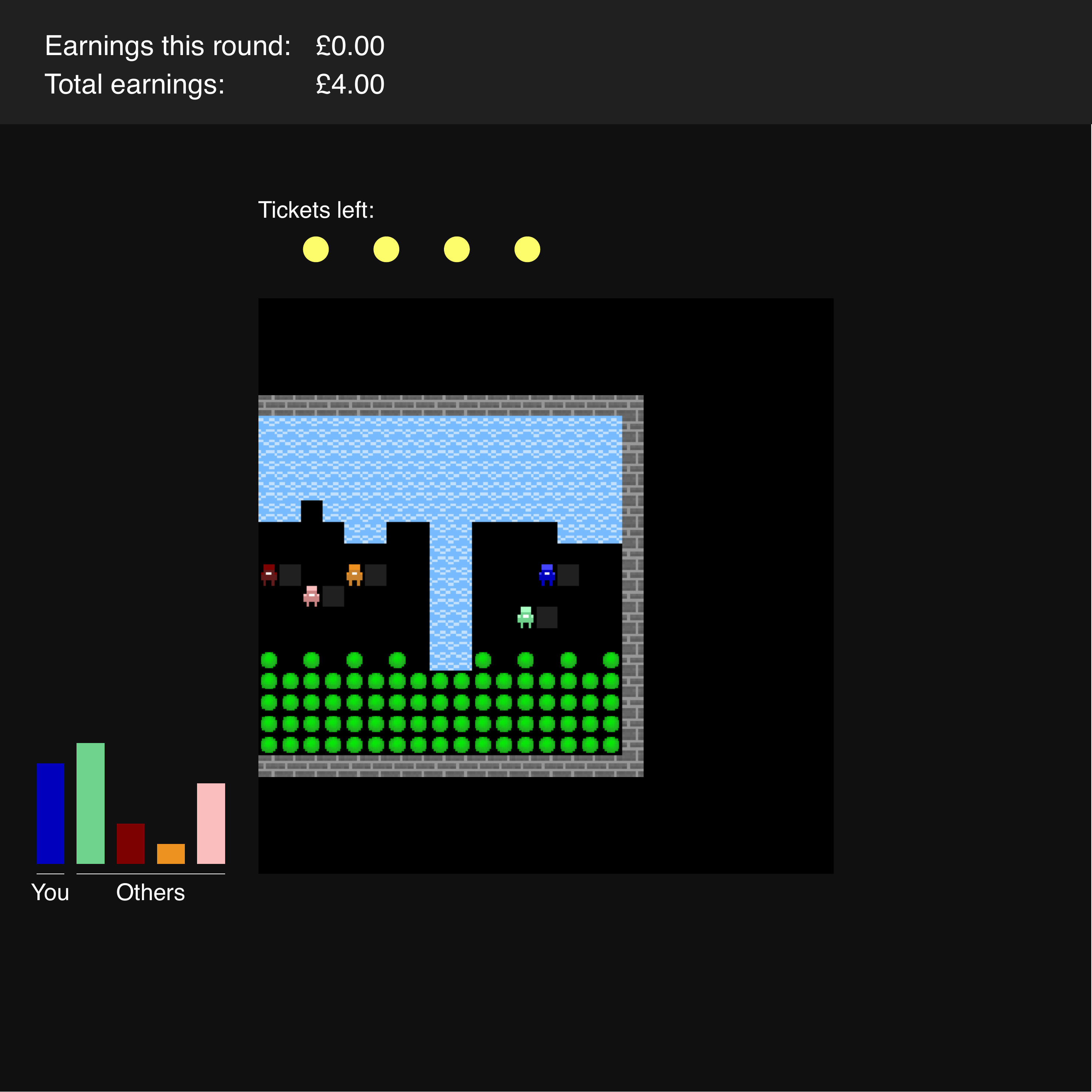}}
    \caption{\small Figure \ref{fig:screenshot}: Participant view of the Cleanup task varied by condition. In both conditions, participants observed the environment through a 27 by 27 window, centered on their avatar (colored blue). Participants also observed their earnings for the current episode, their cumulative earnings through the current episode, and the number of tickets they had available. (a) In the anonymous condition, other participants were represented by lavender avatars. Each participant observed their own contribution level (the blue bar), but did not receive the contribution levels of others in their group. (b) In the identifiable condition, other participants were represented by uniquely colored avatars. Each participant observed their own contribution level, as well as the contribution levels of the others in their group (other colored bars).}
    \label{fig:screenshot}
\end{figure}

During episodes, participants observed the environment through a 27 by 27 window, centered around their avatar (Figure \ref{fig:screenshot}). The size of this observation window allows for a participant to view the entire map by standing in the middle of the map. However, practically speaking, participants spent the majority of the time playing with imperfect observability. In 94.3\% of steps, participants were positioned such that part of the map was obscured from their observation.

Episodes ran for $T=2000$ steps (approximately 2 minutes). Participants were not told the exact length of the episodes. At the end of every episode, each participant’s score for that episode was added to his or her cumulative score for the entire experiment. After the experiment, participants were paid a base payment of £15 and a bonus for their cumulative score at the rate of 0.5 pence per point, up to a maximum bonus of £30.

The University College London Research Ethics Committee conducted ethical review for the project and approved the study protocol (CPB/2013/015). All participants provided informed consent for the study.

The experiment was completed by 120 participants (age: mean $m = 21.5$, standard deviation $sd = 2.3$; gender: 50 male, 70 female), drawn from the University College London psychology department participant pool.\footnote{During one experimental session (i.e., ten participants in the lab for the experiment), one participant’s computer malfunctioned for several episodes. Participants in the dropped session were paid and debriefed as normal. To maintain a balanced design, the entire session was dropped from data analysis and an additional session was scheduled to fulfill the original design. Counting this dropped session, 130 participants were recruited across 13 sessions.} Participants earned an average of £30.85 ($sd =$ £6.44) during the experiment.

\section{Social Dilemma Analysis} \label{sec:social_dilemma_analysis}

Social dilemmas are situations in which there is a tradeoff between short-term individual incentives and long-term collective interest \citep{hughes2018inequity, kollock1998social, rapoport1974prisoner}. In this paper, we study public goods dilemmas, a particular subset of social dilemmas.

In the canonical public goods task \citep{fehr2000cooperation}, $n$ participants receive an initial endowment of $e$ tokens and choose a contribution level $c \in \{0, \dots, e\}$. Subsequently, contributions are pooled and increased by multiplication factor $M$, forming the public good, $G$:
\begin{equation}
   G = M \sum_{j = 1}^{n}{c_j} \label{public_goods_formula} \, .
\end{equation}
The public good is then distributed evenly across all $n$ participants. The collective group payoff (i.e., environmental reward) is consequently described by:
\begin{equation}
   U_{\textrm{total}} = n \cdot e + (M - 1) \sum_{j = 1}^{n}{c_j} \, , \label{group_payoff_formula}
\end{equation}
where the payoff to participant $k$ is described by:
\begin{equation}
   u_k = e - c_k + \frac{M}{n} \sum_{j = 1}^{n}{c_j} \, . \label{individual_payoff_formula}
\end{equation}

In the canonical setting, there is a deterministic relationship between the group's contributions, the size of the public good, and the group payoff. Similarly, the equal distribution of the public good deterministically defines the payoffs for individual participants. As a result, in the traditional public goods task, we can purposefully instantiate a social dilemma---a situation in which there is a tradeoff between individual incentives and the collective interest---by requiring $1 < M < n$.

\begin{figure}[ht]
    \centering
    \subfloat[]{\includegraphics[width=6cm]{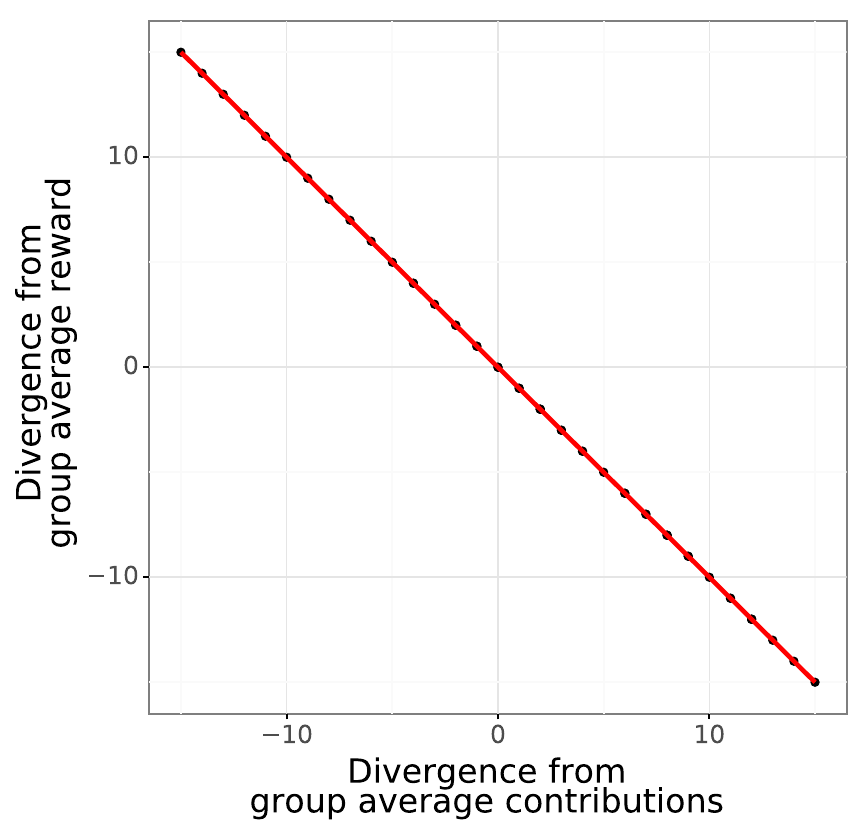}}
    \subfloat[]{\includegraphics[width=6cm]{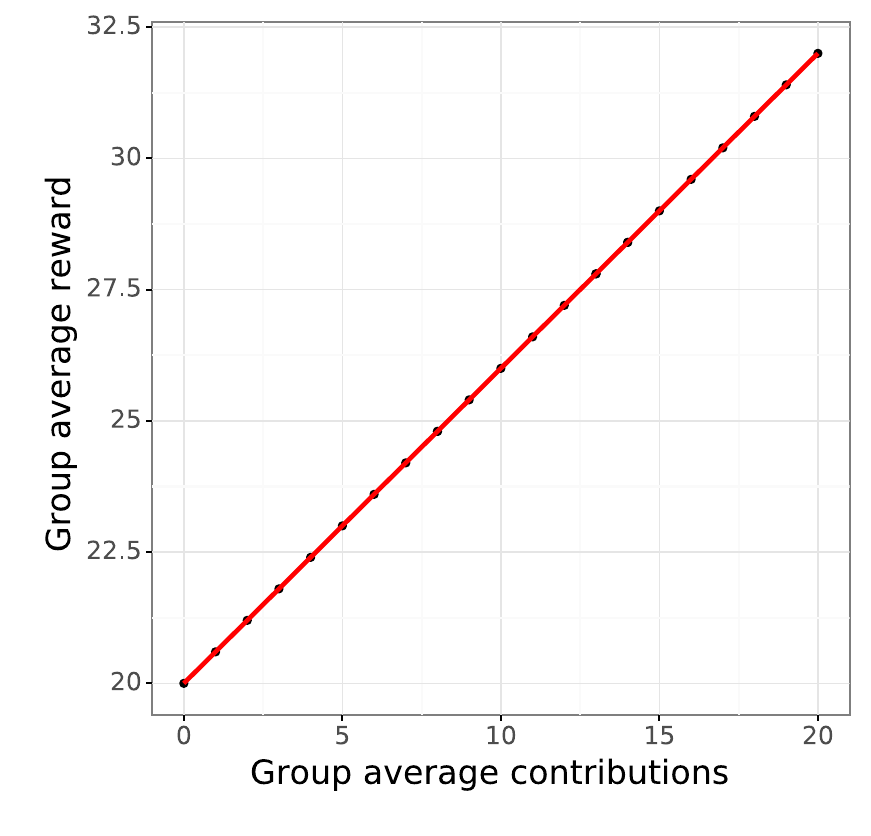}}
    \caption{\small Figure \ref{fig:canonical_dilemma}: The incentive structure of the canonical public goods task \citep{fischbacher2001people} exhibits a distinctive sign reversal between the individual and group level. Contribution level is negatively correlated with reward within groups, but positively correlated between groups. (a) Participants who contribute less relative to their group receive higher reward. (b) However, groups that contribute more earn more reward.}
    \label{fig:canonical_dilemma}
\end{figure}

In the traditional parameterization of the public goods task, $n = 4$, $e = 20$, and $M = 1.6$ \citep{fehr2000cooperation}. With these parameters, the payoff functions \ref{group_payoff_formula} and \ref{individual_payoff_formula} respectively reflect a negative relationship between own contributions and own payoff, \textit{ceteris paribus}, and a positive relationship between group contributions and group payoff (Figure \ref{fig:canonical_dilemma}). This sign reversal is the defining characteristic of individual- and group-level incentive structures for social dilemmas.\footnote{The sign reversal between the individual and group level is conceptually related to Simpson's paradox, a well-studied statistical phenomenon whereby the direction of an association at the population-level reverses within the subgroups comprising that population (as described in \cite{simpson1951interpretation}; see also \cite{kievit2013simpson, pearson1899vi}).}

In stochastic, temporally and spatially extended tasks (cf., \cite{littman1994markov}), it is not straightforward to instantiate social-dilemma incentive structures with environmental rules \citep{leibo2017multi}. In our work, given the variation in environmental parameters between the human behavioral experiment and the model, it was especially important to confirm that both environment versions constituted a social dilemma.

Using linear regression, we characterize the extent to which empirical data match the sign-reversal pattern observed in the canonical case---what we might call a \textit{linear social dilemma} structure (cf., \cite{zelmer2003linear}). We conduct this analysis with two separate regressions. The first examines the individual-level incentive structure experienced by group members. The second examines the effects of collective behavior on group welfare. To match a linear social dilemma structure, we expect to see a negative relationship between cooperation and payoff at the individual level, but a positive relationship between cooperation and welfare at the group level.

\begin{gather}
   Y_j = \beta_0 + \beta_1 \cdot (c_j - \bar{c}) + \epsilon \, \\
   Y_g = \beta_0 + \beta_1 \cdot \bar{c} + \epsilon \, .
\end{gather}

In addition to this linear-dilemma analysis, we follow recent explorations in the multi-agent field by producing empirical Schelling diagrams \citep{perolat2017multi, hughes2018inequity}. Empirical Schelling diagrams are a data-driven approach to analyzing Markov games which characterize payoffs based on group policy composition. Visualizing payoff structures in this way has various benefits, including the opportunity to inspect whether the task reflects a social dilemma and the ability to identify game-theoretic concepts such as Nash equilibria and Pareto-optimal outcomes \citep{schelling1973hockey}. Schelling diagrams rely upon the ability to dichotomize the policy space (e.g., into cooperation and defection policies), as well as the ability to categorize individuals based on their observed behavioral trajectories.

Given the well-established use of punishment to decrease the payoffs of free riders \citep{henrich2006costly, gachter2008long}, for these analyses we use apple consumption to model payoff and welfare.

\subsection{Human Behavioral Experiment}

\begin{figure}[ht]
    \centering
    \subfloat[]{\includegraphics[width=6cm]{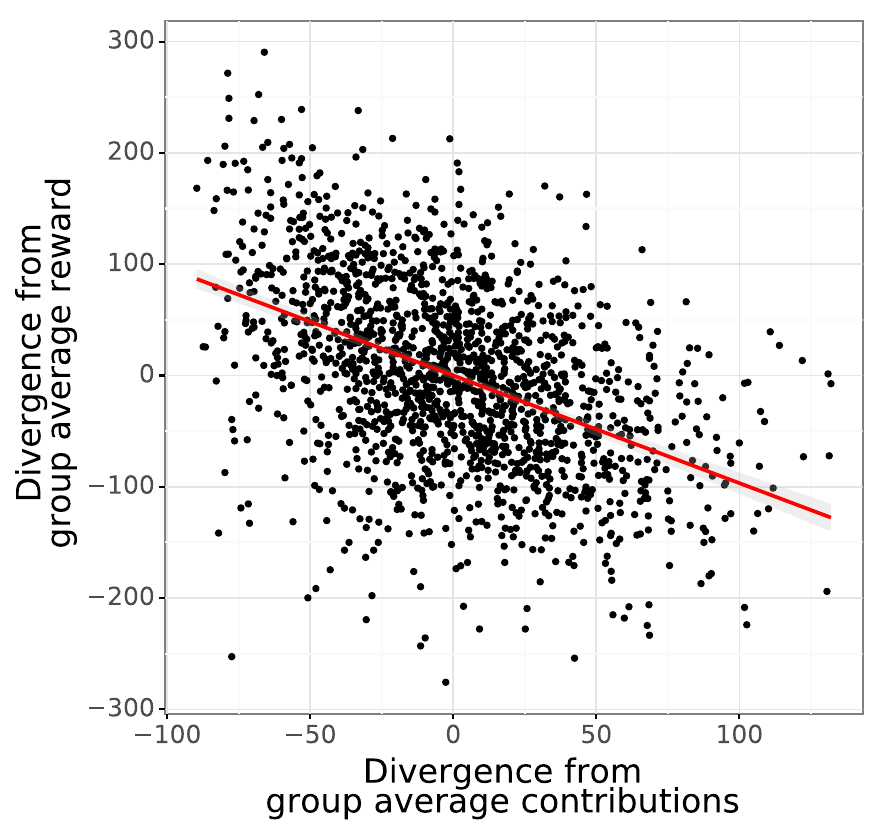}}
    \subfloat[]{\includegraphics[width=6cm]{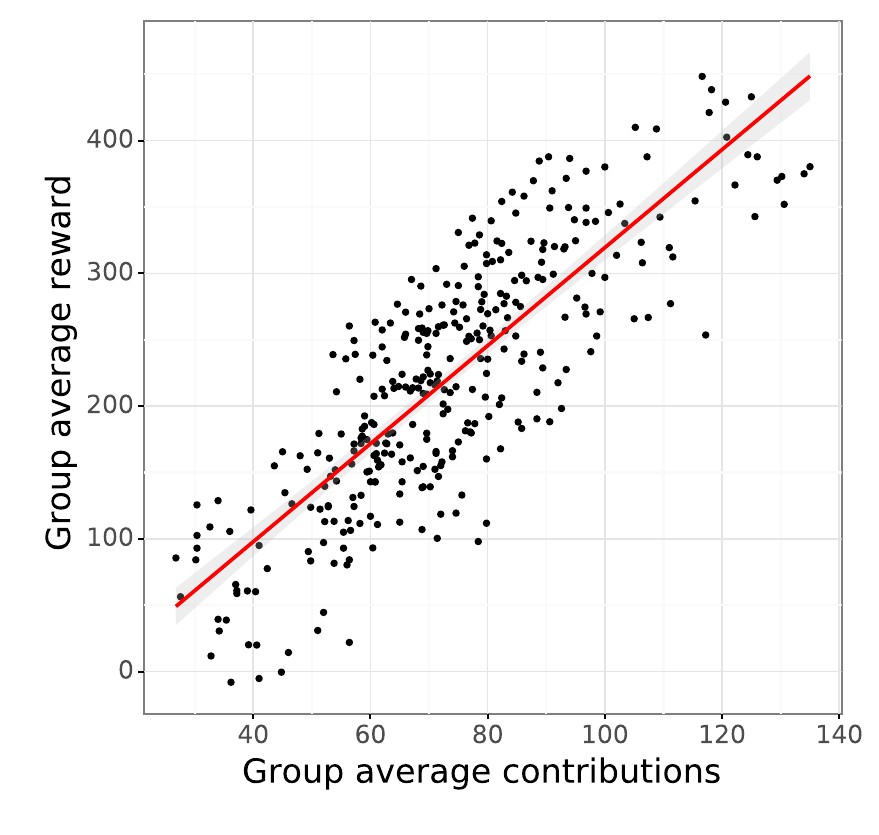}}
    \caption{\small Figure \ref{fig:empirical_dilemma}: Empirical inspection reveals a linear social dilemma structure for humans completing the Cleanup task. Contribution level is negatively correlated with reward within groups, but positively correlated between groups.  Errors bands reflect 95\% confidence intervals. (a) Participants who contribute less relative to their group receive higher reward, $\beta = -0.97$, 95\% CI $[-1.06, -0.88]$, $p < 0.0001$. (b) However, groups that contribute more generate higher welfare, $\beta = 3.69$, 95\% CI $[3.42, 3.97]$, $p < 0.0001$.}
    \label{fig:empirical_dilemma}
\end{figure}

At the individual level, the amount that a participant's contributions exceeded or fell below their group's average contribution level had a significant negative relationship with the amount their payoff exceeded or fell below the group average payoff, $\beta = -0.97$, 95\% CI $[-1.06, -0.88]$, $p < 0.0001$ (Figure \ref{fig:empirical_dilemma}a). Participants who contributed more relative to their peers collected fewer apples than their peers.

At the group level, average contribution level was significantly and positively associated with average apple consumption, $\beta = 3.69$, 95\% CI $[3.42, 3.97]$, $p < 0.0001$ (Figure \ref{fig:empirical_dilemma}b). In contrast with the negative individual-level association, groups that collectively contribute more also collectively consume more apples.

Taken together, these two effects support the existence of a linear social dilemma incentive structure within Cleanup.

\begin{figure}[ht]
    \centering
    \includegraphics[width=8.75cm]{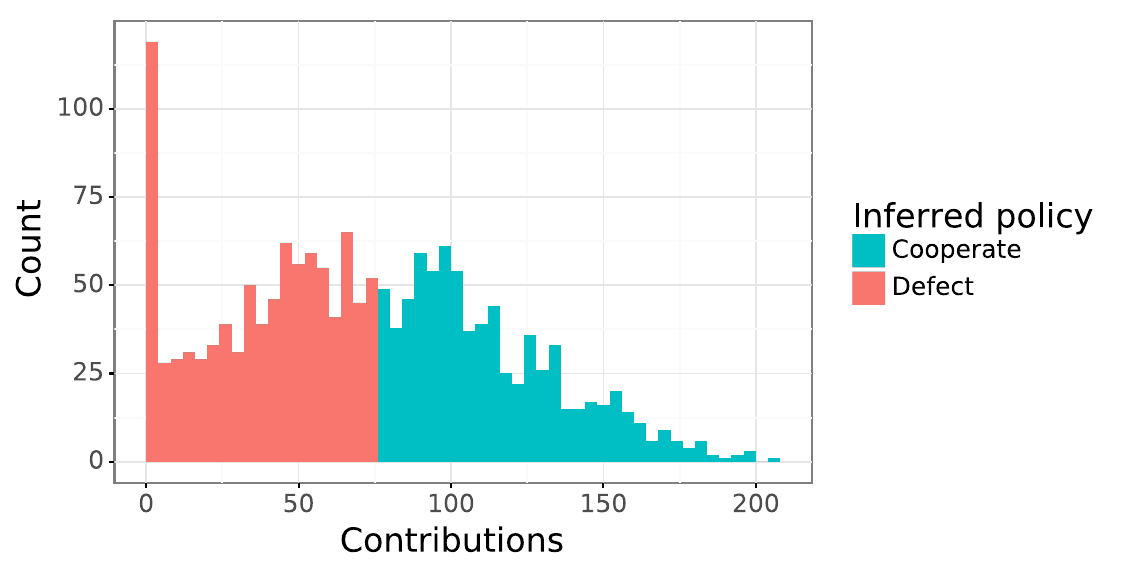}
    \caption{\small Figure \ref{fig:jenks_contributions}: Individual contribution levels were variable across participants and episodes. We use the Jenks natural breaks method to dichotomize this distribution. This results in the categorization of contribution levels below 76 as ``defect'' policies and contribution levels at or above 76 as ``cooperate'' policies.}
    \label{fig:jenks_contributions}
\end{figure}

We sought to verify this finding by generating an empirical Schelling diagram for the human behavioral experiment. In effect, we map empirical data onto a policy space by categorizing observed behavioral trajectories into ``cooperate'' trajectories and ``defect'' trajectories. We chose to dichotomize participant contribution levels using the Jenks optimization method. This method identifies cutoff points which minimize within-category variance and maximize between-category variance, given a number of categories to establish \citep{jenks1967data}. The Jenks method identifies 76 contribution steps as the natural breakpoint dichotomizing the distribution of participant contribution levels (Figure \ref{fig:jenks_contributions}).

\begin{figure}[ht]
    \centering
    \includegraphics[width=8.75cm]{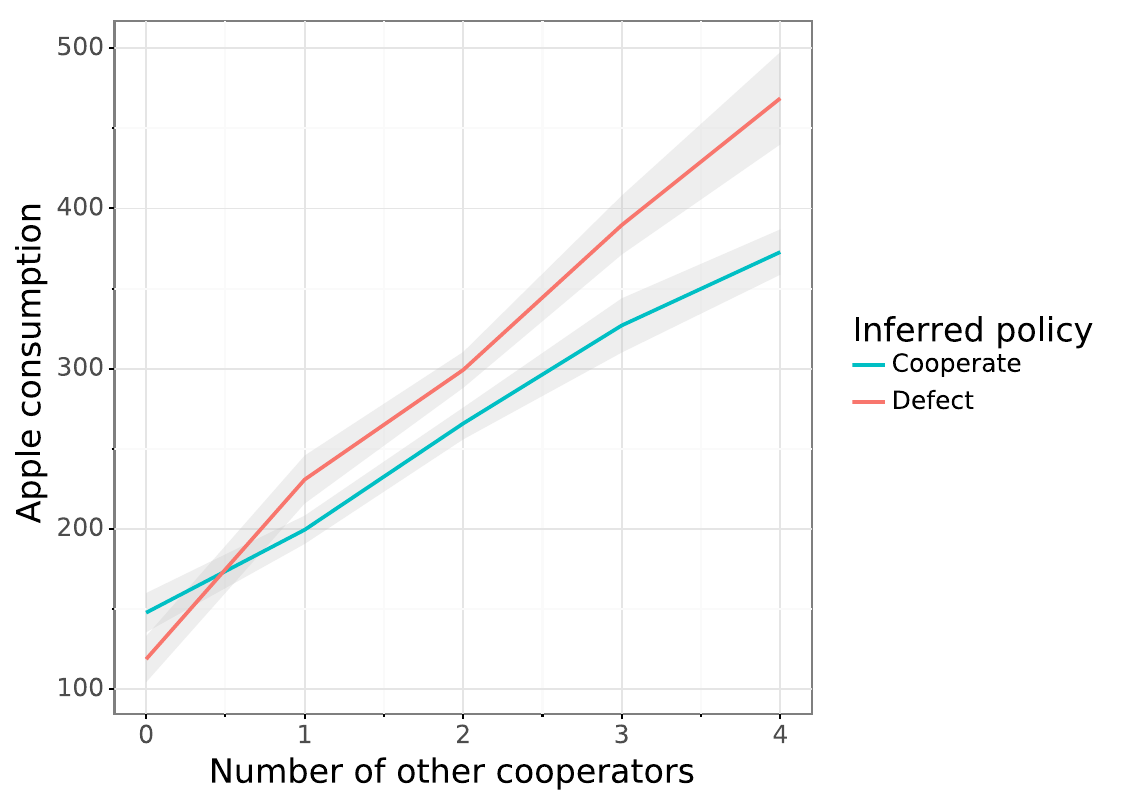}
    \caption{\small Figure \ref{fig:human_schelling}: An inspection of the empirical Schelling diagram indicates that the Cleanup tasks meets \cite{hughes2018inequity}'s definition of a social dilemma.  Errors bands reflect 95\% confidence intervals.}
    \label{fig:human_schelling}
\end{figure}

We classified participants as cooperating in a given episode if they contributed more than this threshold and as defecting if they contributed less. Subsequently, we tabulated the number of cooperators and defectors in each episode for every group. As before, we examine apple consumption as the relevant payoff for participants. Average apple consumption is calculated separately among cooperators and defectors for each episode. To construct the Schelling diagram, average apple consumption for cooperators and defectors is plotted against the count of cooperators in each episode (Figure \ref{fig:human_schelling}).

\cite{hughes2018inequity} delineate the following conditions to define a binary-choice social dilemma:

\begin{enumerate}
    \item Mutual cooperation is preferred over mutual defection: $R_c(N) > R_d(0)$.
    \item Mutual cooperation is preferred to being exploited by defectors: $R_c(N) > R_c(0)$.
    \item Either...
    \begin{enumerate}
        \item Mutual defection is preferred to being exploited (fear): $R_d(i) > R_c(i)$ for sufficiently small $i$,
        \item Or exploiting a cooperator is preferred to mutual cooperation (greed). $R_d(i) > R_c(i)$ for sufficiently large $i$.
    \end{enumerate}
\end{enumerate}

This definition can be translated to a frequentist framework by instantiating each of these conditions with a one-sided independent \textit{t}-test. These \textit{t}-tests produce the following results with the experimental Cleanup data:

\begin{itemize}
    \item Condition 1 is met. The payoff to cooperators under mutual cooperation ($m = 372.2$, $sd = 35.23$) is significantly higher than the payoff to defectors under mutual defection ($m = 109.6$, $sd = 28.6$), $t(36.4) = 25.9$, $p_1 < 0.0001$.
    \item Condition 2 is met. The payoff to cooperators under mutual cooperation ($m = 372.2$, $sd = 35.23$) is significantly higher than the payoff to cooperators when all other group members defect ($m = 150.0$, $sd = 56.7$), $t(63.4) = 22.8$, $p_2 < 0.0001$.
    \item Condition 3a is not met. The payoff to defectors under mutual defection ($m = 109.6$, $sd = 28.6$) is not significantly higher than the payoff to cooperators when all other group members defect ($m = 150.0$, $sd = 56.7$), $t(44.4) = -4.2$, $p_{3a} = 1.00$.
    \item Condition 3b is met. The payoff to defectors when all other group members cooperate ($m = 465.5$, $sd = 84.8$) is higher than the payoff to cooperators under mutual cooperation ($m = 372.2$, $sd = 35.23$), $t(43.8) = 5.6$, $p_{3b} < 0.0001$.
\end{itemize}

We synthesize the results of these \textit{t}-tests ($p_1$, $p_2$, $p_{3a}$, $p_{3b}$, respectively) into a single statistical test through two steps. First, we use Fisher's method \citep{fisher1925statistical} to consolidate $p_{3a}$ and $p_{3b}$ while controlling for multiple comparisons. This results in the joint \textit{p}-value $p_3$. Second, we use a maximum $p$-value approach to combine $p_1$, $p_2$, and $p_3$: $p_{\textrm{overall}} = \max \left(p_1, p_2, p_3\right)$.

In the first step, Fisher's method indicates a significant overall result for condition 3, $\chi^2(4) = 30.3$, $p_3 < 0.0001$. In the second step, combining conditions 1, 2, and 3 results in $p_{\textrm{overall}} < 0.0001$. According to this combined significance test, the Cleanup task meets \cite{hughes2018inequity}'s definition of a social dilemma.

\subsection{Computational Model}

\begin{figure}[ht]
    \centering
    \includegraphics[width=8.75cm]{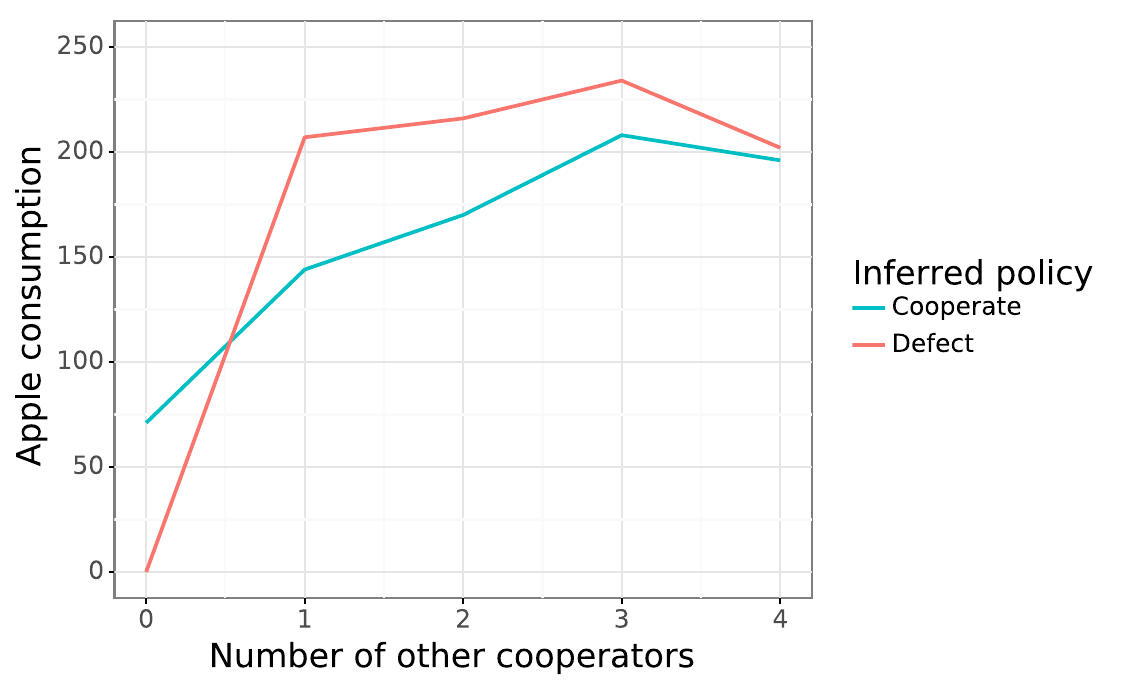}
    \caption{\small Figure \ref{fig:rl_schelling}: Previous work with reinforcement learning agents \cite{hughes2018inequity} demonstrated that the Cleanup task has a social dilemma incentive structure. Reproduced with permission from \cite{hughes2018inequity}.}
    \label{fig:rl_schelling}
\end{figure}

Previous work by \cite{hughes2018inequity} verified that the incentive structure of Cleanup produces social dilemma pressures for reinforcement learning agents (Figure \ref{fig:rl_schelling}). In that work, agents were trained with a specialized protocol. During training, the ability to contribute to the public good was withheld from some agents; a small group reward signal was added to the remaining agents. The former and latter types of agent were classified as defectors and cooperators, respectively. As a result, groups varied in their composition of cooperating and defecting policies. Echoing our findings with the human behavioral analysis, Hughes and colleagues found that the resulting empirical Schelling diagram matched condition 1 ($R_c[4] > R_d[0]$), condition 2 ($R_c[4] > R_c[0]$) and condition 3b ($R_d[4] > R_c[4]$).

Presaging our findings for the human participants, Hughes and colleagues concluded from this pattern that the Cleanup task instantiates a social dilemma for reinforcement learning agents.

\section{Comprehension Check}
\subsection{Human Behavioral Experiment}

Two open-ended questions at the end of the post-experiment questionnaire asked participants whether they attended to the bar display during the episodes when there was one bar and during the episodes when there were five bars. Responses were coded dichotomously as \textit{no, did not attend} or \textit{yes, attended} (Table \ref{tab:mcnemar_manip_check}).

\begin{table}[ht]
    \centering
    \begin{tabular}{c c|c c}
     & \multicolumn{1}{c}{} & \multicolumn{2}{c}{Attended to the} \\
     & \multicolumn{1}{c}{} & \multicolumn{2}{c}{1-bar display?} \\
     & & \multicolumn{1}{c}{Yes} & \multicolumn{1}{c}{No} \\ 
    \cline{2-4}
    Attended to the & \multicolumn{1}{r|}{Yes} & 39 & 74 \\  
    5-bar display? & \multicolumn{1}{r|}{No} & 0 & 7    
    \end{tabular}
    \caption{Frequency of participant \textit{Yes} and \textit{No} responses to the comprehension check question in the post-experiment questionnaire.}
    \label{tab:mcnemar_manip_check}
\end{table}

An exact McNemar’s test indicates a statistically significant difference in the proportion of participants who reported attending to the 1-bar display and the proportion who reported attending to the 5-bar display, $\chi^2 = 74$, $p < 0.0001$. A significantly greater proportion of participants reported attending to the 5-bar display (94.2\%) than reported attending to the 1-bar display (32.5\%).

\begin{figure}[ht]
    \leftskip 14em
    \subfloat[]{\includegraphics[width=8cm]{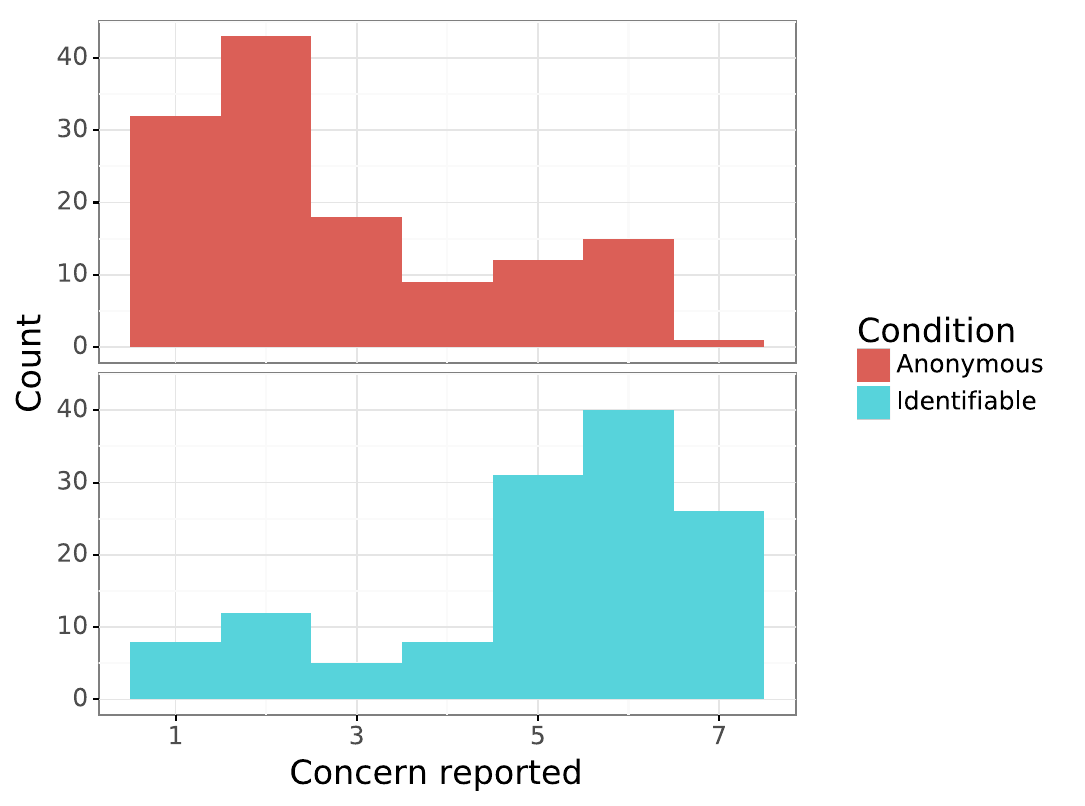}} \\
    \subfloat[]{\includegraphics[width=6cm]{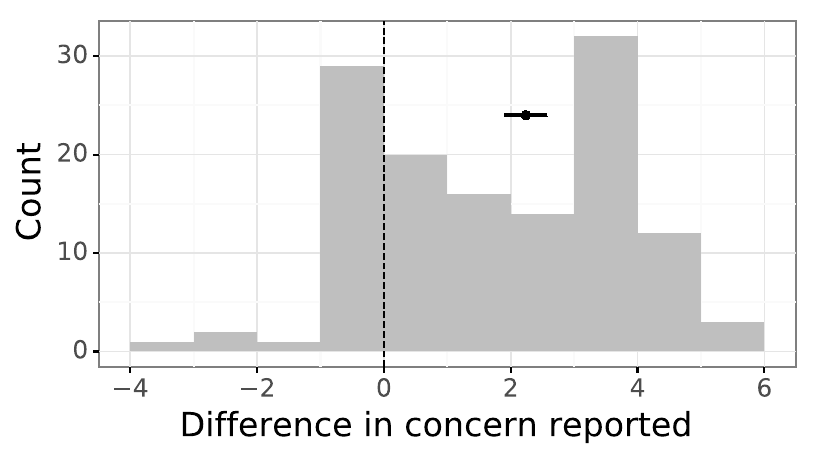}}
    \caption{\small Figure \ref{fig:concern_manip_check}: Above, (a) distributions of participant agreement with the statement ``I was concerned with what other participants thought about how much dirt I was cleaning up'' varied by condition. Below, (b) participants were significantly more likely to report such concern in the identifiable condition than in the anonymous condition. The plotted point reflects the sample mean difference ($m_{\textrm{diff}} = 2.21$), with error bars reflecting the 95\% confidence interval.}
    \label{fig:concern_manip_check}
\end{figure}

Using two 7-point scales, the questionnaire additionally assessed the degree to which participants were concerned about others tracking their behavior during each condition. Participant concern scores were centered on $m = 4.03$ ($sd = 1.81$) in the identifiable condition and on $m = 1.82$ ($sd = 1.71$) in the anonymous condition (Figure \ref{fig:concern_manip_check}). Participants reported being significantly more concerned about others tracking their contribution behavior in the identifiable condition than in the anonymous condition, $m_{\textrm{diff}} = 2.21$ ($95\%$ CI $= [1.85, 2.57]$), $t(119) = 12.25$, $p < 0.0001$ (Figure \ref{fig:concern_manip_check}).

\section{Group Learning Effects}
\subsection{Human Behavioral Experiment}

To evaluate potential learning effects for the behavioral experiment (i.e., whether participants were learning and improving at the task throughout the experiment), we explore between-episode trends in group performance. Following the example set by Fehr and Gachter \citep{fehr2002altruistic}, we conduct paired \textit{t}-tests to test for group learning over episodes. The \textit{t}-tests compared collective return between the first and last episodes for each task condition (separately for groups that experienced the Identifiable condition first or the Anonymous condition first).

\begin{figure}[ht]
    \centering
    \includegraphics[width=8.75cm]{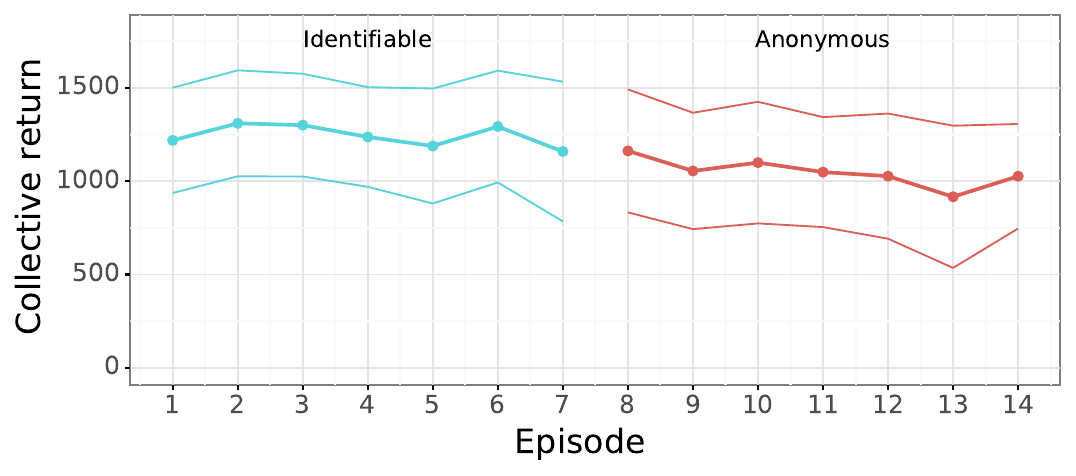}
    \caption{\small Figure \ref{fig:fehr_gachter_a}: Groups that completed the Identifiable condition first did not exhibit significant learning effects over episodes. Following \cite{fehr2000cooperation}, points on inner lines indicate the average collective return earned over all groups that completed the Identifiable task first. Outer lines indicate 95\% confidence intervals. There were no significant differences in group performance between the beginning and end of the Identifiable ($p = 0.59$) and Anonymous episodes ($p = 0.21$).}
    \label{fig:fehr_gachter_a}
\end{figure}

\begin{figure}[ht]
    \centering
    \includegraphics[width=8.75cm]{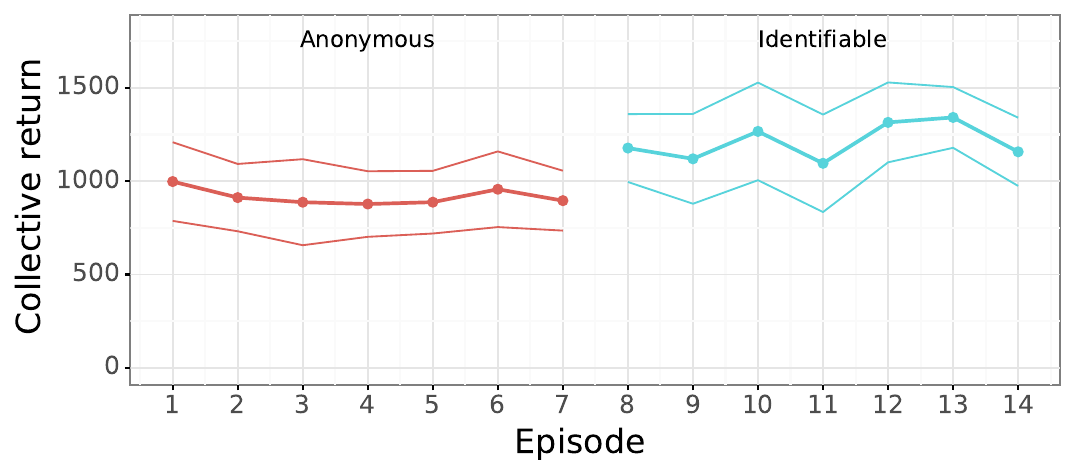}
    \caption{\small Figure \ref{fig:fehr_gachter_b}: Groups that completed the Anonymous condition first did not exhibit significant learning effects over episodes. Following \cite{fehr2000cooperation}, points on inner lines indicate the average collective return earned over all groups that completed the Identifiable task first. Outer lines indicate 95\% confidence intervals. There were no significant differences in group performance between the beginning and end of the Anonymous ($p = 0.38$) and Identifiable episodes ($p = 0.75$).}
    \label{fig:fehr_gachter_b}
\end{figure}

Groups that completed the Identifiable condition first did not exhibit a significant change in performance over Identifiable episodes, $t(11) = 0.6$, $p = 0.59$, or over Anonymous episodes, $t(11) = 1.3$, $p = 0.21$ (Figure \ref{fig:fehr_gachter_a}). Similarly, groups that completed the Anonymous condition first did not exhibit a significant change in performance over Anonymous episodes, $t(11) = 0.9$, $p = 0.38$, or over Identifiable episodes, $t(11) = 0.3$, $p = 0.75$ (Figure \ref{fig:fehr_gachter_b}).

\section{Main Group Effects} \label{sec:main_effects}
\subsection{Computational Model}

The reinforcement learning agents in the computational model did not update their policy during the evaluation stage of the experiment. As a result, the order in which groups experienced the conditions did not influence their behavior. We consequently evaluate the effect of condition on each social outcome metric $Y_g$ with a one-way, repeated-measure analysis of variance (ANOVA):

\begin{equation} \label{eqn:agent_anova}
   Y_g = \beta_0 + \beta_1 \cdot \textrm{Condition} + \mu_g + \epsilon \, .
\end{equation}

We first conduct a one-way ANOVA for collective return. There was a significant effect of condition on collective return, $F(1,\:311) = 2030.3$, $p < 0.0001$. In the model, groups earned significantly more in the identifiable condition (604 points on average) than in the anonymous condition (339 points on average).

We next conduct a one-way ANOVA for group contribution level. There was a significant effect of condition on group contribution level, $F(1,\:311) = 1090.7$, $p < 0.0001$. In the model, groups cleaned significantly more in the identifiable condition (for 422 steps on average) in the identifiable condition than in the anonymous condition (299 steps on average).

\subsection{Human Behavioral Experiment}

The human behavioral experiment took a counterbalanced, within-participant design: each participant was exposed to all experimental conditions, and the ordering of conditions was balanced across participants. Half of the participant groups completed the identifiable condition first and the anonymous condition second, and the other half completed the anonymous condition first and the identifiable condition second.

We would like to evaluate the effect of condition (identifiability versus anonymity) on several outcome metrics. In the human behavioral experiment, the counterbalanced design indicates the use of two-way ANOVA. To facilitate comparison with the model results, in the main text we report the main effects from the two-way ANOVA. Here we present the full models, including the main effect of task and the interaction effect. For this experiment, a two-way, repeated-measures ANOVA was used to assess the effect of task condition (identifiability or anonymity) and task number (first task or second task) on each social outcome metric $Y_g$:

\begin{equation}
   Y_g = \beta_0 + \beta_1 \cdot \textrm{Condition} + \beta_2 \cdot \textrm{Task} + \beta_3 \cdot \textrm{Condition} \times \textrm{Task} + \mu_g + \epsilon \, .
\end{equation}

\begin{figure}[ht]
    \centering
    \includegraphics[width=8cm]{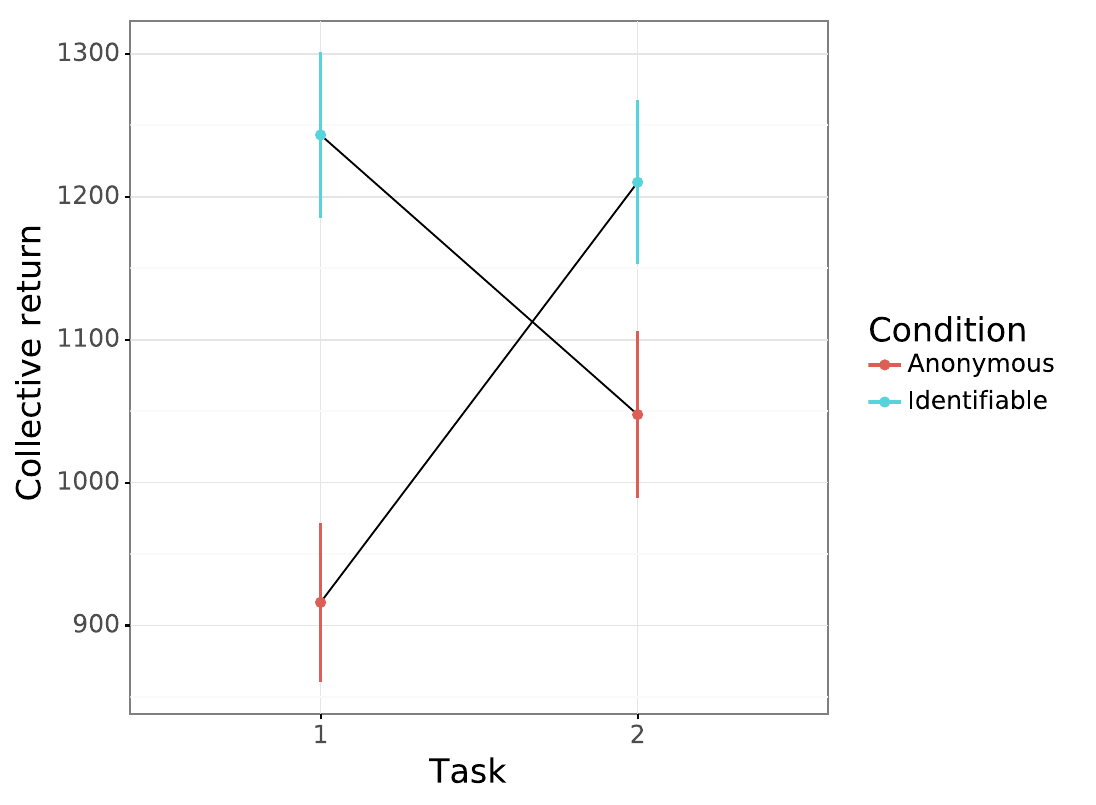}
    \caption{\small Figure \ref{fig:anova_collective_return}: We use a two-way, repeated-measures ANOVA to evaluate the effects of condition and task number on collective return. As expected, there was a significant main effect of condition on collective return. The main effect of task number and the interaction effect between condition and task number were not significant. Error bars reflect 95\% confidence intervals.}
    \label{fig:anova_collective_return}
\end{figure}

We first conduct a two-way ANOVA for collective return (Figure \ref{fig:anova_collective_return}). There was a significant main effect of condition on collective return, $F(1,\:310) = 89.4$, $p < 0.0001$. The main effect of task number was non-significant, $F(1,\:310) = 3.6$, $p = 0.059$. The interaction effect was also non-significant, $F(1,\:22) = 0.2, p = 0.63$. Groups earned significantly more in the identifiable condition (1227 points on average) than in the anonymous condition (982 points on average).

\begin{figure}[ht]
    \centering
    \includegraphics[width=8cm]{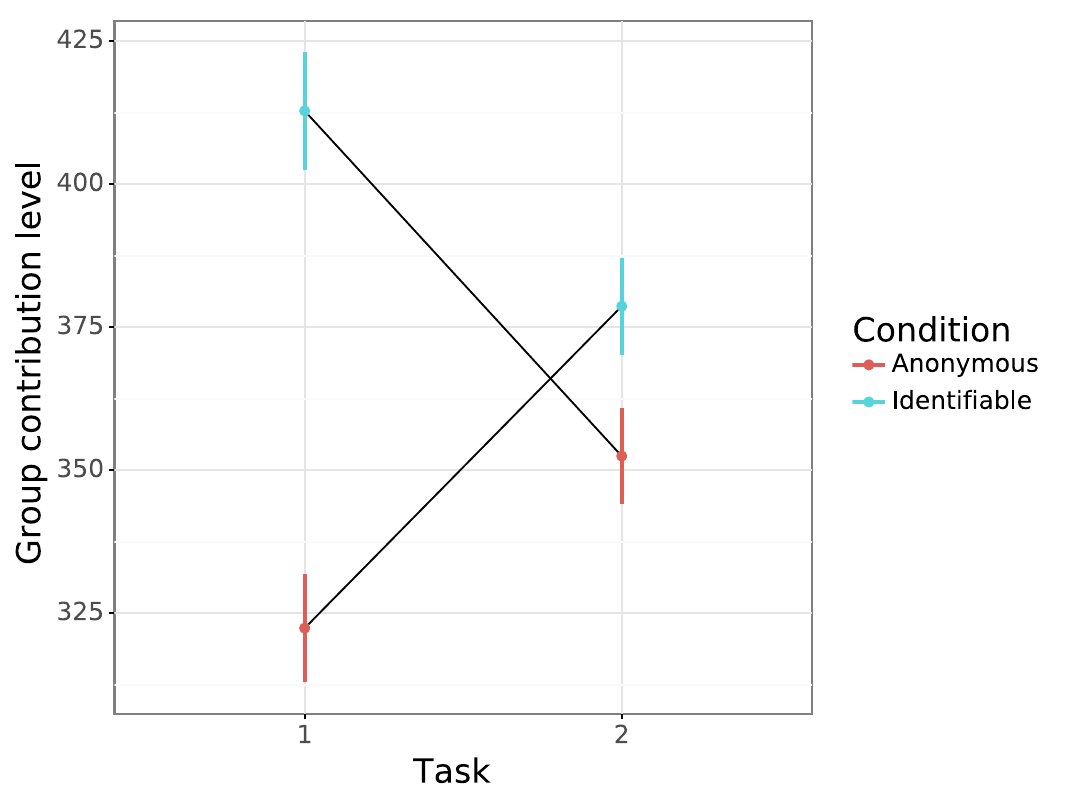}
    \caption{\small Figure \ref{fig:anova_contributions}: We use a two-way, repeated-measures ANOVA to evaluate the effects of condition and task number on group contribution level. As expected, there was a significant main effect of condition on group contribution level. The main effect of task number and the interaction effect between condition and task number were not significant. Error bars reflect 95\% confidence intervals.}
    \label{fig:anova_contributions}
\end{figure}

We next conduct a two-way ANOVA for total contribution level (Figure \ref{fig:anova_contributions}). There was a significant main effect of condition on total contribution level, $F(1,\:310) = 199.4$, $p < 0.0001$. The main effect of task number was non-significant, $F(1,\:310) = 0.2$, $p = 0.62$. The interaction effect was also non-significant, $F(1,\:22) = 0.67$, $p = 0.42$. Groups cleaned significantly more in the identifiable condition (for 396 steps on average) in the identifiable condition than in the anonymous condition (337 steps on average).

\section{Spatial Coordination Analysis}

To explore the use of spatial coordination strategies in Cleanup, we estimate the extent of territoriality groups used to coordinate their investments in the public good. Our measure of territoriality relies on beta diversity, a measure of compositional heterogeneity across locations originally developed by ecologists \citep{koleff2003measuring, whittaker1960vegetation}. Conceptually, we calculate a metric that communicates the degree to which group members' ``territories'' overlapped with each other. We start by converting group member trajectories to presence-absence data for each location in the map. Each location $l_k$ that is visited at least once is recorded in a vector $\mathbf{l} = \{l_1, ... l_{N_l}\}$, of length $N_l$, where $N_l$ denotes the number of distinct locations that were visited. Each group member $j$ that visits location $l_k$ is also recorded, resulting in a corresponding vector $\mathbf{j}_{l_k}$ of length $n_{l_k}$, where $n_{l_k}$ denotes the number of distinct group members who visited location $l_k$. We use the presence-absence data for group members' movements within the river region to calculate alpha, gamma, and beta diversities. Alpha diversity is defined as the number of group members who were present at the average location:
\begin{equation}
   \alpha_d = \frac{1}{N_l} \sum_{l_k \in \pmb{l}}{n_{l_k}} \, .
\end{equation}
Gamma diversity is defined as the number of unique group members who were present over all locations:
\begin{equation}
   \gamma_d = \left| \bigcup_{l_k \in \pmb{l}} \pmb{j}_{l_k} \right| \, .
\end{equation}
Beta diversity, the effective number of different group compositions, is defined as the ratio of gamma to alpha diversity:
\begin{equation}
   \beta_d = \frac{\gamma_d}{\alpha_d} \, .
\end{equation}

Beta diversity is lower bounded by $1$ (representing a single group composition for all visited locations) and upper bounded by whichever of $\gamma_d$ and $N_l$ is lower (representing completely non-overlapping locations visited by group members or completely different compositions per location, respectively). To account for the variability of the upper bound across episodes, we calculate a normalized beta diversity:
\begin{equation}
   {\beta_d}' = \frac{\beta_d}{\textrm{min}\left( \gamma_d, N_l \right)} \, .
\end{equation}

We use the normalized beta diversity as a measure of territoriality. A territoriality of 0 indicates that group members' territories were identical. A territoriality of 1, in contrast, indicates that group members' territories were entirely disjoint.

\subsection{Computational Model}

We conduct a one-way ANOVA to assess the effect of the intrinsic motivation for reputation on group territoriality. There was a significant effect of condition on territoriality, $F(1,\:311) = 432$, $p < 0.0001$. In the model, groups exhibited significantly less territoriality in the identifiable condition (with an average territoriality score of 0.42) than in the anonymous condition (with an average score of 0.58).

\subsection{Human Behavioral Experiment}

\begin{figure}[ht]
    \centering
    \includegraphics[width=8cm]{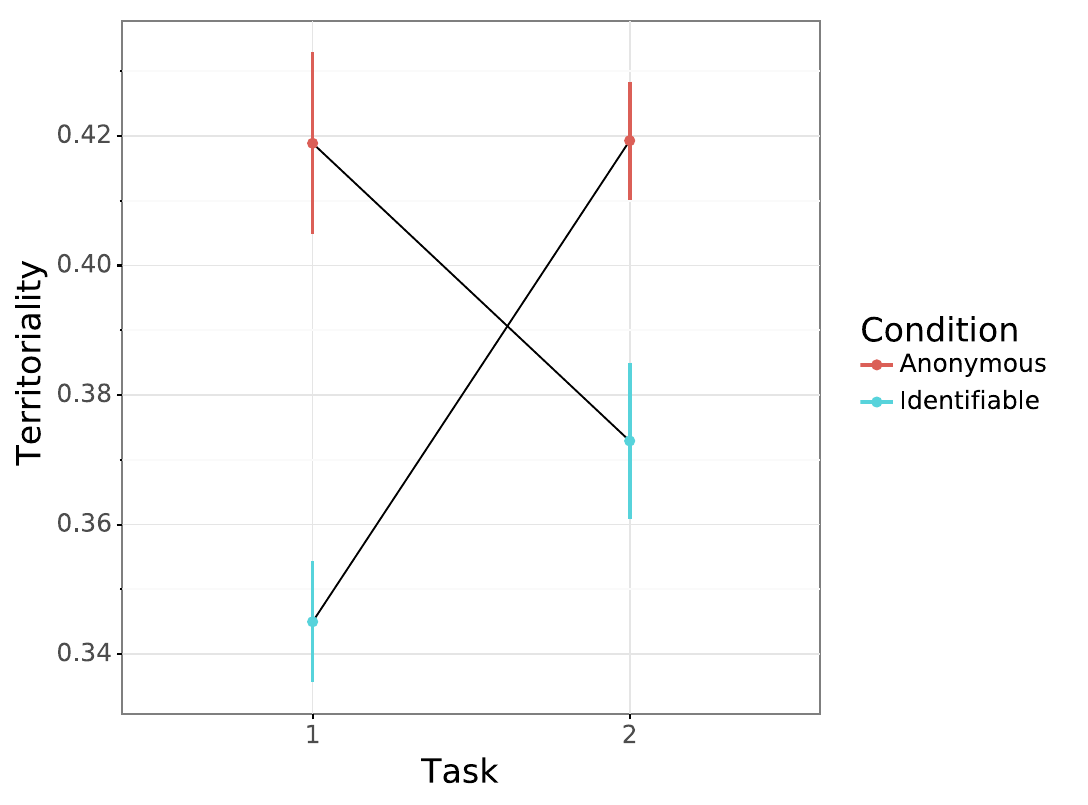}
    \caption{\small Figure \ref{fig:anova_territoriality}: We use a two-way, repeated-measures ANOVA to evaluate the effects of condition and task number on group territoriality. There was a significant main effect of condition on group territoriality. The main effect of task number was also significant, while the interaction effect between condition and task number was not significant. Error bars reflect 95\% confidence intervals.}
    \label{fig:anova_territoriality}
\end{figure}

We conduct a two-way ANOVA to assess the effect of the intrinsic motivation for reputation on group territoriality (Figure \ref{fig:anova_territoriality}). As before, we highlight the main effect of condition in the main text to facilitate comparison with the model results, and here expand on the other terms of the two-way ANOVA. There was a significant main effect of condition on territoriality, $F(1,\:310) = 138.4$, $p < 0.0001$. The main effect of task number was also significant, $F(1,\:310) = 7.7$, $p = 0.0059$. The interaction effect was non-significant, $F(1,\:22) = 0.2$, $p = 0.67$. Groups were significantly less territorial in the identifiable condition (with an average territoriality score of 0.36) than in the anonymous condition (with an average score of 0.42).

\section{Temporal Coordination Analysis}

For this analysis, our aim is to understand whether groups organized their behavior using temporal coordination strategies. Toward this end, we develop {two measures of temporal coordination: a measure of group turn taking and a measure of temporal consistency for group contributions.}

Turn taking is calculated based on the ordering of group member ``turns'' entering the river to clean pollution. For each episode, we record the sequence $\mathbf{S}$ of group members entering the river. For each of the identities in this sequence of turns, we calculated a recency value reflecting the number of turns that had occurred since the group member's last turn in the river:

\begin{table}[ht]
    \centering
    \begin{tabular}{c|c}
    Turns since group & Recency value\\
    member $j$'s last turn & for this turn \\
    \cline{1-2}
    $0$ & $1$ \\
    $1$ & $0.75$ \\
    $2$ & $0.50$ \\
    $3$ & $0.25$ \\
    $4+$ & $0$ \\ 
    \end{tabular}
    \caption{Mapping between the number of turns that occurred since group member $j$'s last turn in the river and the assigned recency value.}
    \label{tab:turn_recency}
\end{table}

Turn-taking scores are generated by averaging the recency values for each turn in $\mathbf{S}$, taking the additive inverse of the average, and then adding a constant of 1 to the subsequent metric (see Table \ref{tab:turn_recency}). A turn-taking score of 0 represents an episode where a single group member took all turns in the river. A turn-taking score of 1 reflects an episode where all group members rotated into the river, such that four turns pass between each of group member $j$'s turns in the river.

\begin{figure}
    \centering
    \includegraphics[width=8cm]{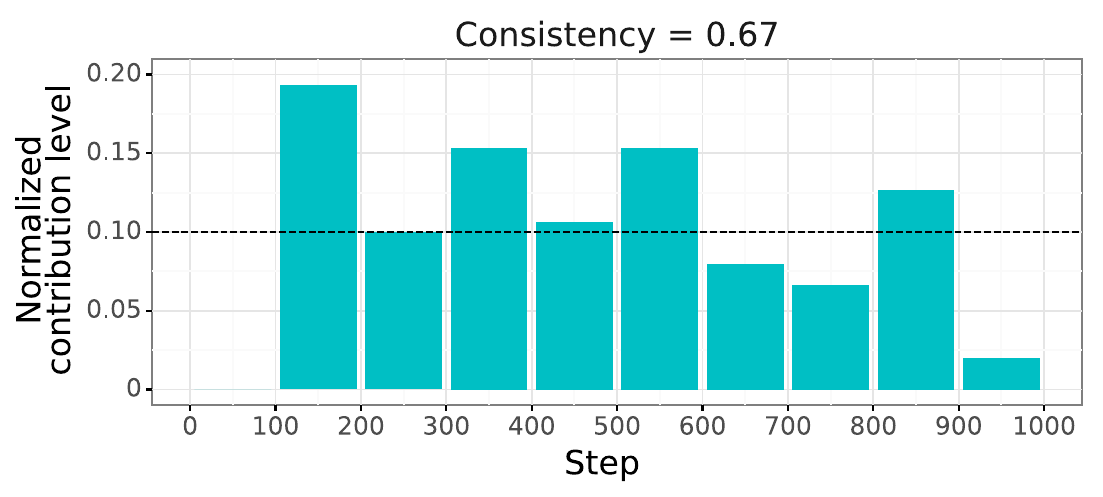} \\
    \includegraphics[width=8cm]{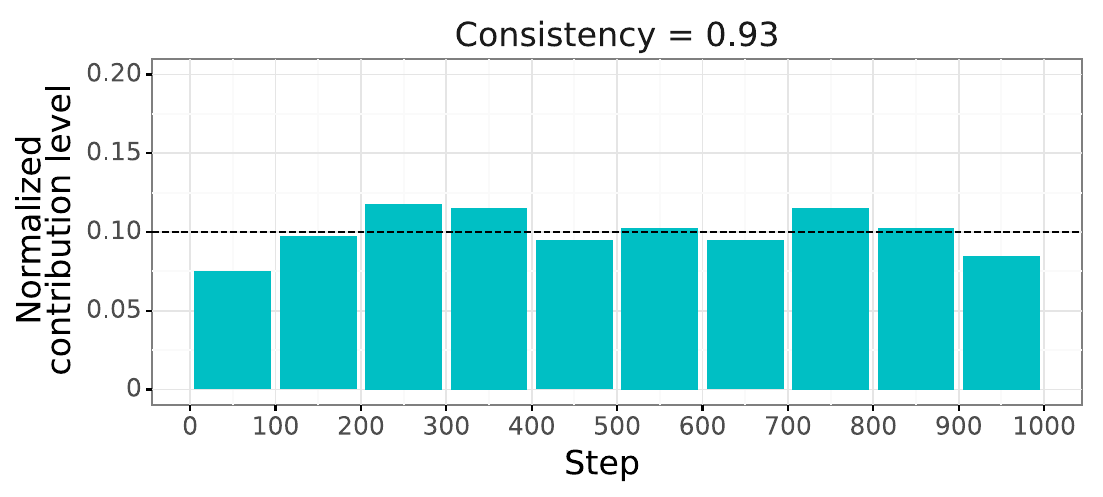}
    \caption{\small Figure \ref{fig:temp_consis}: Example patterns of contribution density over time from the model, showing two groups from the computational model. In the first example, the group achieved low temporal consistency ($\textrm{consistency} = 0.67$). In the second, the group achieved high temporal consistency ($\textrm{consistency} = 0.93$).}
    \label{fig:temp_consis}
\end{figure}

Consistency is calculated by binning the full sequence of contributions from an episode into a number of granular periods. Here we estimate consistency using $t = 10$ periods for each episode (see Figure \ref{fig:temp_consis}).

Contributions within these periods are summed, forming a vector $\mathbf{c_T}$ of binned contributions. A consistency score is then calculated for this binned contribution vector, measuring equality over the temporal dimension:

\begin{equation}
   \textrm{Gini}(\mathbf{c_T}) = \frac{\displaystyle \sum_{j=1}^t \sum_{k=1}^t |c_j - c_k|}{\displaystyle 2 n^2 \bar{c}} \, ,
\end{equation}
\begin{equation}
   \label{eqn:consistency}
   \textrm{Consistency}(\mathbf{c_T}) = 1 - \textrm{Gini}(\mathbf{c_T}) \, .
\end{equation}

Groups that concentrate all of their contribution efforts in a short span of time exhibit low temporal consistency, whereas groups that evenly apportion their contribution efforts over time manifest high temporal consistency (Figure \ref{fig:temp_consis}). Maintaining a high level of contribution consistency requires a group to coordinate which members will clean the river at any given time.

\subsection{Computational Model}

We conduct a one-way ANOVA to assess the effect of the intrinsic motivation for reputation on group turn taking. There was a significant effect of condition on turn taking, $F(1,\:311) = 758.3$, $p < 0.0001$. In the model, groups exhibited significantly more turn taking in the identifiable condition (with an average turn-taking score of 0.75) than in the anonymous condition (with an average score of 0.56).

{In the main text, we evaluate the relationship between turn taking and group performance with a linear regression, averaging observations by group:}

\begin{equation}
   \textrm{Collective Return} = \beta_0 + \beta_1 \cdot \textrm{Turn-Taking Score} + \epsilon \, .
\end{equation}

{With this regression, the model predicted a significant relationship between turn taking and collective return, $\beta = 1030.3$, 95\% CI $[908.1, 1152.5]$, $p < 0.0001$. Turn taking was positively associated with collective return, such that groups that relied more heavily on a rotation scheme tended to achieve higher scores.}

\begin{figure}[ht]
    \centering
    \includegraphics[width=12cm]{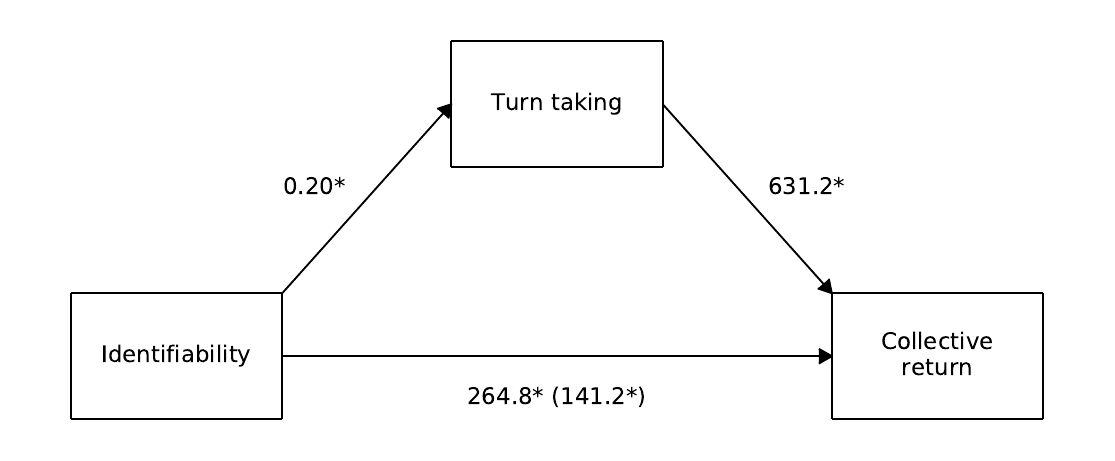}
    \caption{\small Figure \ref{fig:agent_turn_taking}: Mediation analysis of artificial agent group data revealed a significant indirect effect of identifiability on collective return, mediated by group turn taking. * $p < 0.05$.}
    \label{fig:agent_turn_taking}
\end{figure}

{We conduct a mediation analysis \citep{rucker2011mediation} to estimate the indirect effect of identifiability on collective return through group turn-taking (Figure \ref{fig:agent_turn_taking}). The analysis reveals a significant and positive indirect effect of identifiability on collective return through group turn-taking, $AB = 123.6$, 95\% CI $[110.7, 138.7]$, $p < 0.0001$. Furthermore, the positive association between identifiability and collective return ($C = 264.8$, 95\% CI $[248.4, 279.4]$, $p < 0.0001$) is reduced after accounting for turn taking ($C' = 141.2$, 95\% CI $[110.7, 155.4]$, $p < 0.0001$).}

\begin{figure}[ht]
    \centering
    \includegraphics[height=6cm]{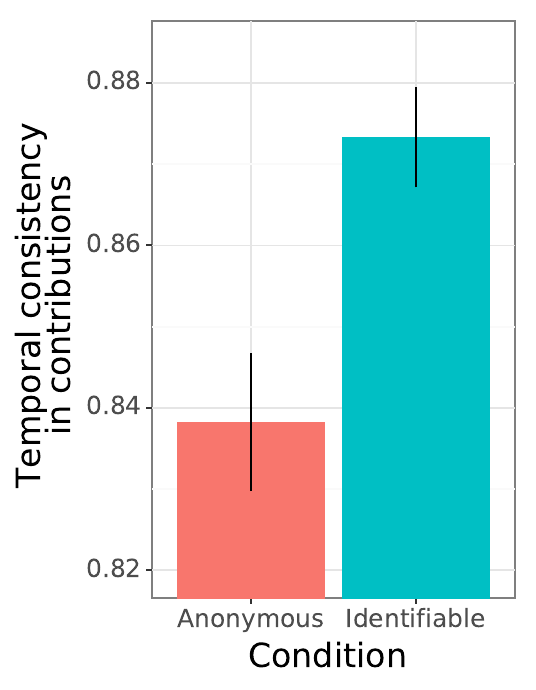}
    \caption{\small Figure \ref{fig:agent_anova_consistency}: We use a repeated-measures ANOVA to evaluate the effect of condition on the temporal consistency of group contributions. Condition exerted a significant effect on contribution consistency. Error bars reflect 95\% confidence intervals.}
    \label{fig:agent_anova_consistency}
\end{figure}

{To further test our findings, we replicate these analyses with the temporal consistency measure. We conduct a repeated-measures ANOVA to assess the effect of the intrinsic motivation for reputation on group contribution consistency over time. Condition exerted a significant effect on temporal consistency, $F(1,\:311) = 81.1$, $p < 0.0001$ (Figure \ref{fig:agent_anova_consistency}). In the model, groups acted with greater consistency in the identifiable condition (with an average consistency score of 0.87) than in the anonymous condition (with an average score of 0.84).}

{We next evaluate the relationship between temporal consistency and group performance with a linear regression, averaging observations by group:}

\begin{equation}
   \textrm{Collective Return} = \beta_0 + \beta_1 \cdot \textrm{Contribution Consistency} + \epsilon \, .
\end{equation}

\begin{figure}[ht]
    \centering
    \includegraphics[height=6cm]{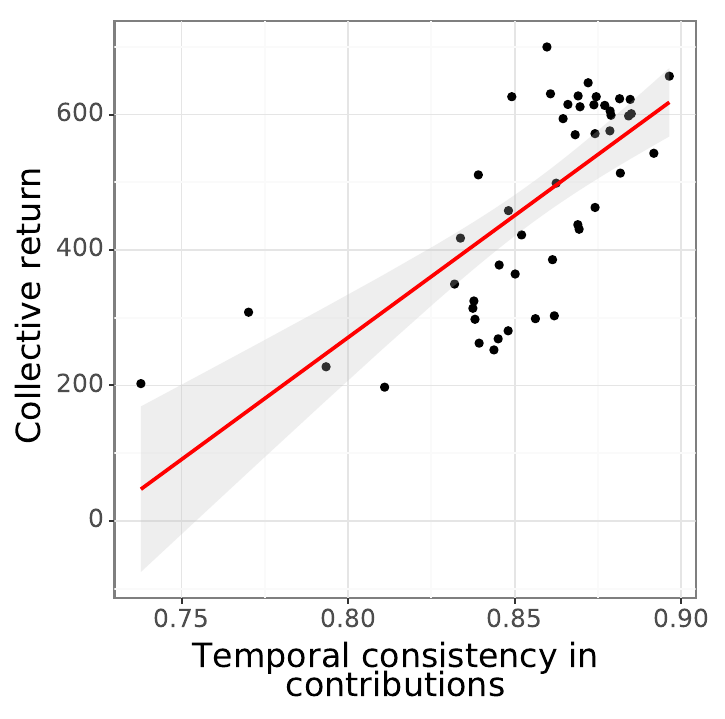}
    \caption{\small Figure \ref{fig:agent_consistency}: In the model, the temporal consistency of group contributions was significantly and positively associated with group performance.}
    \label{fig:agent_consistency}
\end{figure}

{With this regression, the model predicted a significant relationship between contribution consistency and collective return, $\beta = 3601.7$, 95\% CI $[2569.5, 4633.9]$, $p < 0.0001$ (Figure \ref{fig:agent_consistency}). Contribution consistency positively related to collective return, such that groups that provided the public good with greater consistency over time tended to achieve higher scores.}

\subsection{Human Behavioral Experiment}

\begin{figure}[ht]
    \centering
    \includegraphics[width=8cm]{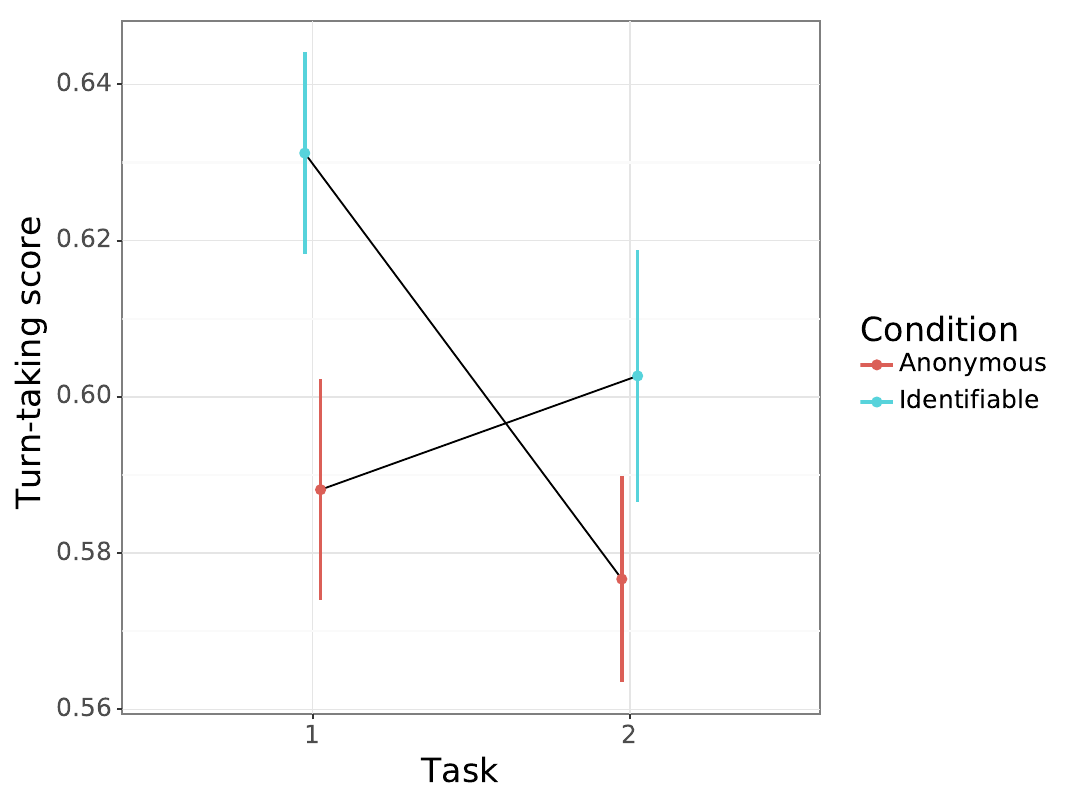}
    \caption{\small Figure \ref{fig:anova_turn_taking}: We use a two-way, repeated-measures ANOVA to evaluate the effects of condition and task number on group turn taking. There was a significant main effect of condition on group turn taking. The main effect of task number was also significant, while the interaction effect between condition and task number was not significant. Error bars reflect 95\% confidence intervals.}
    \label{fig:anova_turn_taking}
\end{figure}

We conduct a two-way ANOVA to assess the effect of the intrinsic motivation for reputation on group turn taking (Figure \ref{fig:anova_turn_taking}). As before, we highlight the main effect of condition in the main text to facilitate comparison with the model results, and here expand on the other terms of the two-way ANOVA. There was a significant main effect of condition on turn taking, $F(1,\:310) = 29.4$, $p < 0.0001$. The main effect of task number was also significant, $F(1,\:310) = 9.8$, $p = 0.0019$. The interaction effect was non-significant, $F(1,\:22) = 0.2$, $p = 0.65$. Groups were significantly more reliant on a turn-taking rotation scheme in the identifiable condition (with an average turn-taking score of 0.62) than in the anonymous condition (with an average score of 0.58).

{In the main text, we analyze the association between turn taking and collective return with a linear regression, averaging observations by group:}

\begin{equation}
   \textrm{Collective Return} = \beta_0 + \beta_1 \cdot \textrm{Turn-Taking Score} + \epsilon \, .
\end{equation}

{Among the human groups, there was a positive relationship between turn taking and collective return, $\beta = 3784.6$, 95\% CI $[1616.8, 5950.4]$, $p = 0.0010$. The use of a turn-taking rotation scheme was positively associated with group performance.}

\begin{figure}[ht]
    \centering
    \includegraphics[width=12cm]{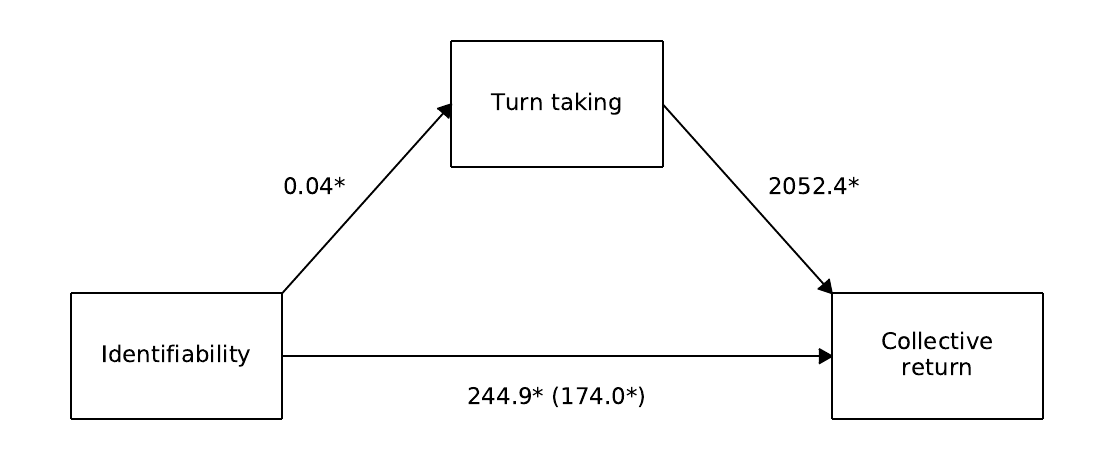}
    \caption{\small Figure \ref{fig:human_turn_taking}: Mediation analysis of human group data revealed a significant indirect effect of identifiability on collective return, mediated by group turn taking. * $p < 0.05$.}
    \label{fig:human_turn_taking}
\end{figure}

{We conduct a mediation analysis to estimate the indirect effect of identifiability on collective return through group turn-taking (Figure \ref{fig:human_turn_taking}). The analysis indicates a significant and positive indirect effect of identifiability on collective return through group turn-taking, $AB = 70.9$, 95\% CI $[37.2, 112.4]$, $p < 0.0001$. In addition, the positive association between identifiability and collective return ($C = 244.9$, 95\% CI $[143.7, 338.6]$, $p < 0.0001$) is reduced after accounting for turn taking ($C' = 174.0$, 95\% CI $[72.4, 272.6]$, $p < 0.0001$).}

\begin{figure}[ht]
    \centering
    \includegraphics[width=8cm]{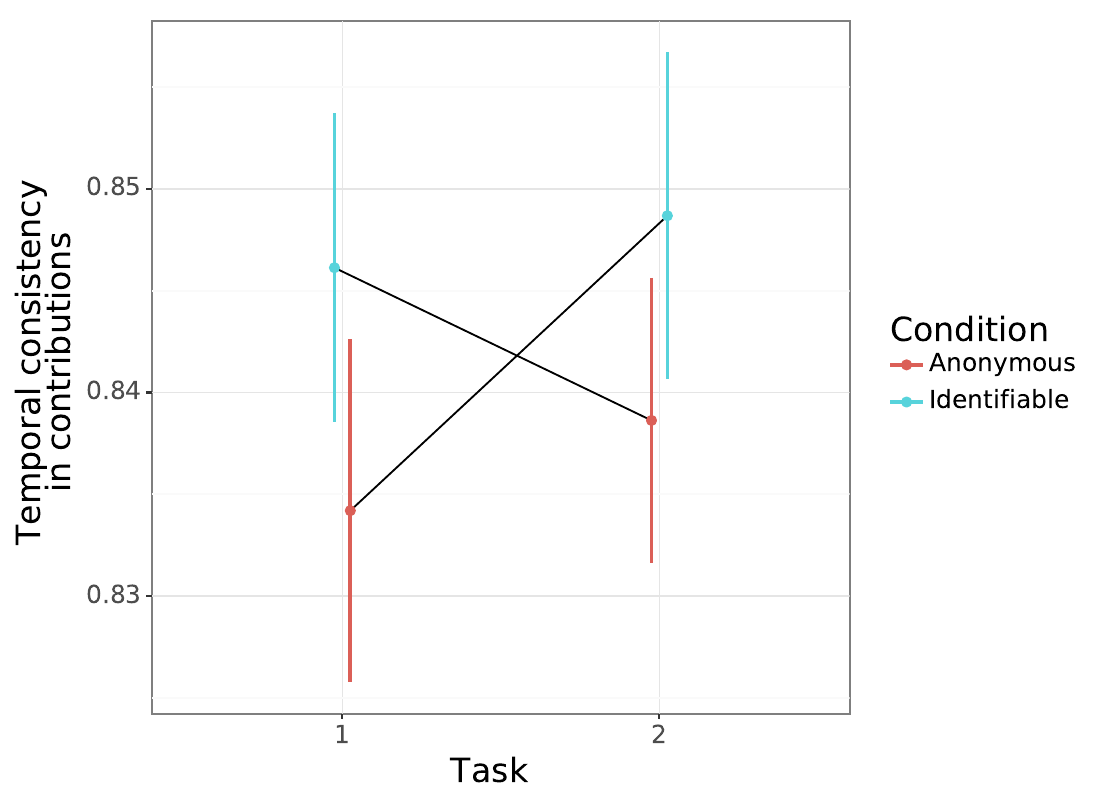}
    \caption{\small Figure \ref{fig:human_anova_consistency}: We use a two-way, repeated-measures ANOVA to evaluate to evaluate the effects of condition and task on temporal consistency in group contributions. There was a significant main effect of condition on temporal consistency. Neither the main effect of task number nor the interaction effect between condition and task number were significant. Error bars reflect 95\% confidence intervals.}
    \label{fig:human_anova_consistency}
\end{figure}

{To further test our findings, we replicate these analyses with the temporal consistency measure. We conduct a two-way, repeated-measures ANOVA to assess the effect of the intrinsic motivation for reputation on group contribution consistency over time (Figure \ref{fig:human_anova_consistency}). There was a significant main effect of condition on temporal consistency, $F(1,\:310) = 9.8$, $p = 0.0019$. The main effect of task number was not significant, $F(1,\:310) = 1.0$, $p = 0.32$. The interaction effect was non-significant, $F(1,\:22) = 0.0$, $p = 0.95$. Groups acted with greater consistency in the identifiable condition (with an average consistency score of 0.85) than in the anonymous condition (with an average score of 0.84).}

{We next evaluate the relationship between temporal consistency and group performance with a linear regression, averaging observations by group:}

\begin{equation}
   \textrm{Collective Return} = \beta_0 + \beta_1 \cdot \textrm{Contribution Consistency} + \epsilon \, .
\end{equation}

\begin{figure}[ht]
    \centering
    \includegraphics[height=6cm]{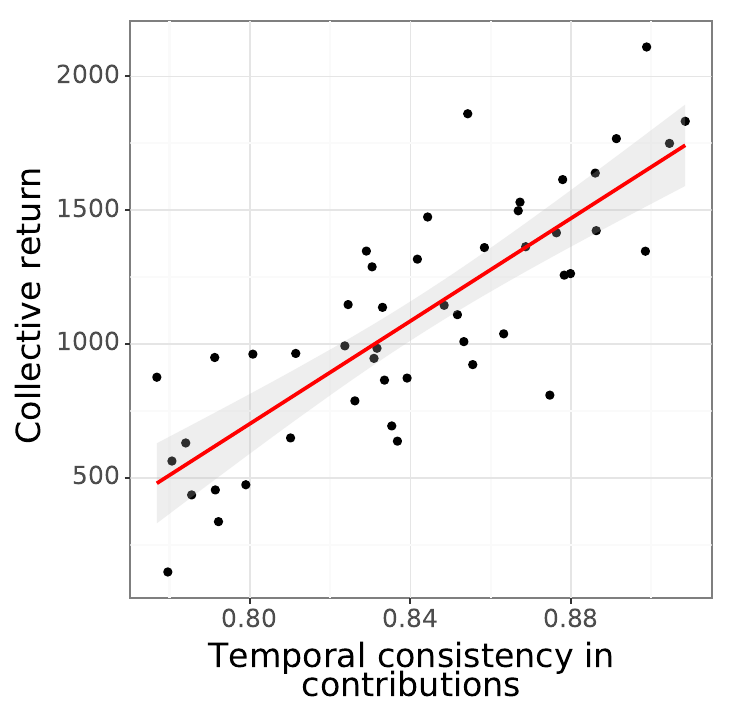}
    \caption{\small Figure \ref{fig:human_consistency}: Among human groups, the temporal consistency of group contributions was significantly and positively associated with group performance.}
    \label{fig:human_consistency}
\end{figure}

{Contribution consistency was significantly and positively related to collective return, $\beta = 9596.6$, 95\% CI $[7537.1, 11656.0]$, $p < 0.0001$ (Figure \ref{fig:human_consistency}). Contribution consistency positively correlated with group performance, such that groups that provided the public good with greater consistency over time tended to achieve higher scores.}

\begin{comment}
\section{Videos}

\subsection{Supporting Video 1}

This video shows human behavior in the identifiable condition.

\subsection{Supporting Video 2}

This video shows human behavior in the anonymous condition.
\end{comment}

%----------------------
% \newpage

%--------------------------------------------
\subsection*{Code availability}

The Clean Up environment is available through Melting Pot \citep{leibo2021scalable}.

 % End materials and methods section

\subsection*{Acknowledgments}
We thank Lucy Campbell-Gillingham, Dorothy Chou, Julia Cohen, Tom Eccles, Richard Everett, Stephen Gaffney, Ian Gemp, Demis Hassabis, Koray Kavukcuoglu, Zeb Kurth-Nelson, Vicky Langston, Brian McWilliams, Matthew Phillips, Oliver Smith, Tayfun Terzi, Gregory Thornton, and Sasha Vezhnevets for feedback and support.

\subsection*{Author contributions}
J.Z.L. conceived of the overall research direction; K.R.M. managed the project; A.G.C. and C.B. designed and coded the Clean Up task; K.R.M., E.H., T.O.Z., M.J.C., R.K., and J.Z.L. designed the human behavioral experiment; K.R.M., E.H., and T.O.Z. implemented the human behavioral research protocol; K.R.M. and T.O.Z. collected data for the human behavioral experiment; K.R.M., E.H., and J.Z.L. designed and coded the reinforcement learning agents and ran the reinforcement learning experiments; K.R.M., E.H., and T.O.Z. analyzed the data, with M.J.C., R.K., T.G., and M.B. providing substantial assistance; K.R.M., E.H., and J.Z.L. wrote the manuscript, with substantial assistance from T.O.Z., M.J.C., R.K., T.G., and M.B.

\printbibliography

@article{roberts1998competitive,
  title={Competitive altruism: from reciprocity to the handicap principle},
  author={Roberts, Gilbert},
  journal={Proceedings of the Royal Society of London. Series B: Biological Sciences},
  volume={265},
  number={1394},
  pages={427--431},
  year={1998},
  publisher={The Royal Society}
}

@article{trawick2001successfully,
  title={Successfully governing the commons: Principles of social organization in an Andean irrigation system},
  author={Trawick, Paul B.},
  journal={Human Ecology},
  volume={29},
  number={1},
  pages={1--25},
  year={2001},
  publisher={Springer}
}

@article{janssen2010introducing,
  title={Introducing ecological dynamics into common-pool resource experiments},
  author={Janssen, Marco A.},
  journal={Ecology and Society},
  volume={15},
  number={2},
  year={2010},
  publisher={JSTOR}
}

@article{cohen2016bandwidth,
  title={What is the bandwidth of perceptual experience?},
  author={Cohen, Michael A. and Dennett, Daniel C. and Kanwisher, Nancy},
  journal={Trends in Cognitive Sciences},
  volume={20},
  number={5},
  pages={324--335},
  year={2016},
  publisher={Elsevier}
}

@article{simpson1951interpretation,
  title={The interpretation of interaction in contingency tables},
  author={Simpson, Edward H.},
  journal={Journal of the Royal Statistical Society: Series B (Methodological)},
  volume={13},
  number={2},
  pages={238--241},
  year={1951},
  publisher={Wiley Online Library}
}

@article{cronbach1957two,
  title={The two disciplines of scientific psychology},
  author={Cronbach, Lee J.},
  journal={American Psychologist},
  volume={12},
  number={11},
  pages={671},
  year={1957},
  publisher={American Psychological Association}
}

@article{kievit2013simpson,
  title={Simpson's paradox in psychological science: A practical guide},
  author={Kievit, Rogier and Frankenhuis, Willem Eduard and Waldorp, Lourens and Borsboom, Denny},
  journal={Frontiers in Psychology},
  volume={4},
  pages={513},
  year={2013},
  publisher={Frontiers}
}

@article{pearson1899vi,
  title={VI. Mathematical contributions to the theory of evolution.—VI. Genetic (reproductive) selection: Inheritance of fertility in man, and of fecundity in thoroughbred racehorses},
  author={Pearson, Karl and Lee, Alice and Bramley-Moore, Leslie},
  journal={Philosophical Transactions of the Royal Society of London. Series A, Containing Papers of a Mathematical or Physical Character},
  number={192},
  pages={257--330},
  year={1899},
  publisher={The Royal Society London}
}

@article{fischbacher2001people,
  title={Are people conditionally cooperative? Evidence from a public goods experiment},
  author={Fischbacher, Urs and G{\"a}chter, Simon and Fehr, Ernst},
  journal={Economics Letters},
  volume={71},
  number={3},
  pages={397--404},
  year={2001},
  publisher={Elsevier}
}

@article{zelmer2003linear,
  title={Linear public goods experiments: A meta-analysis},
  author={Zelmer, Jennifer},
  journal={Experimental Economics},
  volume={6},
  number={3},
  pages={299--310},
  year={2003},
  publisher={Springer}
}

@article{gachter2008long,
  title={The long-run benefits of punishment},
  author={G{\"a}chter, Simon and Renner, Elke and Sefton, Martin},
  journal={Science},
  volume={322},
  number={5907},
  pages={1510--1510},
  year={2008},
  publisher={American Association for the Advancement of Science}
}

@book{fisher1925statistical,
  title={Statistical Methods for Research Workers},
  author={Fisher, Ronald Aylmer},
  year={1925},
  publisher={Oliver and Boyd}
}

@incollection{littman1994markov,
  title={Markov games as a framework for multi-agent reinforcement learning},
  author={Littman, Michael L.},
  booktitle={Machine learning proceedings 1994},
  pages={157--163},
  year={1994},
  publisher={Elsevier}
}

@article{tieleman2012lecture,
  title={Lecture 6.5-RMSProp: Divide the gradient by a running average of its recent magnitude},
  author={Tieleman, Tijmen and Hinton, Geoffrey},
  journal={Coursera: Neural networks for machine learning},
  volume={4},
  number={2},
  pages={26--31},
  year={2012}
}

@inproceedings{mckee2020social,
  title={Social diversity and social preferences in mixed-motive reinforcement learning},
  author={McKee, Kevin R and Gemp, Ian and McWilliams, Brian and Du{\'e}{\~n}ez-Guzm{\'a}n, Edgar A and Hughes, Edward and Leibo, Joel Z},
  year={2020},
  booktitle={Proceedings of the 19th International Conference on Autonomous Agents and Multiagent Systems},
  organization={International Foundation for Autonomous Agents and Multiagent Systems}
}

@article{fehr2000cooperation,
  title={Cooperation and punishment in public goods experiments},
  author={Fehr, Ernst and G{\"a}chter, Simon},
  journal={American Economic Review},
  volume={90},
  number={4},
  pages={980--994},
  year={2000}
}

@article{janssen2010lab,
  title={Lab experiments for the study of social-ecological systems},
  author={Janssen, Marco A. and Holahan, Robert and Lee, Allen and Ostrom, Elinor},
  journal={Science},
  volume={328},
  number={5978},
  pages={613--617},
  year={2010},
  publisher={American Association for the Advancement of Science}
}

@incollection{hughes2018inequity,
  title={Inequity aversion improves cooperation in intertemporal social dilemmas},
  author={Hughes, Edward and Leibo, Joel Z. and Philips, Matthew G and Tuyls, Karl and Du{\'e}{\~n}ez-Guzm{\'a}n, Edgar A. and Casta{\~n}eda, Antonio Garc{\'\i}a and Dunning, Iain and Zhu, Tina and McKee, Kevin R. and Koster, Raphael and Roff, Heather and Graepel, Thore},
  booktitle={Advances in Neural Information Processing Systems},
  pages={3330--3340},
  year={2018}
}

@book{sutton1998introduction,
  title={Reinforcement Learning: An Introduction},
  author={Sutton, Richard S. and Barto, Andrew G.},
  year={1998},
  publisher={MIT Press}
}

@article{silver2017mastering,
  title={Mastering the game of Go without human knowledge},
  author={David Silver and Julian Schrittwieser and Karen Simonyan and Ioannis Antonoglou and Aja Huang and Arthur Guez and Thomas Hubert and Lucas Baker, Matthew Lai and Adrian Bolton and Yutian Chen and Timothy Lillicrap and Fan Hui and Laurent Sifre and George van den Driessche and Thore Graepel and Demis Hassabis},
  journal={Nature},
  volume={550},
  number={7676},
  pages={354--359},
  year={2017}
}

@article{nowak1998evolution,
  title={Evolution of indirect reciprocity by image scoring},
  author={Nowak, Martin A. and Sigmund, Karl},
  journal={Nature},
  volume={393},
  number={6685},
  pages={573--577},
  year={1998},
  publisher={Nature Publishing Group}
}

@book{axelrod1984evolution,
  title={The Evolution of Cooperation},
  author={Axelrod, Robert M.},
  isbn={9780465021215},
  lccn={lc83045255},
  series={Basic Books},
  year={1984},
  publisher={Basic Books}
}

@book{poteete2010working,
  title={Working Together: Collective Action, the Commons, and Multiple Methods in Practice},
  author={Poteete, Amy R. and Janssen, Marco A. and Ostrom, Elinor},
  year={2010},
  publisher={Princeton University Press}
}

@inproceedings{singh2005intrinsically,
  title={Intrinsically motivated reinforcement learning},
  author={Singh, Satinder P. and Barto, Andrew G. and Chentanez, Nuttapong},
  booktitle={Advances in Neural Information Processing Systems},
  year={2005}
}

@article{saxe2008love,
  title={For love or money: A common neural currency for social and monetary reward},
  author={Saxe, Rebecca and Haushofer, Johannes},
  journal={Neuron},
  volume={58},
  number={2},
  pages={164--165},
  year={2008},
  publisher={Elsevier}
}

@article{izuma2008processing,
  title={Processing of social and monetary rewards in the human striatum},
  author={Izuma, Keise and Saito, Daisuke N. and Sadato, Norihiro},
  journal={Neuron},
  volume={58},
  number={2},
  pages={284--294},
  year={2008},
  publisher={Elsevier}
}

@article{wedekind2000cooperation,
  title={Cooperation through image scoring in humans},
  author={Wedekind, Claus and Milinski, Manfred},
  journal={Science},
  volume={288},
  number={5467},
  pages={850--852},
  year={2000},
  publisher={American Association for the Advancement of Science}
}

@article{engelmann2009indirect,
  title={Indirect reciprocity and strategic reputation building in an experimental helping game},
  author={Engelmann, Dirk and Fischbacher, Urs},
  journal={Games and Economic Behavior},
  volume={67},
  number={2},
  pages={399--407},
  year={2009},
  publisher={Elsevier}
}

@article{hardy2006nice,
  title={Nice guys finish first: The competitive altruism hypothesis},
  author={Hardy, Charlie L. and van Vugt, Mark},
  journal={Personality and Social Psychology Bulletin},
  volume={32},
  number={10},
  pages={1402--1413},
  year={2006},
  publisher={SAGE}
}

@article{balliet2009social,
  title={{Social Value Orientation} and cooperation in social dilemmas: A meta-analysis},
  author={Balliet, Daniel and Parks, Craig and Joireman, Jeff},
  journal={Group Processes \& Intergroup Relations},
  volume={12},
  number={4},
  pages={533--547},
  year={2009},
  publisher={SAGE}
}

@article{fehr2002altruistic,
  title={Altruistic punishment in humans},
  author={Fehr, Ernst and G{\"a}chter, Simon},
  journal={Nature},
  volume={415},
  number={6868},
  pages={137--140},
  year={2002},
  publisher={Nature Publishing Group}
}

@article{leibo2019autocurricula,
  title={Autocurricula and the emergence of innovation from social interaction: A manifesto for multi-agent intelligence research},
  author={Leibo, Joel Z. and Hughes, Edward and Lanctot, Marc and Graepel, Thore},
  journal={arXiv:1903.00742},
  year={2019}
}

@incollection{rapoport1974prisoner,
  title={Prisoner’s Dilemma—Recollections and observations},
  author={Rapoport, Anatol},
  booktitle={Game Theory as a Theory of a Conflict Resolution},
  pages={17--34},
  year={1974},
  publisher={Springer}
}

@article{henrich2006costly,
  title={Costly punishment across human societies},
  author={Henrich, Joseph P. and McElreath, Richard and Barr, Abigail and Ensminger, Jean and Barrett, Clark and Bolyanatz, Alexander and Cardenas, Juan Camilo and Gurven, Michael and Gwako, Edwins and Henrich, Natalie and Lesorogol, Carolyn and Marlowe, Frank and Tracer, David and Ziker, John},
  journal={Science},
  volume={312},
  number={5781},
  pages={1767--1770},
  year={2006},
  publisher={American Association for the Advancement of Science}
}

@inproceedings{perolat2017multi,
  title={A multi-agent reinforcement learning model of common-pool resource appropriation},
  author={Perolat, Julien and Leibo, Joel Z. and Zambaldi, Vinicius and Beattie, Charles and Tuyls, Karl and Graepel, Thore},
  booktitle={Advances in Neural Information Processing Systems},
  pages={3643--3652},
  year={2017}
}

@article{schelling1973hockey,
  title={Hockey helmets, concealed weapons, and daylight saving: A study of binary choices with externalities},
  author={Schelling, Thomas C.},
  journal={Journal of Conflict Resolution},
  volume={17},
  number={3},
  pages={381--428},
  year={1973},
  publisher={SAGE}
}

@article{jenks1967data,
  author={Jenks, George F.},
  year={1967},
  month={01},
  pages={186-190},
  title={The data model concept in statistical mapping},
  volume={7},
  journal={International Yearbook of Cartography}
}

@article{kollock1998social,
  title={Social dilemmas: The anatomy of cooperation},
  author={Kollock, Peter},
  journal={Annual Review of Sociology},
  volume={24},
  number={1},
  pages={183--214},
  year={1998},
  publisher={Annual Reviews}
}

@article{van2010cooperation,
  title={Cooperation for reputation: Wasteful contributions as costly signals in public goods},
  author={van Vugt, Mark and Hardy, Charlotte L.},
  journal={Group Processes \& Intergroup Relations},
  volume={13},
  number={1},
  pages={101--111},
  year={2010},
  publisher={SAGE}
}

@inproceedings{espeholt2018impala,
  title={Impala: Scalable distributed deep-{RL} with importance weighted actor-learner architectures},
  author={Espeholt, Lasse and Soyer, Hubert and Munos, Remi and Simonyan, Karen and Mnih, Volodymir and Ward, Tom and Doron, Yotam and Firoiu, Vlad and Harley, Tim and Dunning, Iain and Legg, Shane and Kavukcuoglu, Koray},
  booktitle={International Conference on Machine Learning},
  pages={1407--1416},
  year={2018}
}

@article{fehr1999theory,
  title={A theory of fairness, competition, and cooperation},
  author={Fehr, Ernst and Schmidt, Klaus M.},
  journal={The Quarterly Journal of Economics},
  volume={114},
  number={3},
  pages={817--868},
  year={1999},
  publisher={MIT Press}
}

@article{rockenbach2006efficient,
  title={The efficient interaction of indirect reciprocity and costly punishment},
  author={Rockenbach, Bettina and Milinski, Manfred},
  journal={Nature},
  volume={444},
  number={7120},
  pages={718},
  year={2006},
  publisher={Nature Publishing Group}
}

@article{ariely2009doing,
  title={Doing good or doing well? Image motivation and monetary incentives in behaving prosocially},
  author={Ariely, Dan and Bracha, Anat and Meier, Stephan},
  journal={American Economic Review},
  volume={99},
  number={1},
  pages={544--55},
  year={2009}
}

@misc{sherif1936psychology,
  title={The Psychology of Social Norms},
  author={Sherif, Muzafer},
  year={1936},
  publisher={Harper}
}

@article{phan2010reputation,
  title={Reputation for reciprocity engages the brain reward center},
  author={Phan, K. Luan and Sripada, Chandra Sekhar and Angstadt, Mike and McCabe, Kevin},
  journal={Proceedings of the National Academy of Sciences},
  volume={107},
  number={29},
  pages={13099--13104},
  year={2010},
  publisher={National Academy of Sciences}
}

@article{lerer2017maintaining,
  title={Maintaining cooperation in complex social dilemmas using deep reinforcement learning},
  author={Lerer, Adam and Peysakhovich, Alexander},
  journal={arXiv preprint arXiv:1707.01068},
  year={2017}
}

@article{hochreiter1997long,
  title={Long short-term memory},
  author={Hochreiter, Sepp and Schmidhuber, J{\"u}rgen},
  journal={Neural Computation},
  volume={9},
  number={8},
  pages={1735--1780},
  year={1997},
  publisher={MIT Press}
}

@inproceedings{leibo2017multi,
  title={Multi-agent Reinforcement Learning in Sequential Social Dilemmas},
  author={Leibo, Joel Z and Zambaldi, Vinicius and Lanctot, Marc and Marecki, Janusz and Graepel, Thore},
  booktitle={Proceedings of the 16th Conference on Autonomous Agents and MultiAgent Systems},
  pages={464--473},
  year={2017}
}

@article{janssen2007robustness,
  title={Robustness of social-ecological systems to spatial and temporal variability},
  author={Janssen, Marco A. and Anderies, John M. and Ostrom, Elinor},
  journal={Society and Natural Resources},
  volume={20},
  number={4},
  pages={307--322},
  year={2007},
  publisher={Taylor \& Francis}
}

@article{miller1992collective,
  title={Collective action and rational choice: Place, community, and the limits to individual self-interest},
  author={Miller, Byron},
  journal={Economic Geography},
  volume={68},
  number={1},
  pages={22--42},
  year={1992},
  publisher={Taylor \& Francis}
}

@article{north1991institutions,
  title={Institutions},
  author={North, Douglass C.},
  journal={Journal of Economic Perspectives},
  volume={5},
  number={1},
  pages={97--112},
  year={1991}
}

@inproceedings{wang2019evolving,
  title={Evolving intrinsic motivations for altruistic behavior},
  author={Wang, Jane X. and Hughes, Edward and Fernando, Chrisantha and Czarnecki, Wojciech M. and Du{\'e}{\~n}ez-Guzm{\'a}n, Edgar A. and Leibo, Joel Z.},
  booktitle={Proceedings of the 18th International Conference on Autonomous Agents and Multiagent Systems},
  pages={683--692},
  year={2019},
  organization={International Foundation for Autonomous Agents and Multiagent Systems}
}

@inproceedings{mnih2016asynchronous,
  title={Asynchronous methods for deep reinforcement learning},
  author={Mnih, Volodymyr and Badia, Adria Puigdomenech and Mirza, Mehdi and Graves, Alex and Lillicrap, Timothy and Harley, Tim and Silver, David and Kavukcuoglu, Koray},
  booktitle={International Conference on Machine Learning},
  pages={1928--1937},
  year={2016}
}

@article{berridge2009dissecting,
  title={Dissecting components of reward: `Liking', `wanting', and learning},
  author={Berridge, Kent C. and Robinson, Terry E. and Aldridge, J. Wayne},
  journal={Current Opinion in Pharmacology},
  volume={9},
  number={1},
  pages={65--73},
  year={2009},
  publisher={Elsevier}
}

@book{ostrom2005understanding,
  title={Understanding Institutional Diversity},
  author={Ostrom, Elinor},
  year={2005},
  publisher={Princeton University Press}
}

@article{koleff2003measuring,
  title={Measuring beta diversity for presence--absence data},
  author={Koleff, Patricia and Gaston, Kevin J. and Lennon, Jack J.},
  journal={Journal of Animal Ecology},
  volume={72},
  number={3},
  pages={367--382},
  year={2003},
  publisher={Wiley Online Library}
}

@article{whittaker1960vegetation,
  title={Vegetation of the Siskiyou mountains, Oregon and California},
  author={Whittaker, Robert Harding},
  journal={Ecological Monographs},
  volume={30},
  number={3},
  pages={279--338},
  year={1960},
  publisher={Wiley Online Library}
}

@article{milinski2002reputation,
  title={Reputation helps solve the ‘tragedy of the commons’},
  author={Milinski, Manfred and Semmann, Dirk and Krambeck, Hans-J{\"u}rgen},
  journal={Nature},
  volume={415},
  number={6870},
  pages={424},
  year={2002},
  publisher={Nature Publishing Group}
}

@article{jaderberg2019human,
  title={Human-level performance in 3D multiplayer games with population-based reinforcement learning},
  author={Jaderberg, Max and Czarnecki, Wojciech M. and Dunning, Iain and Marris, Luke and Lever, Guy and Casta{\~n}eda, Antonio Garcia and Beattie, Charles and Rabinowitz, Neil C. and Morcos, Ari S. and Ruderman, Avraham and Sonnerat, Nicolas and Green, Tim and Deason, Louise and Leibo, Joel Z. and Silver, David and Hassabis, Demis and Kavukcuoglu, Koray and Graepel, Thore},
  journal={Science},
  volume={364},
  number={6443},
  pages={859--865},
  year={2019},
  publisher={American Association for the Advancement of Science}
}

@article{baker2019emergent,
  title={Emergent tool use from multi-agent autocurricula},
  author={Baker, Bowen and Kanitscheider, Ingmar and Markov, Todor and Wu, Yi and Powell, Glenn and McGrew, Bob and Mordatch, Igor},
  journal={arXiv preprint arXiv:1909.07528},
  year={2019}
}

@article{barclay2004trustworthiness,
  title={Trustworthiness and competitive altruism can also solve the “tragedy of the commons”},
  author={Barclay, Pat},
  journal={Evolution and Human Behavior},
  volume={25},
  number={4},
  pages={209--220},
  year={2004},
  publisher={Elsevier}
}

@article{barclay2006partner,
  title={Partner choice creates competitive altruism in humans},
  author={Barclay, Pat and Willer, Robb},
  journal={Proceedings of the Royal Society B: Biological Sciences},
  volume={274},
  number={1610},
  pages={749--753},
  year={2006},
  publisher={The Royal Society London}
}

@article{elsawah2020eight,
  title={Eight grand challenges in socio-environmental systems modeling},
  author={Elsawah, Sondoss and Filatova, Tatiana and Jakeman, Anthony J. and Kettner, Albert J. and Zellner, Moira L. and Athanasiadis, Ioannis N. and Hamilton, Serena H and Axtell, Robert L. and Brown, Daniel G. and Gilligan, Jonathan M. and others},
  journal={Socio-Environmental Systems Modelling},
  volume={2},
  pages={16226--16226},
  year={2020}
}

@book{camerer2003behavioral,
  title={Behavioral Game Ttheory: Experiments in Strategic Interaction},
  author={Camerer, Colin F.},
  year={2003},
  publisher={Princeton University Press}
}

@article{vinyals2019grandmaster,
  title={Grandmaster level in {StarCraft} {II} using multi-agent reinforcement learning},
  author={Vinyals, Oriol and Babuschkin, Igor and Czarnecki, Wojciech M and Mathieu, Micha{\"e}l and Dudzik, Andrew and Chung, Junyoung and Choi, David H and Powell, Richard and Ewalds, Timo and Georgiev, Petko and others},
  journal={Nature},
  pages={1--5},
  year={2019},
  publisher={Nature Publishing Group}
}

@inproceedings{jaques2019social,
  title={Social influence as intrinsic motivation for multi-agent deep reinforcement learning},
  author={Jaques, Natasha and Lazaridou, Angeliki and Hughes, Edward and Gulcehre, Caglar and Ortega, Pedro and Strouse, DJ and Leibo, Joel Z. and De Freitas, Nando},
  booktitle={International Conference on Machine Learning},
  pages={3040--3049},
  year={2019}
}

@article{eccles2019learning,
  title={Learning reciprocity in complex sequential social dilemmas},
  author={Eccles, Tom and Hughes, Edward and Kram{\'a}r, J{\'a}nos and Wheelwright, Steven and Leibo, Joel Z},
  journal={arXiv preprint arXiv:1903.08082},
  year={2019}
}

@inproceedings{foerster2018learning,
  title={Learning with opponent-learning awareness},
  author={Foerster, Jakob and Chen, Richard Y and Al-Shedivat, Maruan and Whiteson, Shimon and Abbeel, Pieter and Mordatch, Igor},
  booktitle={Proceedings of the 17th International Conference on Autonomous Agents and Multiagent Systems},
  pages={122--130},
  year={2018},
  organization={International Foundation for Autonomous Agents and Multiagent Systems}
}

@article{ostrom1993coping,
  title={Coping with asymmetries in the commons: Self-governing irrigation systems can work},
  author={Ostrom, Elinor and Gardner, Roy},
  journal={Journal of Economic Perspectives},
  volume={7},
  number={4},
  pages={93--112},
  year={1993}
}

@book{mollinga2003waterfront,
  title={On the Waterfront: Water Distribution, Technology and Agrarian Change in a South Indian Canal Irrigation System},
  author={Mollinga, Peter P.},
  year={2003},
  publisher={Orient Blackswan}
}

@article{kramar2020should,
  title={Should I tear down this wall? Optimizing social metrics by evaluating novel actions},
  author={Kram{\'a}r, J{\'a}nos and Rabinowitz, Neil and Eccles, Tom and Tacchetti, Andrea},
  journal={arXiv preprint arXiv:2004.07625},
  year={2020}
}

@article{dafoe2020open,
  title={Open problems in {Cooperative} {AI}},
  author={Dafoe, Allan and Hughes, Edward and Bachrach, Yoram and Collins, Tantum and McKee, Kevin R. and Leibo, Joel Z. and Larson, Kate and Graepel, Thore},
  journal={arXiv preprint arXiv:2012.08630},
  year={2020}
}

@article{mckee2022quantifying,
  title={Quantifying the effects of environment and population diversity in multi-agent reinforcement learning},
  author={McKee, Kevin R and Leibo, Joel Z and Beattie, Charlie and Everett, Richard},
  journal={Autonomous Agents and Multi-Agent Systems},
  volume={36},
  number={1},
  pages={1--16},
  year={2022},
  publisher={Springer}
}

@article{rucker2011mediation,
  title={Mediation analysis in social psychology: Current practices and new recommendations},
  author={Rucker, Derek D. and Preacher, Kristopher J. and Tormala, Zakary L. and Petty, Richard E.},
  journal={Social and Personality Psychology Compass},
  volume={5},
  number={6},
  pages={359--371},
  year={2011},
  publisher={Wiley Online Library}
}

@article{schill2019more,
  title={A more dynamic understanding of human behaviour for the {Anthropocene}},
  author={Schill, Caroline and Anderies, John M and Lindahl, Therese and Folke, Carl and Polasky, Stephen and C{\'a}rdenas, Juan Camilo and Cr{\'e}pin, Anne-Sophie and Janssen, Marco A. and Norberg, Jon and Schl{\"u}ter, Maja},
  journal={Nature Sustainability},
  volume={2},
  number={12},
  pages={1075--1082},
  year={2019},
  publisher={Nature Publishing Group}
}

@book{weibull1997evolutionary,
  title={Evolutionary game theory},
  author={Weibull, J{\"o}rgen W},
  year={1997},
  publisher={MIT Press}
}

@inbook{tesfatsion2021,
  title={Agent-based computational economics: Overview and brief history},
  year={2023},
  author={Tesfatsion, Leigh},
  booktitle={Artificial Intelligence, Learning and Computation in Economics and Finance},
  publisher={Springer}
}

@book{luce1957games,
  title={Games and decisions: Introduction and critical survey},
  author={Luce, R. Duncan and Raiffa, Howard},
  year={1957},
  publisher={Courier Corporation}
}

@article{koster2022spurious,
  title={Spurious normativity enhances learning of compliance and enforcement behavior in artificial agents},
  author={K{\"o}ster, Raphael and Hadfield-Menell, Dylan and Everett, Richard and Weidinger, Laura and Hadfield, Gillian K and Leibo, Joel Z.},
  journal={Proceedings of the National Academy of Sciences},
  volume={119},
  number={3},
  year={2022},
  publisher={National Academy of Sciences}
}

@inproceedings{leibo2021scalable,
  title={Scalable evaluation of multi-agent reinforcement learning with {M}elting {P}ot},
  author={Leibo, Joel Z. and Due{\~n}ez-Guzman, Edgar A and Vezhnevets, Alexander and Agapiou, John P. and Sunehag, Peter and Koster, Raphael and Matyas, Jayd and Beattie, Charlie and Mordatch, Igor and Graepel, Thore},
  booktitle={International Conference on Machine Learning},
  pages={6187--6199},
  year={2021},
  organization={PMLR}
}

@article{christofferson2022get,
  title={Get it in writing: Formal contracts mitigate social dilemmas in multi-agent {RL}},
  author={Christofferson, Phillip J. K. and Haupt, Andreas A. and Hadfield-Menell, Dylan},
  journal={arXiv preprint arXiv:2208.10469},
  year={2022}
}

@article{baker2020emergent,
  title={Emergent reciprocity and team formation from randomized uncertain social preferences},
  author={Baker, Bowen},
  journal={Advances in Neural Information Processing Systems},
  volume={33},
  pages={15786--15799},
  year={2020}
}

@article{shapley1953stochastic,
  title={Stochastic games},
  author={Shapley, Lloyd S.},
  journal={Proceedings of the National Academy of Sciences},
  volume={39},
  number={10},
  pages={1095--1100},
  year={1953},
  publisher={National Academy of Sciences}
}

@article{giardini2021competitive,
  title={Gossip and competitive altruism support cooperation in a {Public Good} game},
  author={Giardini, Francesca and Vilone, Daniele and Sanchez, Angel and Antonioni, Alberto},
  journal={Proceedings of the Royal Society B: Biological Sciences},
  volume={376},
  number={20200303},
  year={2021},
  publisher={The Royal Society}
}

@article{vinitsky2023learning,
  title={A learning agent that acquires social norms from public sanctions in decentralized multi-agent settings},
  author={Vinitsky, Eugene and K{\"o}ster, Raphael and Agapiou, John P and Du{\'e}{\~n}ez-Guzm{\'a}n, Edgar A and Vezhnevets, Alexander S and Leibo, Joel Z},
  journal={Collective Intelligence},
  volume={2},
  number={2},
  pages={26339137231162025},
  year={2023},
  publisher={SAGE Publications Sage UK: London, England}
}

@book{schelling1960strategy,
  title={The strategy of conflict},
  author={Schelling, Thomas C.},
  year={1960},
  publisher={Harvard University Press}
}

@book{sugden2004economics,
  title={The economics of rights, co-operation and welfare},
  author={Sugden, Robert},
  year={1986},
  publisher={Springer}
}

@book{zahavi1999handicap,
  title={The handicap principle: A missing piece of {Darwin's} puzzle},
  author={Zahavi, Amotz and Zahavi, Avishag},
  year={1999},
  publisher={Oxford University Press}
}

@article{janssen2006empirically,
  title={Empirically based, agent-based models},
  author={Janssen, Marco A. and Ostrom, Elinor},
  journal={Ecology and Society},
  volume={11},
  number={2},
  year={2006},
  publisher={JSTOR}
}

@article{hassabis2017artificial,
  title={Artificial intelligence: Chess match of the century},
  author={Hassabis, Demis},
  journal={Nature},
  volume={544},
  number={7651},
  pages={413--414},
  year={2017},
  publisher={Nature Publishing Group}
}

@article{wong2022deep,
  title={Deep multiagent reinforcement learning: Challenges and directions},
  author={Wong, Annie and B{\"a}ck, Thomas and Kononova, Anna V. and Plaat, Aske},
  journal={Artificial Intelligence Review},
  pages={1--34},
  year={2022},
  publisher={Springer}
}

@article{lansing2005cooperation,
  title={Cooperation, games, and ecological feedback: Some insights from {Bali}},
  author={Lansing, J. Stephen and Miller, JohnH},
  journal={Current Anthropology},
  volume={46},
  number={2},
  pages={328--334},
  year={2005},
  publisher={The University of Chicago Press}
}

@article{bonabeau2002agent,
  title={Agent-based modeling: Methods and techniques for simulating human systems},
  author={Bonabeau, Eric},
  journal={Proceedings of the national academy of sciences},
  volume={99},
  number={suppl\_3},
  pages={7280--7287},
  year={2002},
  publisher={National Acad Sciences}
}

@article{semmann2004strategic,
  title={Strategic investment in reputation},
  author={Semmann, Dirk and Krambeck, Hans-J{\"u}rgen and Milinski, Manfred},
  journal={Behavioral Ecology and Sociobiology},
  volume={56},
  number={3},
  pages={248--252},
  year={2004},
  publisher={Springer}
}

@article{roberts2021benefits,
  title={The benefits of being seen to help others: indirect reciprocity and reputation-based partner choice},
  author={Roberts, Gilbert and Raihani, Nichola and Bshary, Redouan and Manrique, H{\'e}ctor M and Farina, Andrea and Samu, Fl{\'o}ra and Barclay, Pat},
  journal={Philosophical Transactions of the Royal Society B},
  volume={376},
  number={1838},
  pages={20200290},
  year={2021},
  publisher={The Royal Society}
}

@article{gintis2001costly,
  title={Costly signaling and cooperation},
  author={Gintis, Herbert and Smith, Eric Alden and Bowles, Samuel},
  journal={Journal of theoretical biology},
  volume={213},
  number={1},
  pages={103--119},
  year={2001},
  publisher={Elsevier}
}

@article{yang2020learning,
  title={Learning to incentivize other learning agents},
  author={Yang, Jiachen and Li, Ang and Farajtabar, Mehrdad and Sunehag, Peter and Hughes, Edward and Zha, Hongyuan},
  journal={Advances in Neural Information Processing Systems},
  volume={33},
  pages={15208--15219},
  year={2020}
}

@article{radke2022importance,
  title={The importance of credo in multiagent learning},
  author={Radke, David and Larson, Kate and Brecht, Tim},
  journal={arXiv preprint arXiv:2204.07471},
  year={2022}
}

@inproceedings{gemp2022d3c,
  title={D3C: Reducing the Price of Anarchy in Multi-Agent Learning},
  author={Gemp, Ian and McKee, Kevin R and Everett, Richard and Du{\'e}{\~n}ez-Guzm{\'a}n, Edgar and Bachrach, Yoram and Balduzzi, David and Tacchetti, Andrea},
  booktitle={Proceedings of the 21st International Conference on Autonomous Agents and Multiagent Systems},
  pages={498--506},
  year={2022}
}

@article{kolle2023learning,
  title={Learning to Participate through Trading of Reward Shares},
  author={K{\"o}lle, Michael and Matheis, Tim and Altmann, Philipp and Schmid, Kyrill},
  journal={arXiv preprint arXiv:2301.07416},
  year={2023}
}

@inproceedings{kleiman2016coordinate,
  title={Coordinate to cooperate or compete: abstract goals and joint intentions in social interaction},
  author={Kleiman-Weiner, Max and Ho, Mark K and Austerweil, Joseph L and Littman, Michael L and Tenenbaum, Joshua B},
  booktitle={CogSci},
  year={2016}
}

@article{tilbury2022identity,
  title={Identity and Dynamic Teams in Social Dilemmas},
  author={Tilbury, Kyle and Hoey, Jesse},
  journal={arXiv preprint arXiv:2208.03293},
  year={2022}
}

@article{bradley2018does,
  title={Does observability affect prosociality?},
  author={Bradley, Alex and Lawrence, Claire and Ferguson, Eamonn},
  journal={Proceedings of the Royal Society B: Biological Sciences},
  volume={285},
  number={1875},
  pages={20180116},
  year={2018},
  publisher={The Royal Society}
}

@article{manrique2021psychological,
  title={The psychological foundations of reputation-based cooperation},
  author={Manrique, H{\'e}ctor M and Zeidler, Henriette and Roberts, Gilbert and Barclay, Pat and Walker, Michael and Samu, Fl{\'o}ra and Fari{\~n}a, Andrea and Bshary, Redouan and Raihani, Nichola},
  journal={Philosophical Transactions of the Royal Society B},
  volume={376},
  number={1838},
  pages={20200287},
  year={2021},
  publisher={The Royal Society}
}

@article{barclay2013strategies,
  title={Strategies for cooperation in biological markets, especially for humans},
  author={Barclay, Pat},
  journal={Evolution and Human Behavior},
  volume={34},
  number={3},
  pages={164--175},
  year={2013},
  publisher={Elsevier}
}

@inproceedings{tan1993multi,
  title={Multi-agent reinforcement learning: Independent vs. cooperative agents},
  author={Tan, Ming},
  booktitle={Proceedings of the tenth international conference on machine learning},
  pages={330--337},
  year={1993}
}

@article{hirschman2016stylized,
  title={Stylized facts in the social sciences},
  author={Hirschman, Daniel},
  journal={Sociological Science},
  volume={3},
  pages={604--626},
  year={2016}
}

@book{gintis2014bounds,
  title={The bounds of reason: game theory and the unification of the behavioral sciences-revised edition},
  author={Gintis, Herbert},
  year={2014},
  publisher={Princeton university press}
}

@article{gazzard2011unlocking,
  title={Unlocking the gameworld: The rewards of space and time in videogames},
  author={Gazzard, Alison},
  journal={Game Studies},
  volume={11},
  number={1},
  pages={9--13},
  year={2011}
}

@book{lawler2018introduction,
  title={Introduction to stochastic processes},
  author={Lawler, Gregory F},
  year={2018},
  publisher={CRC Press}
}

@article{mckee2021multi,
  title={A multi-agent reinforcement learning model of reputation and cooperation in human groups},
  author={McKee, Kevin R and Hughes, Edward and Zhu, Tina O and Chadwick, Martin J and Koster, Raphael and Castaneda, Antonio Garcia and Beattie, Charlie and Graepel, Thore and Botvinick, Matt and Leibo, Joel Z},
  journal={arXiv preprint arXiv:2103.04982},
  year={2021}
}

\end{document}